\documentclass[aps,prd,twocolumn,showpacs,amsmath,amssymb,nofootinbib,floatfix,superscriptaddress,10pt]{revtex4-2}

\usepackage{hyperref}
\usepackage{nameref}

\usepackage[dvipdfmx]{graphicx}
\usepackage{dcolumn}
\usepackage{bm}
\usepackage{comment}
\usepackage{color}
\usepackage[rgb]{xcolor}
\usepackage[normalem]{ulem}
\usepackage{booktabs}
\usepackage{multirow}
\hypersetup{colorlinks=true}

\newcommand{\mnras}{Mon.~Not.~Roy.~Astron.~Soc.}
\newcommand{\physrep}{Phys.~Rep.}
\newcommand{\jcap}{JCAP}
\newcommand{\pasj}{Publ.~Astron.~Soc.~Japan}

\newcommand{\aap}{Astronomy~\&~Astrophysics}
\newcommand{\aj}{The~Astronomical~J.}

\newcommand{\ssr}{Space~Science~Rev.}
\newcommand{\bx}{{\mathbf{x}}}
\newcommand{\by}{{\mathbf{y}}}
\newcommand{\bk}{{\mathbf{k}}}

\newcommand{\be}{{\mathbf{e}}}
\newcommand{\bq}{{\mathbf{q}}}

\newcommand{\bQ}{{\mathbf{Q}}}
\newcommand{\mybm}{{\mathbf{m}}}

\newcommand{\br}{{\mathbf{r}}}

\newcommand{\bs}{{\mathbf{s}}}
\newcommand{\bt}{{\mathbf{t}}}
\newcommand{\bu}{{\mathbf{u}}}

\newcommand{\hk}{\hat{k}}

\newcommand{\hq}{\hat{q}}

\newcommand{\hn}{\hat{n}}

\newcommand{\hz}{\hat{z}}

\newcommand{\ha}{\hat{a}}
\newcommand{\hb}{\hat{b}}

\newcommand{\hbx}{\hat{\mathbf{x}}}
\newcommand{\hby}{\hat{\mathbf{y}}}
\newcommand{\hbk}{\hat{\mathbf{k}}}

\newcommand{\hbq}{\hat{\mathbf{q}}}

\newcommand{\hbn}{\hat{\mathbf{n}}}

\newcommand{\hbz}{\hat{\mathbf{z}}}

\newcommand{\hba}{\hat{\mathbf{a}}}
\newcommand{\hbb}{\hat{\mathbf{b}}}

\newcommand{\hMpci}{h\mathrm{Mpc}^{-1}}
\newcommand{\hiMpc}{h^{-1}\mathrm{Mpc}}

\newcommand{\hiMsun}{h^{-1}M_\odot}

\newcommand{\TF}{\mathrm{TF}}

\newcommand{\delD}{\delta^\mathrm{D}}
\newcommand{\delK}{\delta^\mathrm{K}}

\newcommand{\calO}{\mathcal{O}}
\newcommand{\calN}{\mathcal{N}}
\newcommand{\calP}{\mathcal{P}}
\newcommand{\calG}{\mathcal{G}}
\newcommand{\calI}{\mathcal{I}}
\newcommand{\calJ}{\mathcal{J}}
\newcommand{\calF}{\mathcal{F}}

\newcommand{\calM}{\mathcal{M}}
\newcommand{\calL}{\mathcal{L}}
\newcommand{\calY}{\mathcal{Y}}
\newcommand{\calU}{\mathcal{U}}

\newcommand{\calR}{\mathcal{R}}

\newcommand{\eq}[1]{Eq.~(\ref{#1})}
\newcommand{\eqs}[2]{Eqs.~(\ref{#1}) and (\ref{#2})}

\newcommand{\avrg}[1]{\left\langle #1 \right\rangle}

\newcommand{\Om}{\Omega_{\mathrm{m}}}
\newcommand{\rmd}{{\mathrm{d}}}
\newcommand{\rmg}{\mathrm{g}}

\newcommand{\rmI}{\mathrm{I}}
\newcommand{\rmII}{\mathrm{II}}
\newcommand{\rmG}{\mathrm{G}}
\newcommand{\rmGG}{\mathrm{GG}}
\newcommand{\rmGI}{\mathrm{GI}}
\newcommand{\rmIG}{\mathrm{IG}}

\newcommand{\rmc}{\mathrm{c}}
\newcommand{\rmm}{\mathrm{m}}
\newcommand{\rmF}{\mathrm{F}}

\definecolor{olivegreen}{rgb}{0,0.6,0}
\definecolor{mycyan}{rgb}{0.13, 0.62, 0.8}
\definecolor{deepcarminepink}{rgb}{0.94, 0.19, 0.22}

\newcommand{\mpa}{Max-Planck-Institut f\"ur Astrophysik, Karl-Schwarzschild-Str. 1, 85748 Garching, Germany}
\newcommand{\lmu}{Ludwig-Maximilians-Universität München, Schellingstr. 4, 80799 München, Germany}
\newcommand{\ipmu}{Kavli Institute for the Physics and Mathematics of the Universe (WPI), \\
The University of Tokyo Institutes for Advanced Study (UTIAS), The University of Tokyo, Chiba 277-8583, Japan}

\begin{document}

\title{ 
Parity Violation in Galaxy Shapes: Primordial Non-Gaussianity
}

\author{Toshiki Kurita}\email{ktosh@mpa-garching.mpg.de}
\affiliation{\mpa}
\author{Drew Jamieson}
\affiliation{\mpa}
\author{Eiichiro Komatsu}
\affiliation{\mpa}
\affiliation{\lmu}
\affiliation{\ipmu}
\author{Fabian Schmidt}
\affiliation{\mpa}

\date{\today}
\begin{abstract}
We present a comprehensive study of galaxy intrinsic alignment (IA) as a probe of parity-violating primordial non-Gaussianity (PNG). 
Within the effective field theory (EFT) framework, we show that the parity-odd IA power spectrum is sensitive to the collapsed limit of the parity-odd primordial trispectrum. 
For a $U(1)$-gauge inflationary model, the IA power spectrum is proportional to the power spectrum of the curvature perturbation, $P_\zeta(k) \propto k^{-3}$. 
However, the proportionality constants contain not only the PNG amplitude but also undetermined EFT bias parameters. 
We use $N$-body simulations to determine the bias parameters for dark matter halos. 
Using these bias parameters, we forecast IA's constraining power, assuming data from the Dark Energy Spectroscopic Instrument (DESI) and the Rubin Observatory Legacy Survey of Space and Time (LSST). 
We find that the IA power spectrum can improve the current limits on the amplitude of parity-violating PNG derived from galaxy four-point correlation and CMB trispectrum analyses. 
Moreover, galaxy shapes are complementary to these probes as they are sensitive to different scales and trispectrum configurations.
Beyond galaxy shapes, we develop a new method to generate initial conditions for simulations and forward models from the parity-odd trispectrum with an enhanced collapsed limit. 
\end{abstract}
\maketitle
\tableofcontents
\section{Introduction} 

Parity symmetry is one of the fundamental symmetries of nature, but whether it holds on cosmological scales remains an open question \citep{Komatsu2022:Review_CMBpol,Shiraishi2016:PV_CMBTriSpec_U1Gauge,Cahn+2023:PVinGC}.
While parity is maximally violated in the Standard Model of elementary particles and fields through the weak interaction \citep{Lee&Yang1956:PV_weak,Wu+1957:PV_experiment}, the cosmological standard model, $\Lambda$CDM, which describes the evolution of the universe dominated by dark matter and dark energy \citep{Weinberg2008:Cosmology}, is generally assumed to be parity-conserving. 
However, if cosmic inflation \citep{Starobinsky1980:Inflation,Guth1981:Inflation,Sato1981:Inflation,Linde1982:Inflation,Albrecht&Steinhardt1982:Inflation}, dark matter, or dark energy originates from physics beyond the Standard Model, it is natural to ask whether parity symmetry might also be violated on cosmological scales. 
Probing potential parity violation thus offers a unique window into the physics of the early universe and the nature of its dark components. 

Recent observations have provided a variety of cosmological probes to search for parity violation. 
One key probe is the polarization of the cosmic microwave background (CMB) \citep{Komatsu2022:Review_CMBpol}. 
Since the CMB polarization is described as a spin-2 field on the sky, it can be decomposed into eigenstates of parity, called $E$ and $B$ modes \citep{Zaldarriaga1997:CMBpol_EB,Kamionkowski&Kosowsky1997:CMBpol_EB}.
Parity violation can be probed in statistical correlations that involve an odd number of $B$-mode components, which are parity-odd by construction.
In particular, the $EB$ power spectrum represents the lowest-order statistic sensitive to parity violation, and recent analyses have reported evidence for a non-zero signal potentially originating from cosmic birefringence \citep{Minami&Komatsu2020:Birefringence}. 
One promising class of models to explain the observed cosmic birefringence involves axion-like particles with a Chern-Simons coupling \citep{Carroll+1990:Cosmic_birefrengence,Carroll&Field1991:Cosmic_birefrengence,Harari&Sikivie1992:Cosmic_birefrengence,Carroll1998:Cosmic_birefrengence}. 

If the statistics of primordial curvature perturbations do not obey Gaussian statistics, i.e., if there is a primordial non-Gaussianity (PNG) \citep{Bartolo+2004:PNG_review}, then higher-order statistics beyond two-point functions can be used to probe a violation of parity symmetry in the primordial universe. 
Recent studies have also investigated parity violation in the CMB temperature and $E$-mode polarization trispectra \citep{Philcox&Shiraishi2024:Constraints_PV_CMBTE,Philcox2025:CMB_Trispectrum_III}. 
Although $T$ and $E$ are parity-even scalar fields, their four-point correlations can acquire parity-odd components \citep{Shiraishi2016:PV_CMBTriSpec_U1Gauge}. 
Parity-violating physics during inflation \citep[e.g.,][]{Sorbo2011:PVinCMB_pseudo} can imprint parity-odd components in the trispectrum of the primordial curvature perturbations, i.e., as a parity-violating PNG, at the leading order \citep{Shiraishi2016:PV_CMBTriSpec_U1Gauge}. 

Complementary constraints come from the large-scale structure (LSS) of the universe, where parity-odd components of the galaxy density four-point correlation function have been explored \citep{Cahn+2023:PVinGC}. 
Recent observational analyses have placed constraints on parity violation using the parity-odd galaxy four-point function \citep{Philcox2022:PV4PCF_BOSS,Hou+2023:PV4PCF_BOSS,Krolewski+2024:NoEvidence_PV_BOSS,Philcox&Ereza2025:Cov_PV_BOSS,Slepian+2025:DESIY1_PV4PCF}, with some studies reporting a detection. 

In addition to galaxy clustering, intrinsic galaxy shapes provide another observable for probing the late-time universe. 
The cosmological correlations of these intrinsic shapes are known as galaxy intrinsic alignment (IA) \citep{Croft&Metzler2000:IA_dawn,Catelan+2001:IA_dawn,Crittenden+2002:IA_dawn,Jing&Suto2002:IA_dawn}. 
Because IA can contaminate weak lensing measurements by mimicking part of the cosmic shear signal \citep{Hirata&Seljak2004:IA_LA}, much of the early work on IA focused on its role as a systematic effect in weak lensing surveys \citep[see][for a review]{Joachimi+2015:IA_review,Kiessling+2015:IA_review,Kirk+2015:IA_review,Troxel&Ishak2015:IA_review,Lamman+2024:IAGuide}. 
More recently, however, IA has also been recognized as a potential source of cosmological information, and several theoretical studies have explored how IA measurements could be used to constrain cosmological models \citep[][]{Schmidt&Jeong2012:LSSwithGW_2,Chisari&Dvorkin2013:IA_PNG,Schmidt+2015:IA_PNG,Taruya&Okumura2020:IA_improvement}. 

Since the observed galaxy ellipticities form a spin-2 field on the sky, similar to CMB polarization, the shape field can be decomposed into $E$- and $B$-mode components \citep{Crittenden+2002:Cosmic_shear_EB}. 
Under this decomposition, the galaxy shape two-point statistics are sensitive to parity. 
In particular, several studies have proposed using the $EB$ cross power spectrum of galaxy shapes as a probe of parity violation \citep{Schmidt&Jeong2012:LSSwithGW_2,Schmidt+2014:LSSwithGW_3,Biagetti&Orlando2020:PVGW_IA,Akitsu+2022:SU_sim_GW,Philcox+2024:NewPhys_GalaxyShape,Okumura&Sasaki2024:PVGW_PrjTensor,Mikura+2025:GWPol_PrjTidal}. 
These studies are based on the so-called tensor fossil effects, where primordial gravitational waves generated during the early universe modulate small-scale matter fluctuations, leaving imprints in the initial conditions of large-scale structure formation \citep{Dodelson+2003:metric_shear,Masui&Pen2010:GWFossil,Donghui&Marc2012:fossil,Schmidt&Jeong2012:LSSwithGW_2,Schmidt+2014:LSSwithGW_3,Masui+2017:PVGW_2D3D}. 
If the primordial gravitational waves are chiral, they can induce a non-zero parity-odd $EB$ power spectrum in galaxy shapes\footnote{There are also studies that aim to use galaxy spins (i.e., the orientation of galaxy angular momentum), rather than shapes, as probes of primordial parity violation, based on both simulations \citep{Yu+2020:PV_Spin_Sim,Coulton+2024:Quijote-Odd,Shim+2025:PVIC_VectorFossil} and observations \citep{Motloch+2021:PV_Spin_ELUCID,Motloch+2022a:PV_Spin_ELUCID,Motloch+2022b:PV_Spin_ELUCID}. 
In addition, constraints on cosmic birefringence using joint galaxy shape and polarization information have been proposed \citep{Yin+2025:GalaxyShapePolarization_Birefringence}. 
These approaches are closely related to, and complementary with, studies based on intrinsic galaxy shapes. }. 

In this work, by contrast, we investigate a different class of parity-violating signals in galaxy shapes. 
Rather than focusing on the tensor fossil effects associated with chiral gravitational waves, we explore parity-violating imprints in the scalar sector, specifically, signatures embedded in the statistics of curvature perturbations. 
At the lowest order, such parity-odd signals arise from the four-point correlation function, or trispectrum, of the curvature perturbation. 
We show, for the first time, how the parity-odd trispectrum of curvature perturbations can be probed through galaxy shape statistics. 
In particular, we investigate both the \textit{three-dimensional power spectrum} of projected shapes and the \textit{angular power spectrum} of projected shapes. 
The former leverages the combination of projected ellipticities observed in imaging surveys and the three-dimensional galaxy positions measured in spectroscopic surveys, thereby offering access to rich three-dimensional information beyond currently standard probes. 
The latter is a well-studied probe and major workhorse for photometric galaxy surveys, i.e., for galaxy samples without spectroscopic redshift measurements. 

The paper is structured as follows:  
In Section~\ref{sec:theory}, we derive an analytic expression for the parity-odd power spectrum of galaxy shapes based on an effective field theory (EFT) framework, and present theoretical predictions for specific models of parity-odd primordial trispectra.  
In Section~\ref{sec:nbody_sim}, we develop a new method to generate initial conditions that realize a given parity-odd primordial trispectrum, perform $N$-body simulations with these initial conditions, and confirm that the simulation results are consistent with the EFT predictions.  
In Section~\ref{sec:projection_effects}, we forecast the constraining power of IA on parity-violating models using actual galaxy survey parameters. 

To help guide the reader, we highlight two key results of this work here:  
First, as shown in Fig.~\ref{fig:EFT_fit}, the EFT predictions derived in Section~\ref{sec:theory} are in good agreement with the $N$-body simulation results from Section~\ref{sec:nbody_sim}.  
In particular, for parity-odd trispectra that are enhanced in the collapsed limit, such as the one predicted by the $U(1)$-gauge model, the IA power spectrum shows strong sensitivity to the underlying parity-violating signal.  
Second, in Figs.~\ref{fig:fisher_1d_DESI} and \ref{fig:fisher_1d_LSST}, we present forecasted $1\sigma$ constraints on the amplitude parameters of parity-odd trispectra expected from the IA signal in future observations. 
We find that the constraints on the amplitude of parity-violating PNG derived from IA measurements alone are competitive with, and in some regimes stronger than, current bounds from galaxy four-point correlation \citep{Philcox2022:PV4PCF_BOSS} and CMB trispectrum \citep{Philcox&Shiraishi2024:Constraints_PV_CMBTE} analyses.

We use the convention for the Fourier and inverse Fourier transform as
\begin{align*}
    f(\bk) \equiv \int \rmd \bx f(\bx) e^{-i\bk\cdot\bx},~
    f(\bx) \equiv \int \frac{\rmd\bk}{(2\pi)^3} f(\bk) e^{i\bk\cdot\bx},
\end{align*} 
and the abbreviations for the momentum integral:
\begin{align*}
    \int_\bk \equiv \int \frac{\rmd\bk}{(2\pi)^3},~
    \int_k \equiv \int_0^\infty \frac{k^2\rmd k}{2\pi^2},~
    \int_{\hbk} \equiv \int \frac{\rmd \Omega_{\hbk}}{4\pi}. 
\end{align*}
To save space, the notation $X_\bk\equiv X(\bk)$ and $\bk_{1\cdots n} \equiv \bk_1+\cdots+\bk_n$ will be used at times, and the time (redshift) dependence of the fields and functions will be suppressed when irrelevant. 
We also introduce the prime notation to reduce the Dirac delta for the momentum conservation: 
\begin{align*}
    (2\pi)^3\delD_{\bq_{1\cdots n}}\avrg{X_{\bq_1}\cdots X_{\bq_n}}' \equiv \avrg{X_{\bq_1}\cdots X_{\bq_n}}. 
\end{align*} 
Throughout this paper, repeated indices are implicitly summed over. 
We use the publicly available Boltzmann code \texttt{CLASS} \citep{Blas+2011:CLASS_II} to compute the linear transfer function with the Planck 2018 \citep{Planck2018_cosmo} baseline cosmology model 2.20.\footnote{\texttt{base\_plikHM\_TTTEEE\_lowl\_lowE\_lensing\_post\_BAO\_Pantheon} (\url{https://wiki.cosmos.esa.int/planck-legacy-archive/images/b/be/Baseline_params_table_2018_68pc.pdf})}

\section{Theory} 
\label{sec:theory}
In this section, we begin by introducing the primordial parity-odd trispectrum, which is the main focus of our study and will be explored through galaxy shape statistics. 
We then define the spatial distribution of galaxy shapes as a rank-two tensor field and examine key properties of the galaxy shape power spectrum  under parity transformation, demonstrating that galaxy shapes can be sensitive to parity violation even at the level of two-point statistics. 
Next, we review an EFT framework that describes large-scale correlations of galaxy shapes, treating them as a tensorial biased tracer of the underlying matter density field, $\delta$, in the late-time universe.
Within this framework, we derive an analytic expression for the parity-odd power spectrum of galaxy shapes in the presence of the parity-odd trispectrum, calculated at leading order.
Finally, we provide examples of the expected parity-odd signals based on specific parity-odd trispectrum models. 
\subsection{Recap: Parity-odd trispectrum} 
\label{subsec:recap_trisp}
If parity-violating physics existed in the early universe, its signatures would be imprinted at the lowest order in the four-point function of the curvature perturbation $\zeta$ under the assumptions of statistical homogeneity and isotropy \citep{Shiraishi2016:PV_CMBTriSpec_U1Gauge}. 
We define the primordial trispectrum of the Bardeen potential $\Phi = 3\zeta/5$ \citep{Bardeen1980:Bardeen_potential} as 
\begin{align}
    T_\Phi(\bk_1,\bk_2,\bk_3,\bk_4) 
    \equiv 
    \avrg{\Phi_{\bk_1}\Phi_{\bk_2}\Phi_{\bk_3}\Phi_{\bk_4}}_\rmc', \label{eq:def_trispectrum_Phi} 
\end{align}
The trispectrum can be decomposed into the parity-even and -odd parts, classified by their transformation properties under parity: 
\begin{align}
    T_\Phi = T_\Phi^{(+)} + T_\Phi^{(-)}, 
\end{align}
where $\mathbb{P}[T_\Phi^{(\pm)}] = \pm T_\Phi^{(\pm)}$. 
Geometrically speaking, consider a closed tetrahedron configuration guaranteed by statistical homogeneity, for which $T_\Phi^{(+)}$ gives the same amplitude for both the configuration and its mirror image (including any rotations), while $T_\Phi^{(-)}$ gives the opposite amplitude for each configuration.
Hence, $T_\Phi^{(-)}$ is zero when the tetrahedral configuration can be mapped onto its mirror image by a rotation. 
From the reality condition of $\zeta$, $\zeta^*(\bk) = \zeta(-\bk)$, the parity-even and parity-odd components are purely real and imaginary, respectively. 
Due to statistical isotropy, the degrees of freedom in the trispectrum are significantly reduced, meaning it can depend only on the four side lengths and the two diagonals. 
Moreover, the parity-odd component must be proportional to the scalar triple product constructed from three out of the four momenta \citep{Coulton+2024:Quijote-Odd}:  
\begin{align}
    T_\Phi^{(-)} 
    = i \left[\bk_1\cdot\left(\bk_2\times\bk_3\right) \right] 
    \tau_-(k_1,k_2,k_3,k_4,k_{12},k_{14}), 
    \label{eq:def_tau-}
\end{align}
where $k_{12}=|\bk_1+\bk_2|$ and $k_{14}=|\bk_1+\bk_4|$ are the two diagonals. 
The function $\tau_-$ is totally antisymmetric under permutations of momenta and can generally be decomposed as follows using a model-dependent function $f_-$:  
\begin{align}
    \tau_- 
    &= 
    \sum_{\sigma \in S_4} \mathrm{sgn}(\sigma)  \nonumber\\
    &\times
    f_-(k_{\sigma(1)},k_{\sigma(2)},k_{\sigma(3)},k_{\sigma(4)},k_{\sigma(1)\sigma(2)},k_{\sigma(1)\sigma(4)}), \label{eq:def_f-}
\end{align}
where $S_4$ is the symmetric group of degree four, and $\mathrm{sgn}(\sigma)$ is the sign of a permutation $\sigma \in S_4$.  
$k_{\sigma(1)\sigma(2)} = \left| \bk_{\sigma(1)}+\bk_{\sigma(2)} \right|$ 
and 
$k_{\sigma(1)\sigma(4)} = \left| \bk_{\sigma(1)}+\bk_{\sigma(4)} \right|$ 
represent the two independent diagonals of a tetrahedron after a permutation $\sigma$. 
Thus, specifying a parity-violating model corresponds to determining the explicit form of $f_-$ when studying the parity-odd trispectrum. 

The linear matter field is related to $\Phi$ in the matter-dominated era as
\begin{align}
    \delta^{(1)}(\bk,z) = \calM(k,z) \Phi(\bk),
\end{align}
where 
\begin{align}
    \calM(k,z) \equiv \frac{2k^2\mathcal{T}(k)D(z)}{3\Om H_0^2},
    \label{eq:def_Mk}
\end{align}
with $\mathcal{T}$ being the transfer function and $D$ being the linear growth factor normalized to the scale factor in the matter-dominated era. 
We will drop the redshift dependence on $\delta$ and $\calM$ hereafter. 
Therefore, the trispectrum of the linear matter field is written as
\begin{align}
    T_\delta(\bk_1,\bk_2,\bk_3,\bk_4) 
    &\equiv 
    \avrg{\delta_{\bk_1}^{(1)} \delta_{\bk_2}^{(1)} \delta_{\bk_3}^{(1)} \delta_{\bk_4}^{(1)}}_\rmc' \label{eq:def_trispectrum_delta}\\
    &=
    \prod_{i=1}^4\left[\calM(k_i)\right]T_\Phi(\bk_1,\bk_2,\bk_3,\bk_4). \nonumber
\end{align}

Throughout this paper, we consider two specific trispectrum models as working examples: the ``squeezed-type trispectrum'' and the ``collapsed-type trispectrum,'' as introduced below. 
We will investigate in detail how these models impact galaxy shape statistics within the EFT framework in Section~\ref{subsec:case_study}.

\subsubsection{Squeezed-type trispectrum} 
First, we employ the model proposed by Ref.~\cite{Coulton+2024:Quijote-Odd}:
\begin{align}
    f_-
    = -g_- 
    k_1^\alpha k_2^\beta k_3^\gamma 
    P_\phi(k_1) P_\phi(k_2) P_\phi(k_3), 
    \label{eq:def_f-_squeezed}
\end{align}
where $g_-$ is the amplitude parameter, and $\alpha$, $\beta$, and $\gamma$ are distinct integers satisfying $\alpha + \beta + \gamma = -3$ for scale-invariant initial conditions. 
We adopt the specific choice $(\alpha, \beta, \gamma) = (-2, -1, 0)$ as in Ref.~\cite{Coulton+2024:Quijote-Odd}.
$P_\phi(k) = \avrg{\phi_\bk \phi_{\bk'}}'$ denotes the primordial power spectrum. 
This model is motivated by its simplicity and ease of implementation in the initial conditions of numerical simulations, rather than being derived from a specific, currently existing inflationary model. 
The shape of the trispectrum resembles the well-known $g_\mathrm{NL}$-type trispectrum, as seen in the terms $P_\phi(k_1) P_\phi(k_2) P_\phi(k_3)$, and it exhibits a large amplitude in the \textit{squeezed} limit where one of the wavevectors along an edge of the tetrahedron is taken to be small \citep{Jamieson+2024:POP,Hou+2024:PVinBAO}.

\subsubsection{Collapsed-type trispectrum}
A template that can describe a specific class of parity-odd trispectra was introduced in Ref.~\cite{Shiraishi2016:PV_CMBTriSpec_U1Gauge}: 
\begin{align}
    f_-
    =
    \frac{25}{9}\sum_{n\geq0} 
    &d_n^\mathrm{odd} 
    \left[
    \calL_n\left(\mu_{13}\right)
    + (-1)^n
    \calL_n\left(\mu_1\right)
    + 
    \calL_n\left(\mu_3\right)
    \right] \nonumber\\
    &\times
    \frac{P_\phi(k_1)}{k_1}\frac{P_\phi(k_3)}{k_3} \frac{P_\phi(k_{12})}{k_{12}} , 
    \label{eq:def_dn_odd_template} 
\end{align}
with 
$\mu_{13} \equiv  \hbk_1\cdot\hbk_3$, $\mu_1 \equiv \hbk_{12}\cdot\hbk_1$, and $\mu_3 \equiv \hbk_{12}\cdot\hbk_3$. 
The sum of the three Legendre polynomials determines the angular dependence, and $d_n^\mathrm{odd}$ is the amplitude parameter for each order $n$. 
The trispectrum depends on the diagonal, and it shows the same scale dependence as the well-known $\tau_\mathrm{NL}$-type trispectrum, as seen from the form $P_\phi(k_1)P_\phi(k_3)P_\phi(k_{12})$, which is enhanced in the \textit{collapsed} limit (or the \textit{double-hard} limit), where one of the diagonal momenta of the tetrahedron, e.g., $k_{12}$, is taken to be small compared to the two edge momenta, each of which belongs to a different triangle connected by $k_{12}$, e.g., $k_1$ and $k_3$.\footnote{The collapsed limit of the parity-odd trispectrum involves taking the limit for the magnitudes of the momenta while preserving the angular information needed to distinguish the mirror image.}
We will refer to this template as the ``$d_n^\mathrm{odd}$ template'' hereafter. 

This template was originally introduced as a generalization of the parity-odd component of the trispectrum predicted by an inflationary model in which the inflaton field $\varphi$ couples to a $U(1)$ gauge field through an interaction term of the form $\mathcal{L}\supset 1/4 f(\varphi)(-F^2+\gamma F\tilde{F})$ with $\gamma\ne0$, as computed in Ref.~\cite{Shiraishi2016:PV_CMBTriSpec_U1Gauge}.  
In this model, often referred to as the ``$U(1)$-gauge model'', the parity-odd component of the trispectrum is given by  
\begin{align}
    f_-
    = - \frac{25}{3}A_\mathrm{gauge} 
    \calF\left(\mu_{13},\mu_1,\mu_3\right)
    \frac{P_\phi(k_1)}{k_1}\frac{P_\phi(k_3)}{k_3} \frac{P_\phi(k_{12})}{k_{12}}, 
    \label{eq:def_f-_collapsed}
\end{align}
where $A_\mathrm{gauge}$ is the amplitude parameter and 
\begin{align}
    \calF\left(\mu_{13},\mu_1,\mu_3\right)
    \equiv
    1 - \mu_{13} + \mu_1 - \mu_3,
    \label{eq:def_calF}
\end{align} 
is the angular function. 
We can show that the $U(1)$-gauge model corresponds to the specific case of the $d_n^\mathrm{odd}$ template (Eq.~\ref{eq:def_dn_odd_template}) where 
\begin{align}
    d_0^\mathrm{odd} = -d_1^\mathrm{odd}/3 = -A_\mathrm{gauge},~
    d_{n\geq2}^\mathrm{odd} = 0. 
    \label{eq:dn_odd_parametrization}
\end{align} 
Recent analyses of the 4PCF of galaxy clustering \citep{Philcox2022:PV4PCF_BOSS} and the trispectra of CMB temperature fluctuations and $E$-mode polarization \citep{Philcox&Shiraishi2024:Constraints_PV_CMBTE} have placed constraints on $A_\mathrm{gauge}$, $d_0^\mathrm{odd}$ and $d_1^\mathrm{odd}$.
We discuss the current constraints in detail in Section~\ref{sec:projection_effects}. 
In the following, we will focus on the $U(1)$-gauge model as an example of the collapsed-type parity-odd trispectrum.

\subsection{Definition of galaxy shapes} 
We approximate a three-dimensional shape of a galaxy or dark matter halo, which we label as ``$\rmg$'', as a triaxial ellipsoid estimated from its inertia tensor
\begin{align}
    I_{ij,\rmg} = \frac{1}{N_w}\int \rmd\br \rho_\rmg(\br) w(r) r_i r_j, 
    \label{eq:def_Iij}
\end{align} 
where $\rho_\rmg$ is a density or luminosity profile of the object and $\br$ is the position relative to the center. 
$w$ is an arbitrary radial weighting function, and the normalization factor is defined as 
$N_w \equiv \int \rmd\br \rho_\rmg(\br) w(r)$. 
Since $I_{ij}$ is symmetric, it has six degrees of freedom: one for the trace part (i.e., size of object), and five for the trace-free part (i.e., two for ellipsoidal distortions and three for orientation). 
Given a set of inertia tensors of a target sample $\{I_{ij,\rmg}\}_{\rmg=1,\cdots}$, we formally define the (density-weighted) inertia tensor \textit{field} by assigning the values at each position as 
\begin{align}
    I_{ij}(\bx) \equiv \frac{1}{\bar{n}_\rmg} \sum_\rmg I_{ij,\rmg} \delD(\bx-\bx_\rmg), 
    \label{eq:def_Iij_field}
\end{align} 
where $\bar{n}_\rmg$ is the mean number density of the sample. 
In this paper we only focus on the trace-free part because the trace part is a rotational scalar and thus parity-insensitive at the level of two point statistics as we will see later. 
We then define the (trace-free) \textit{shape field} from \eq{eq:def_Iij_field} as its dimensionless fluctuation from the ensemble average: 
\begin{align}
    S_{ij}(\bx) 
    &\equiv 
    \TF
    \left[ 
    \frac{I_{ij}(\bx) - \avrg{I_{ij}}}{\avrg{\mathrm{Tr}I}}
    \right] \\
    &= 
    \frac{1}{\avrg{\mathrm{Tr}I}}
    \left(I_{ij}(\bx) - \frac{\delK_{ij}}{3}\mathrm{Tr}I\right), 
    \label{eq:def_Sij}
\end{align} 
where ``Tr'' denotes the trace part: $\mathrm{Tr}X \equiv X_{kk}$ and we use the notation ``TF'' to extract the trace-free part: $\TF\left[ X_{ij} \right] \equiv X_{ij} - \mathrm{Tr}X\delK_{ij}/3$. 
We have used $\avrg{I_{ij}} = \avrg{\mathrm{Tr}I}\delK_{ij}/3$ due to statistical isotropy at the second equation. 

Note that we will initially focus on three-dimensional shapes in theoretical modeling (Section \ref{sec:theory}) and simulation measurements (Section \ref{sec:nbody_sim}), rather than two-dimensional shapes, which are the realistic observables when considering projections on the sky. 
This choice is motivated by the fact that IA is fundamentally a spatially three-dimensional physical phenomenon. 
The effects of projection will be addressed in detail in Section \ref{sec:projection_effects}.
\subsection{Symmetry properties of tensor spectrum} 
\label{subsec:symmetry} 
\subsubsection{Parity} 
\label{subsubsec:parity} 
Let us start by considering how statistics transform under parity transformations and deriving the consequences of parity invariance/violation without the usual assumptions of statistical homogeneity and isotropy to clarify the properties that hold under parity alone. 
The power spectrum of a shape field is defined as
\begin{align}
    \avrg{S_{ij}(\bk)S_{kl}^*(\bk')} \equiv \tilde{P}_{ij,kl}(\bk,\bk').
    \label{eq:def_Pijkl_wo_homogeneity}
\end{align}
Without assuming statistical homogeneity, the power spectrum includes off-diagonal components ($\bk \neq \bk'$). 
The tilde notation distinguishes this from the diagonal power spectrum that holds under statistical homogeneity, which we will define in \eq{eq:def_Pijkl}. 
Note that this spectrum is a complex quantity in general. 
Since the shape field $S_{ij}(\bx)$ is a real-valued tensor field, we have a reality condition in Fourier space: $S_{ij}^*(\bk)=S_{ij}(-\bk)$. 
Therefore, by taking the complex conjugate of \eq{eq:def_Pijkl_wo_homogeneity}, we have
\begin{align}
    \tilde{P}_{ij,kl}^*(\bk,\bk') = \tilde{P}_{ij,kl}(-\bk,-\bk'). 
    \label{eq:reality_condition_wo_homogeneity}
\end{align}

We next consider the parity transformation. 
Since the shape tensor is a tensor of rank two by definition in \eq{eq:def_Iij}, it transforms as $\mathbb{P}[S_{ij}(\bk)]=S_{ij}(-\bk)$, which defines the transformation of the power spectrum as
\begin{align}
    \mathbb{P}\left[\tilde{P}_{ij,kl}(\bk,\bk')\right] = \tilde{P}_{ij,kl}(-\bk,-\bk'). 
    \label{eq:def_Pijkl_parity_trs}
\end{align}
With the reality condition \eq{eq:reality_condition_wo_homogeneity}, we find that applying a parity transformation to the power spectrum is equivalent to taking its complex conjugate. 
If the power spectrum is parity even, i.e., $\mathbb{P}[\tilde{P}_{ij,kl}] = \tilde{P}_{ij,kl}$, then combined \eq{eq:reality_condition_wo_homogeneity}, we have $\tilde{P}_{ij,kl}^*=\tilde{P}_{ij,kl}$, which implies $\tilde{P}_{ij,kl}$ is purely real. 
In other words, the parity-violating signal, corresponding to $\mathbb{P}[\tilde{P}_{ij,kl}] \neq \tilde{P}_{ij,kl}$, appears in the \textit{imaginary} part of the power spectrum. 
To emphasize, the properties of the real and imaginary parts of the auto power spectrum under parity hold independently of statistical homogeneity and isotropy for any real-valued tensor field of any rank. 

\subsubsection{Homogeneity} 
\label{subsubsec:homogeneity} 
In order to extract the diagonal components, non-zero under statistical homogeneity, we begin by considering the transpose, which introduces the momentum exchange identity. 
The exchange of two momenta in \eq{eq:def_Pijkl_wo_homogeneity}, $\bk\leftrightarrow\bk'$, leads to the following relation associated with the exchange of the index pairs $(i,j)$ and $(k,l)$\footnote{Since $S_{ij}$  is symmetric and traceless, it inherits symmetries under the exchange of $i(k)$ and $j(l)$: 
$P_{ji,kl}=P_{ij,lk}=P_{ji,lk}=P_{ij,kl}$, as well as traceless conditions: $P_{ii,kl}=P_{ij,kk}=0$. 
This symmetry and tracelessness are intrinsic properties of the observables of interest (i.e., distortion of galaxy or halo shapes), and are independent of statistical symmetries such as parity, homogeneity, or isotropy.}:
\begin{align}
    \tilde{P}_{kl,ij}(\bk,\bk') 
    = \tilde{P}_{ij,kl}^*(\bk',\bk). 
    \label{eq:exchange_major_wo_homogeneity}
\end{align} 
Particularly for the diagonal components ($\bk=\bk'$), from this and the reality condition \eq{eq:reality_condition_wo_homogeneity}, we obtain the equivalence between taking the complex conjugate and swapping the index pairs: 
$\tilde{P}_{ij,kl}^*=\tilde{P}_{kl,ij}$ for $\bk=\bk'$. 

Under statistical homogeneity, we define the diagonal power spectrum due to the translational invariance as $P_{ij,kl}$,
\begin{align}
    \avrg{S_{ij}(\bk)S_{kl}(\bk')} 
    &\equiv (2\pi)^3 \delD_{\bk+\bk'} P_{ij,kl}(\bk)\label{eq:def_Pijkl} \\
    &= \tilde{P}_{ij,kl}(\bk,-\bk')\notag, 
\end{align}
where the second equality holds from the reality condition of $S_{ij}$. 
We immediately obtain $P_{ij,kl}^*=P_{kl,ij}$ from the discussion above.
Therefore, considering parity transformation, we find that the parity-even condition for the diagonal power spectrum, $\mathbb{P}[P_{ij,kl}] = P_{ij,kl}$, corresponds to the symmetry of the index pairs: $P_{kl,ij}=P_{ij,kl}$ along with its reality: $P_{ij,kl}^*=P_{ij,kl}$. 

In this work, on the other hand, we will focus on the parity-violating signal, 
$\mathbb{P}\left[P_{ij,kl}\right] = -P_{ij,kl}$. 
In this case, the parity-odd component appears in the \textit{imaginary} part of the power spectrum, which is equivalent to appearing in the \textit{antisymmetric} part under the exchange of indices, $(i,j)$ and $(k,l)$: $P_{kl,ij} = -P_{ij,kl}$\footnote{In general, one could also have a mixed-parity spectrum, where both symmetric and antisymmetric components coexist, so that the spectrum is neither purely parity-even nor purely parity-odd. }. 
Note that, since scalar fields (rank-zero tensors) lack tensorial structure, their power spectrum is always real, making them insensitive to parity violation at the level of two-point statistics.
In contrast, for any tensor with rank greater than zero, parity-violating signatures can be imprinted in the antisymmetric components of the power spectrum (or two-point correlation function).
\subsubsection{Isotropy} 
\label{subsubsec:isotropy} 
The Cartesian expression of the power spectrum, $P_{ij,kl}$, involves many indices and appears cumbersome to handle. 
To simplify the treatment, we introduce the helicity decomposition following Ref.~\cite{Vlah+2020:IA_EFT}, an efficient mathematical framework for isolating independent degrees of freedom under the assumption of statistical isotropy. 
Let $S_{ij}$ be a real, symmetric, trace-free tensor field. 
In Fourier space, its components can be decomposed into five orthogonal states 
\begin{align}
    S_{ij}(\bk) \equiv \sum_{m=-2}^{2} S_m(\bk) \calY_{m,ij}(\hbk), 
    \label{eq:helisity_decomposition}
\end{align}
where $\calY_{m}$ $(m=0,\pm 1,\pm 2)$ are the basis tensors of rank two with helicity $m$, defined as 
\begin{align}
    \calY_{0,ij}(\hbk) &= \sqrt{\frac{3}{2}} \left(\hk_i\hk_j-\frac{\delK_{ij}}{3}\right),~ \label{eq:Yij_0}\\
    \calY_{\pm1,ij}(\hbk) &= \sqrt{\frac{1}{2}} \left(\hk_ie_{\pm,j} + \hk_je_{\pm,i}\right),~ \label{eq:Yij_pm1}\\ 
    \calY_{\pm2,ij}(\hbk) &= e_{\pm,i} e_{\pm,j},~\label{eq:Yij_pm2}
\end{align}
with $e_{\pm,i}$ being a complex unit vector defined for each mode $\bk$ as
\begin{align}
    e_{\pm,i} (\hbk)
    \equiv 
    \mp\frac{1}{\sqrt{2}}
    \left( \ha_i \pm i\hb_i \right), 
    \label{eq:def_epm}
\end{align}
where the basis $\{\hba,\hbb,\hbk\}$ forms a right-handed orthonormal coordinate system. 
We adopt the polar coordinate basis $\{\hba, \hbb, \hbk\} = \{\be_\theta(\hbk), \be_\phi(\hbk), \hbk\}$\footnote{It is not necessary to explicitly specify the directions of the basis vectors $\{\hba, \hbb\}$ on the plane perpendicular to $\hbk$ on the theoretical side. 
However, on the simulation side, when implementing helicity explicitly, we need to define these directions. 
The orthonormality relation is 
$\calY_m^* \cdot \calY_{m'}=\delK_{mm'}$ 
with the complex conjugate being
$\calY_m^*(\hbk) = (-1)^{m} \calY_{-m}(\hbk) = \calY_m(-\hbk)$, where the second equation follows from 
$e_{\pm,i}(-\hbk) = -e_{\mp,i}(\hbk) = e_{\pm,i}^*(\hbk)$.}. 
In particular, when $\hbk=\hbz$, this reproduces the cartesian basis 
$\{\hbx, \hby, \hbz\}$. 
Hereafter we will omit the dependence on $\hbk$ in $e_{\pm,i}$ when irrelevant. 
Since the helicity components in \eq{eq:helisity_decomposition} transform as 
$S_m \rightarrow e^{-im\psi} S_m$ 
under a rotation by an angle $\psi$ around $\hbk$, we have the helicity spectra under statistical isotropy (and homogeneity): 
\begin{align}
    \avrg{S_m(\bk)S_{m'}(\bk')} \equiv (2\pi)^3 \delD_{\bk+\bk'} \delK_{mm'} P_m(k). 
    \label{eq:def_Pm_helicity}
\end{align}
The reality condition of $S$ yields $P_m^*=P_{-m}$. 
Thus, helicity modes only correlate with modes of the same helicity under isotropy, and there are five degrees of freedom in the power spectrum of a real, symmetric, trace-free tensor. 
From \eq{eq:def_Pijkl} and \eq{eq:def_Pm_helicity}, we isolate the independent helicity components from the cartesian components of the power spectrum as
\begin{align}
    P_m(k) = \left[ \calY_{m,ij}(\hbk) \calY_{m,kl}(-\hbk)\right]^* P_{ij,kl}(\bk). 
\end{align}

So far, we have not used the properties related to parity transformation. 
As mentioned before, the parity-even (odd) component correspond to the symmetric (antisymmetric) part of the cartesian components with respect to the index pairs, $(i,j)$ and $(k,l)$. 
By further symmetrizing (and antisymmetrizing) the basis tensors, we extract parity-even $(+)$ and -odd $(-)$ components separately as 
\begin{align}
    P_\pm^{(\lambda)}(k) 
    &\equiv \left[\Lambda^{(\lambda,\pm)}_{ij,kl}(\hbk)\right]^* P_{ij,kl}(\bk), \label{eq:def_projection} \\
    &= \frac{1}{2}\left( P_m(k) \pm P_{-m}(k)\right) 
\end{align}
with $\lambda=|m|$ $(\lambda=0,1,2)$, where the projection tensors are defined by 
\begin{align}
    \Lambda^{(\lambda,\pm)}_{ij,kl}(\hbk) 
    \equiv 
    \frac{1}{2}\left[ \calY_{\lambda,ij}(\hbk) \calY_{\lambda,kl}(-\hbk) \pm \calY_{\lambda,ij}(-\hbk) \calY_{\lambda,kl}(\hbk) \right], 
    \label{eq:def_projection_tensors}
\end{align}
satisfying the orthogonal relation: 
$\Lambda^{(\lambda,s)*} \cdot \Lambda^{(\lambda',s')} = \mathcal{N}_\lambda\delK_{\lambda\lambda'}\delK_{ss'}$ 
with the normalization factor
\begin{align}
    \mathcal{N}_\lambda \equiv \frac{1+\delK_{\lambda0}}{2}, 
    \label{eq:N_lambda}
\end{align}
and $s\in\{\pm\}$. 
By explicitly writing these out, we obtain the following: 
\begin{align}
    \Lambda^{(0,+)}_{ij,kl} &= \frac{3}{2} \left( \hk_i\hk_j - \frac{1}{3}\delK_{ij} \right) \left( \hk_k\hk_l - \frac{1}{3}\delK_{kl} \right), \label{eq:Lambda0_def}\\
    \Lambda^{(1,+)}_{ij,kl} &= \frac{1}{4} \left( \mathcal{P}_{ik}\hk_j\hk_l + \mathcal{P}_{il}\hk_j\hk_k + \mathcal{P}_{jk}\hk_i\hk_l + \mathcal{P}_{jl}\hk_i\hk_k \right), \label{eq:Lambda1+_def}\\
    \Lambda^{(2,+)}_{ij,kl} &= \frac{1}{4} \left( \mathcal{P}_{ik}\mathcal{P}_{jl} + \mathcal{P}_{il}\mathcal{P}_{jk} - \mathcal{P}_{ij}\mathcal{P}_{kl} \right), 
    \label{eq:Lambda2+_def}
\end{align} 
for the parity-even projectors and 
\begin{align}
    \Lambda^{(1,-)}_{ij,kl} 
    &= -\frac{i}{4} 
    \bigl(
    \hk_i\hk_k\varepsilon_{jlm}\hk_m +\hk_i\hk_l\varepsilon_{jkm}\hk_m \nonumber\\
    &\quad\quad\quad\quad+\hk_j\hk_k\varepsilon_{ilm}\hk_m +\hk_j\hk_l\varepsilon_{ikm}\hk_m 
    \bigr), 
    \label{eq:Lambda1-_def} \\
    \Lambda^{(2,-)}_{ij,kl}
    &= -\frac{i}{4} \left( \varepsilon_{ikm}\hk_m \mathcal{P}_{jl} + \varepsilon_{jlm}\hk_m \mathcal{P}_{ik} \right),
    \label{eq:Lambda2-_def}
\end{align} 
for the parity-odd projectors. 
We have defined the usual projector onto the plane normal to $\bk$ as 
$\mathcal{P}_{ij} \equiv \delK_{ij} - \hk_i\hk_j$. 
To derive these expressions, we have used the relation: 
$e_{+,i} e_{-,j} = - \mathcal{P}_{ij}/2 + i\varepsilon_{ijm}\hk_m/2$. 
In our convention, the original Cartesian tensor power spectrum is expanded as $P_{ij,kl} = \sum_{\lambda,s} \calN_{\lambda}^{-1} \Lambda^{(\lambda,s)}_{ij,kl} P_s^{(\lambda)}$. 

Note that $\Lambda^{(\lambda,s)}$ is, by construction, consistent with statistical isotropy, as it is built from the isotropic tensors $\delK_{ij}$ and $\varepsilon_{ijk}$ together with the unit wavevector $\hk_i$. 
In particular, we observe that the parity-odd projectors contain an odd number of $i\varepsilon_{ijk}$, which guarantees that they are purely imaginary and equivalently carry the antisymmetric structure as expected. 
Another interpretation is that this projection extracts the scalar coefficients of independent \textit{pseudotensors} under statistical isotropy (Eqs.~\ref{eq:Lambda1-_def} and \ref{eq:Lambda2-_def}). 
The fact that the parity-odd projectors are pseudotensors follows directly from the fact that $\varepsilon_{ijk}$ is a pseudotensor. 

Now, the five helicity components are rearranged into three parity-even and two parity-odd components, with the latter defined as the difference between the helicity power spectra of $\pm m$ ($m\ne0$), i.e., the difference between the power spectra of right- and left-handed helical modes. 
It is worth mentioning that the mode with helicity-$0$, i.e., longitudinal scalar mode, is not sensitive to parity.

The parity-odd projector, $\Lambda^{(\lambda,-)}$, is a key mathematical quantity in this paper, as we will extract the parity-odd components in the galaxy/halo shape power spectrum, $P_-^{(\lambda)}$, from its cartesian expression, $P_{ij,kl}$, by projecting onto this basis both in theoretical calculation and in measurements from simulations.

\subsection{Effective field theory for galaxy shapes} 
\label{subsec:eft_galaxy_shape} 
Up to this point, we have derived the mathematical properties of galaxy shape statistics based on symmetry considerations. 
Moving forward, we will develop a physical model for them. 
In this paper, we adopt an EFT for galaxy shapes \citep{Vlah+2020:IA_EFT}. 
We first briefly review the basics of EFT, then extend the framework to the cases of parity-violating initial conditions. 
\subsubsection{Bias expansion} 
\label{subsubsec:bias_expansion} 
In the EFT framework, a shape tensor field is schematically written as: 
\begin{align} 
    S_{ij} = S_{ij}^\mathrm{local} + S_{ij}^\mathrm{h.d.} + S_{ij}^\mathrm{stoch.}, 
    \label{eq:def_eft_expansion} 
\end{align} 
where 
$S_{ij}^\mathrm{local}$ represents the deterministic terms expressed by local gravitational operators, i.e., at lowest order in derivatives, 
$S_{ij}^\mathrm{h.d.}$ accounts for the higher-derivative terms arising from the nonlocality of galaxy formation, 
and $S_{ij}^\mathrm{stoch.}$ represents the non-deterministic terms from small-scale physics and the discrete nature of galaxies. 

The perturbative bias expansion of the local component, 
$S_{ij}^\mathrm{local} = \sum_{n=1}^\infty S_{ij}^{(n)} $, 
is expressed by an operator basis as 
\begin{align}
    S_{ij}^{(n)} 
    = \sum_{q=1}^{n} \sum_{p=1}^{N_q} c_{q,p} \TF\left[\calO_{ij}^{(q,p)}\right]^{(n)}, 
    \label{eq:def_bias_expansion} 
\end{align}
where $n$ denotes the perturbative order, and the index pair $(q,p)$ specifies the linearly independent operators.
Here $q$ labels the order of the bias operator (sometimes referred to as the generation, not to be confused with the perturbative order $n$), and $p$ enumerates the $N_q$ independent operators of a given order $q$. 
$c_{q,p}$ is the corresponding bias coefficient. 
For example, there are three independent quadratic ($q=2$) operators for a rank-two tensor field, as listed below \citep[see][for details]{Vlah+2020:IA_EFT}. 
Note that the triplet $(n,q,p)$ uniquely determines the kernels of the operators at $n$-th order in perturbation. 
We consider the perturbative expansion up to $n=3$, i.e., terms up to the order of $(\delta^{(1)})^4$ in the statistics. 
This order is required to probe parity violation, as it constitutes the lowest-order scalar statistics sensitive to parity. 
At this order, the leading-order parity-odd \textit{power spectrum} sourced by the primordial parity-odd \textit{trispectrum} appears, as we will show below. 

The first few operators, $\calO_{ij}^{(q,p)}$, are given by 
\begin{align}
    \calO_{ij}^{(1,1)} &= \Pi^{[1]}_{ij},\label{eq:Oij_11}\\
    \calO_{ij}^{(2,1)} &= \Pi^{[2]}_{ij},\label{eq:Oij_21}\\
    \calO_{ij}^{(2,2)} &= \Pi^{[1]}_{ik}\Pi^{[1]}_{kj},\label{eq:Oij_22}\\
    \calO_{ij}^{(2,3)} &= \Pi^{[1]}_{ij}\delta,\label{eq:Oij_23}
\end{align}
where $\Pi_{ij}^{[1]}\equiv\partial^{-2}\partial_i\partial_j\delta$ is the rescaled Hessian matrix of the gravitational potential satisfying $\mathrm{Tr}[\Pi^{[1]}]=\delta$ via the Poisson equation. 
Notice that the first line ($q=1$) \emph{starts} at first order in perturbations, but receives higher-order (second-order, third-order, ...) corrections due to the nonlinear evolution of matter. Similarly, the following lines ($q=2$) \emph{start} at second order, with third- and higher-order contributions. 
When evaluated at first order in perturbation, the trace-free part of the first-order operator (Eq.~\ref{eq:Oij_11}) thus represents the linear tidal field, which corresponds to the ``linear alignment'' model \citep{Hirata&Seljak2004:IA_LA}. 
$\Pi_{ij}^{[q]}~(q>1)$ is the higher generation operator recursively defined by the convective time derivative as \citep{Mirbabayi:2015,Desjacques+2018:bias_review}:
\begin{align}
    \Pi_{ij}^{[q]} = \frac{1}{(q-1)!}\left[(\mathcal{H}f)^{-1}\frac{D}{D\tau}\Pi_{ij}^{[q-1]} - (q-1)\Pi_{ij}^{[q-1]}\right], 
\end{align}
with the conformal Hubble parameter $\mathcal{H}$, the logarithmic growth rate $f$, and the convective time derivative: 
$D/D\tau \equiv \partial/\partial\tau + v^i \partial_i$. 
The physical interpretations of the second-order effects of the four operators in Eqs.~\eqref{eq:Oij_21}--\eqref{eq:Oij_23} are, briefly, dependence on the formation history of the tidal field, tidal torquing, and density weighting, respectively.

We define the Fourier-space kernels of the operators at the $n$-th order as 
\begin{align}
    &\left[\calO_{ij}^{(q,p)}\right]^{(n)}(\bk)
    = 
    \int_{\bq_1,\cdots,\bq_n} (2\pi)^3\delD_{\bk-\bq_{1\cdots n}} \nonumber\\
    &\quad\quad\quad\quad\times \left[K_{ij}^{\calO^{(q,p)}}\right]^{(n)}(\bq_1,\cdots,\bq_n) \delta_{\bq_1}^{(1)}\cdots\delta_{\bq_n}^{(1)}, 
    \label{eq:def_kernels} 
\end{align}
where their explicit forms are given by (up to the second order) 
\begin{align}
    \left[K_{ij}^{\calO^{(1,1)}}\right]^{(1)} &= \hq_i\hq_j, 
    \label{eq:kernel_nqp_111} \\
    \left[K_{ij}^{\calO^{(1,1)}}\right]^{(2)} &= F_2(\bq_1,\bq_2) \hq_{12,i}\hq_{12,j}, 
    \label{eq:kernel_nqp_211} \\
    \left[K_{ij}^{\calO^{(2,1)}}\right]^{(2)} &= \frac{5}{7}(1-\mu_{12}^2) \hq_{12,i}\hq_{12,j} + \mu_{12} \hq_{1,(i}\hq_{2,j)}, 
    \label{eq:kernel_nqp_221} \\
    \left[K_{ij}^{\calO^{(2,2)}}\right]^{(2)} &= \mu_{12} \hq_{1,(i}\hq_{2,j)}, 
    \label{eq:kernel_nqp_222} \\
    \left[K_{ij}^{\calO^{(2,3)}}\right]^{(2)} &= \frac{1}{2}\left( \hq_{1,i}\hq_{1,j} + \hq_{2,i}\hq_{2,j} \right), 
    \label{eq:kernel_nqp_223} 
\end{align}
with 
$F_2$ being the second-order gravitational evolution kernel in the standard perturbation theory \citep{Bernardeau+2002:PT_review}, 
$\mu_{mn} \equiv \hbq_m\cdot\hbq_n$ being the cosine of angle between two momenta, and 
we have defined the notation for symmetrization: $x_{(i}y_{j)} \equiv (x_iy_j+x_jy_i)/2$. 
We omit the full expressions for the third-order terms $S_{ij}^{(3)}$ because, as we will see shortly, the first- and third-order terms do not contribute to the parity-odd signals at the leading order due to the symmetries. 

So far, we have used a bias expansion assuming that galaxy and halo formation is spatially local. 
However, since gravitational collapse occurs over a finite region, we need to account for nonlocal effects at certain scales of interest \citep{Matsubara1999:HD_nonlocal_bias,Coles&Erdogdu2007:HD_nonlocal_bias}. 
These effects can be captured by including higher derivative operators. 
Up to the third order, only one operator is relevant:
\begin{align}
    \calO_{ij}^{(\mathrm{h.d.})} = R_*^2\partial^2\calO_{ij}^{(1,1)}, \label{eq:Oij_hd}
\end{align}
with $R_*$ being a typical halo formation scale. 
This term is suppressed by a factor of $\mathcal{O}(R_*^2 k^2)$ relative to the linear operator $\mathcal{O}_{ij}^{(1,1)}$, effectively counting as a third-order operator because the additional $k^2$ dependence becomes degenerate with the counterterm that accounts for backreaction from small-scale matter density fluctuations in the EFT of large-scale structure (EFTofLSS) at this order \citep[see, e.g.,][for details]{Baumann+2012:EFTofLSS}. 

In summary, the power spectrum of a shape field (Eq.~\ref{eq:def_Pijkl}) is perturbatively expanded up to the order of $(\delta^{(1)})^3$ (excluding stochastic terms for now): 
\begin{align}
     P_{ij,kl} =  P_{ij,kl}^{(11)} + P_{ij,kl}^{(22)} + P_{ij,kl}^{(13)} + P_{ij,kl}^{(1\,\mathrm{h.d.})} + P_{ij,kl}^{(\mathrm{ctr.})}, \label{eq:Pijkl_expansion}
\end{align}
where the last term represents appropriate counterterms from the loop integrals. 
Note that the higher-derivative term shares the same angular structure as the tree-level spectrum by definition in \eq{eq:Oij_hd}, differing only in its $k^2$-dependence. 
Furthermore, we extract the independent helicity components by projecting onto the basis tensors, as defined in \eq{eq:def_projection}. 

In Ref.\cite{Vlah+2020:IA_EFT}, these next-to-leading order corrections (including stochastic contributions) were derived for Gaussian initial conditions \citep[see also][]{Bakx+2023:EFTofIAvsSims,Chen&Kokron2024:LEFTofIA}. 
In this case, all the information is contained in the primordial power spectrum, which is parity-even, as mentioned in Section~\ref{subsec:symmetry}.
Since gravitational evolution preserves parity, any parity-odd signal remains zero. 
Thus, the only parity-even components projected onto $\Lambda^{(\lambda,+)}$ ($\lambda=0,1,2$) defined in Eqs.~\eqref{eq:Lambda0_def}--\eqref{eq:Lambda2+_def} are non-zero for Gaussian initial conditions. 
In this work, on the other hand, we focus on non-Gaussian initial conditions, particularly those that violate parity.
The corresponding signals appear in the parity-odd components obtained by projecting onto $\Lambda^{(\lambda,-)}$ ($\lambda=1,2$) defined in \eqs{eq:Lambda1-_def}{eq:Lambda2-_def}.

\begin{figure}
    \centering
    \includegraphics[width=0.95\columnwidth]{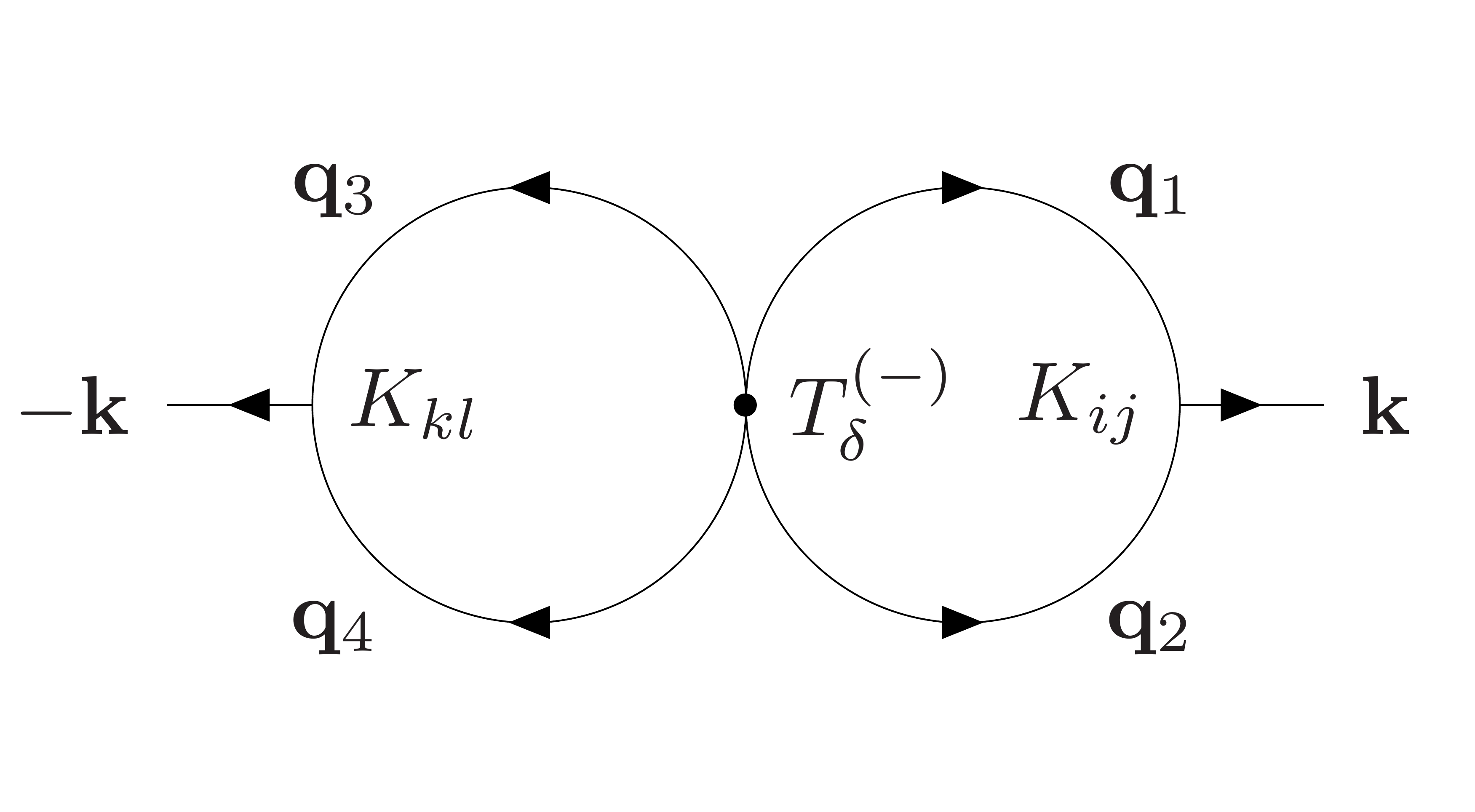}
    \caption{A diagram of the parity-odd power spectrum sourced by the parity-odd trispectrum. 
    }
    \label{fig:diagram_P22}
\end{figure}
\subsubsection{Parity-odd power spectrum} 
\label{subsubsec:pop_spectrum} 
Here, we present a useful fact that can simplify future calculations for parity-odd signals. 
Recall that the first-generation operator (Eq.~\ref{eq:Oij_11}) is expressed in Fourier space as
\begin{align}
    \calO_{ij}^{(1,1)}(\bk) = \hk_i\hk_j \delta(\bk). 
    \label{eq:Oij_11_Fourier}
\end{align}
This represents a pure longitudinal scalar mode \emph{at any order in perturbations}, i.e. helicity-$0$ mode in terms of the helicity basis (Eqs.~\ref{eq:Yij_0}--\ref{eq:Yij_pm2}). 
From \eq{eq:def_Pm_helicity}, the two-point correlations involving this operator contribute only to a helicity-$0$ power spectrum under statistical isotropy, which is parity-insensitive. 
Therefore, $P^{(11)}$, $P^{(13)}$ and $P^{(1\,\mathrm{h.d.})}$ in \eq{eq:Pijkl_expansion} are all parity-insensitive\footnote{$P_{ij,kl}^{(1n)}$ ($n\geq1$) is parity-insensitive in general.}. 
Thus, we only need to focus on $P^{(22)}$, which consists of the following operator spectra with the loop integral: 
\begin{align}
    P_-^{(\lambda)}
    &\supset 
    c_{q,p}c_{q',p'} P_{\calO^{(q,p)}\calO^{(q',p')}}^{(\lambda,-)}, 
\end{align}
with 
\begin{align}
    &P_{\calO\calO'}^{(\lambda,-)}
    \equiv 
    \int_{\bq_1,\cdots,\bq_4} (2\pi)^3\delD_{\bk-\bq_{12}} (2\pi)^3\delD_{\bk+\bq_{34}} \nonumber\\
    &\quad\quad\times
    K_{\calO\calO'}^{(\lambda,-)}(\bq_1,\bq_2,\bq_3,\bq_4)
    T_\delta^{(-)} (\bq_1,\bq_2,\bq_3,\bq_4), 
    \label{eq:P-_22}
\end{align}
where $T_\delta^{(-)}$ is the underlying parity-odd matter trispectrum (Eq.~\ref{eq:def_trispectrum_delta}), and we have defined the projected kernel: 
\begin{align}
    &K_{\calO\calO'}^{(\lambda,s)}(\bq_1,\bq_2,\bq_3,\bq_4) \nonumber\\
    &\equiv 
    \left[\Lambda^{(\lambda,s)}_{ij,kl}(\hbk)\right]^* 
    \Bigl[K_{ij}^{\calO}\Bigr]^{(2)}(\bq_1,\bq_2)
    \Bigl[K_{kl}^{\calO'}\Bigr]^{(2)}(\bq_3,\bq_4). 
    \label{eq:projected_kernels}
\end{align}
We show a diagrammatic representation of the parity-odd power spectrum sourced by the parity-odd trispectrum through the (22)-type loop integral in Fig.~\ref{fig:diagram_P22}. 
Note that when deriving \eq{eq:P-_22}, we used the fact that the parity-odd projection is orthogonal to the parity-even statistics (such as the Gaussian part and parity-even non-Gaussian components) contained in the 4th-order moment of $\delta^{(1)}$, i.e., they vanish after the loop integration due to their transformation properties under parity $\bq_i\rightarrow -\bq_i$.

We can further simplify $P^{(22)}$ by using the degeneracies among the second-order kernels (Eqs.~\ref{eq:kernel_nqp_211}--\ref{eq:kernel_nqp_223}) derived in Ref.~\citep{Vlah+2020:IA_EFT}. 
The detailed calculations are provided in Appendix~\ref{app:operator_degeneracy}. Briefly, the helicity-1 component of the parity-odd spectrum depends only on the combination $c_{2,1} + c_{2,2} + c_{2,3}$, while the helicity-2 component depends only on $c_{2,1} + c_{2,2}$ and $c_{2,3}$ as independent bias coefficients.  
Therefore, by appropriately redefining the linear combinations of the second-generation operators, we obtain the following simplified expressions for each helicity component: 
\begin{align}
    P_-^{(1)} &= c_Q^2 P_{QQ}^{(1,-)}, \label{eq:pop_spectrum_lambda1}\\
    P_-^{(2)} &= c_Q^2 P_{QQ}^{(2,-)} + 2c_Qc_R P_{(QR)}^{(2,-)} + c_R^2 P_{RR}^{(2,-)}, \label{eq:pop_spectrum_lambda2}
\end{align}
where $c_Q\equiv c_{2,1} + c_{2,2} + c_{2,3}$, $c_R\equiv c_{2,3}$, and we have introduced the new operator labels: 
\begin{align}
    Q_{ij}\equiv \calO_{ij}^{(2,2)},~ 
    R_{ij}\equiv \calO_{ij}^{(2,3)} - \calO_{ij}^{(2,2)}. 
    \label{eq:def_Qij_Rij}
\end{align} 
The explicit forms of the corresponding Fourier-space kernels are given by: for helicity $\lambda=1$,
\begin{widetext}
\begin{align}
    K_{QQ}^{(1,-)}(\bq_1,\bq_2,\bq_3,\bq_4)
    &= \frac{i}{4} \left[\hbk\cdot(\hbq_1\times\hbq_3) \right] 
    \left[ \mu_{k1} - \frac{q_1}{q_2}\mu_{k2} \right]
    \left[ \mu_{k3} - \frac{q_3}{q_4}\mu_{k4} \right], 
    \label{eq:prj_kernel_lambda1_QQ}
\end{align}
and for $\lambda=2$, 
\begin{align}
    K_{QQ}^{(2,-)}(\bq_1,\bq_2,\bq_3,\bq_4)
    &= \frac{i}{2} \left[\hbk\cdot(\hbq_1\times\hbq_3) \right] 
    (\mu_{13} - \mu_{k1}\mu_{k3})
    \frac{q_1}{q_2}\mu_{12}
    \frac{q_3}{q_4}\mu_{34}, \label{eq:prj_kernel_lambda2_QQ}\\
    K_{(QR)}^{(2,-)}(\bq_1,\bq_2,\bq_3,\bq_4)
    &= -\frac{i}{8} \left[\hbk\cdot(\hbq_1\times\hbq_3) \right] 
    (\mu_{13} - \mu_{k1}\mu_{k3})
    \frac{k^2}{q_2q_4} \left[\frac{q_1}{q_4}\mu_{12} + \frac{q_3}{q_2}\mu_{34}\right], \label{eq:prj_kernel_lambda2_QR}\\
    K_{RR}^{(2,-)}(\bq_1,\bq_2,\bq_3,\bq_4)
    &= \frac{i}{8} \left[\hbk\cdot(\hbq_1\times\hbq_3) \right] 
    (\mu_{13} - \mu_{k1}\mu_{k3})
    \frac{k^4}{q_2^2 q_4^2}, \label{eq:prj_kernel_lambda2_RR}
\end{align}
\end{widetext}
where $\mu_{kn} \equiv \hbk\cdot\hbq_n$ and the cross term have already been symmetrized: 
$K_{(QR)} \equiv (K_{QR}+K_{RQ})/2$. 
The scalar triple product (and imaginary unit) common to the kernels
combines with the analogous term in the parity-odd trispectrum in \eq{eq:def_tau-}, leading to a nonzero parity-odd signal. 

The angular structure in these kernels determines how the second-order gravitational effects on galaxy shapes capture the underlying parity-violating signal. 
Moreover, according to the two-loop configuration shown in Fig.~\ref{fig:diagram_P22}, we expect that the loop integral over the two internal momenta is dominated by hard modes of the integrand. 
Since the wave vector $\bk$ of interest corresponds to one of the diagonals of the formed tetrahedron, this soft-limit configuration, $k \ll q_1,q_3$, is referred to as the \textit{double-hard} limit or the \textit{collapsed} limit.
We thus expect that the parity-odd spectrum of galaxy shapes is enhanced if the primordial trispectrum has a large amplitude in the collapsed limit. 
The calculations assuming specific models are presented in Section~\ref{subsec:case_study}. 
More generally, the sensitivity of galaxy shapes to parity-violating physics is model-dependent and is essentially determined by the similarity (or \textit{cosine}) between the parity-odd kernels and the trispectrum.

\subsubsection{Stochasticity} 
\label{subsubsec:stochasticity}

We now consider the stochastic term $S_{ij}^{\text{stoch.}}$, which constitutes the final component included in the EFT expansion (Eq.~\ref{eq:def_eft_expansion}).  
This term is necessary to describe the stochastic effects of small-scale fluctuations, as galaxy shapes are influenced by both large-scale perturbations (the deterministic part) and small-scale fluctuations. 
In this paper, we focus only on the leading term, which is represented by adding a stochastic operator $\epsilon_{ij}(\bx)$ to the bias expansion, uncorrelated with the large-scale perturbations: 
\begin{align}
    S_{ij} \supset \epsilon_{ij},
\end{align}
where $\avrg{\calO_{ij}\epsilon_{kl}}=0$ for any deterministic operators $\calO_{ij}$. 
We define the power spectrum of this field as 
\begin{align}
    \avrg{\epsilon_{ij}(\bk)\epsilon_{kl}(\bk')} &\equiv (2\pi)^3 \delD_{\bk+\bk'} N_{ij,kl}(\bk). 
    \label{eq:def_Nijkl}
\end{align} 
Our objective here is to determine the general form of the parity-odd components in this noise power spectrum. 
The parity-even components have already been derived up to leading and next-to-leading order (in derivatives) in Refs.~\cite{Vlah+2020:IA_EFT} and \cite{Chen&Kokron2024:LEFTofIA}, respectively. 
In the following, by extending the method used in Ref.~\cite{Chen&Kokron2024:LEFTofIA}, we derive the parity-odd components, which are imaginary, or equivalently, antisymmetric under the pair exchange $(i,j) \leftrightarrow (k,l)$. 

The stochastic contributions can be considered as the limit of the shape correlation function in the small-separation limit. 
This consideration helps in identifying the nontrivial index structure of $N_{ij,kl}$, especially beyond leading order in the derivative expansion. 
Hence, we first define the two-point correlation function of the stochastic field as
\begin{align}
    \avrg{\epsilon_{ij}(\bx)\epsilon_{kl}(\bx')} &\equiv \xi_{ij,kl}(\br/R_*), 
\end{align}
with $\br\equiv\bx-\bx'$. 
This correlation function, by definition, is expected to have a typical extent on the order of halo radius $R_*$, i.e., $\xi_{ij,kl}(\by)$ is non-zero up to $y\sim 1$ with $\by\equiv\br/R_*$. 
The noise power spectrum is obtained via the Fourier transform of this. 
Due to the locality, we can perturbatively expand the transformation in the regime with $R_*k \ll 1$ as 
\begin{align}
    N_{ij,kl}(\bk) 
    &= \int \rmd\br \xi_{ij,kl}(\br/R_*) e^{-i\bk\cdot\br} \\
    &\simeq R_*^3\int \rmd\by \xi_{ij,kl}(\by) \nonumber\\
    &\quad\times \Bigl( 1 - i(R_*k_a)y_a - \frac{1}{2}(R_*^2k_ak_b)y_ay_b \nonumber\\
    &\quad\quad\quad + \frac{i}{6}(R_*^3k_ak_bk_c)y_ay_by_c + \cdots\Bigr). 
    \label{eq:N_ijkl_expansion}
\end{align}
For the first term, since the term $\propto \int \xi_{ij,kl}$ is an even-rank constant tensor, it must be constructed solely from $\delK_{ij}$. 
This leading order contribution is given by \citep{Vlah+2020:IA_EFT}
\begin{align}
    N_{ij,kl}^{(0)} 
    &= \frac{a_0}{2}
    \left(\delK_{ik}\delK_{jl} + \delK_{il}\delK_{jk} - \frac{2}{3}\delK_{ij}\delK_{kl} \right) \\
    &= a_0
    \left( 
    \Lambda^{(0,+)}_{ij,kl}
    + 2\Lambda^{(1,+)}_{ij,kl}
    + 2\Lambda^{(2,+)}_{ij,kl}
    \right), \label{eq:N_ijkl_even_0}
\end{align} 
where we used the definitions of the parity-even basis tensors (Eqs.~\ref{eq:Lambda0_def}--\ref{eq:Lambda2+_def}) in the second equation. 
This is the only tensor that satisfies the index symmetry and traceless condition inherited from $S_{ij}$ at this order. 
We introduced a dimensionful parameter $a_0$. 
In practice, the dominant contribution to $a_0$ arises from Poisson shape noise due to the discrete nature of galaxies. 
Ref.~\cite{Chen&Kokron2024:LEFTofIA} derived the next-to-leading order contributions, at second order in derivatives $\calO(k^2)$, for the parity-even part from the third term in \eq{eq:N_ijkl_expansion}, and found that there are two independent tensors:
\begin{align}
    N_{ij,kl}^{(2,1)} &= a_{2,1} k^2 N_{ij,kl}^{(0)}, \label{eq:N_ijkl_even_2_1}\\
    N_{ij,kl}^{(2,2)} &= 2a_{2,2} k^2 \left(\Lambda^{(1,+)}_{ij,kl}
    + 4\Lambda^{(2,+)}_{ij,kl}\right). \label{eq:N_ijkl_even_2_2}
\end{align}
These tensors can be derived by considering that the term $\propto k_a k_b \int \xi_{ij,kl} y_a y_b$ must be constructed from combinations of two $\bk$ vectors ($k_i k_j$ or $k^2 \delK_{ij}$) and $\delK_{ij}$, satisfying the required symmetry properties\footnote{With our notation, the parity-even stochastic power spectra are given by
\begin{align}
    P_+^{(0)} &\supset a_0 + a_{2,1}k^2, \label{eq:even_spectrum_lambda0_stoc}\\
    P_+^{(1)} &\supset a_0 + (a_{2,1}+a_{2,2})k^2, \label{eq:even_spectrum_lambda1_stoc}\\
    P_+^{(2)} &\supset a_0 + (a_{2,1}+4a_{2,2})k^2. \label{eq:even_spectrum_lambda2_stoc}
\end{align}}.  

We now apply the same approach to the parity-odd components.
As inferred from the above discussion, we focus on the second and fourth terms in Eq.~\eqref{eq:N_ijkl_expansion}.
Since these terms involve an odd number of $ik_i$ (or $\partial_i$ in real space), constructing an even-rank tensor $N_{ij,kl}$ requires an odd number of $\varepsilon_{ijk}$ (along with the imaginary unit).
At each order in derivatives, following similar reasoning, we obtain the following independent pseudotensors:
For the first order in derivatives,
\begin{align} 
    N_{ij,kl}^{(1)} 
    &= -\frac{a_1}{2} 
    \left(
    i\delK_{ik}\varepsilon_{jlm}k_m + 3~\mathrm{perms}
    \right)\nonumber\\
    &= 2a_{1} k \left(\Lambda_{ij,kl}^{(1,-)} + 2\Lambda_{ij,kl}^{(2,-)}\right), 
    \label{eq:N_ijkl_odd_1} 
\end{align}
and for the third order,
\begin{align} 
    N_{ij,kl}^{(3,1)} &= a_{3,1} k^2 N_{ij,kl}^{(1)}, \label{eq:N_ijkl_odd_3_1} \\ 
    N_{ij,kl}^{(3,2)} 
    &= -\frac{a_{3,2}}{2} 
    \left(
    ik_ik_k\varepsilon_{jlm}k_m + 3~\mathrm{perms}
    \right)\nonumber\\
    &= 2a_{3,2} k^3 \Lambda^{(1,-)}_{ij,kl}, \label{eq:N_ijkl_odd_3_2} 
\end{align}
where we used the definitions of the parity-odd basis tensors (Eqs.~\ref{eq:Lambda1-_def} and \ref{eq:Lambda2-_def}).
We find that the leading-order stochastic contribution to the parity-odd spectrum is described by a single tensor in \eq{eq:N_ijkl_odd_1}, which is proportional to $k$ and contributes to both helicities.
At next-to-leading order, two independent degrees of freedom appear, as given in Eqs.~\eqref{eq:N_ijkl_odd_3_1} and \eqref{eq:N_ijkl_odd_3_2}, both scaling as $k^3$. Interestingly, the second tensor contributes only to the helicity-1 mode. 
Therefore, in addition to \eqs{eq:pop_spectrum_lambda1}{eq:pop_spectrum_lambda2}, the parity-odd power spectra include stochastic terms up to next-to-leading order in derivatives as
\begin{align}
    P_-^{(1)} &\supset a_1k + (a_{3,1}+a_{3,2})k^3, \label{eq:pop_spectrum_lambda1_stoc} \\
    P_-^{(2)} &\supset 2a_1k + 2a_{3,1}k^3. \label{eq:pop_spectrum_lambda2_stoc} 
\end{align}

The physical significance of the linear-in-$k$ stochastic contribution
can be understood as the local effect of a stochastic torque, i.e., the curl of the white-noise stochastic spectrum, as follows. 
Let $T_{ijkl}$ be the cartesian components of a tensor field $T$. 
We define the curl of $T$ as 
\begin{align}
    \left[ \nabla \times T \right]_{ijkl} 
    \equiv 
    \varepsilon_{lml'} \partial_m T_{ijkl'}. 
    \label{eq:def_curl_tensor}
\end{align}
This is a generalization of the curl of a vector field $A$: 
$\left[ \nabla \times A \right]_i = \varepsilon_{ijk} \partial_j A_k$. 
In Fourier space, \eq{eq:def_curl_tensor} corresponds to 
$i\varepsilon_{lml'} k_m T_{ijkl'}$.
Now suppose that $T(\br)$ describes the small-scale correlation function of a 2-tensor, such as the small-scale tidal field. Since this has the same symmetries as $N_{ij,kl}^{(0)}$ (Eq.~\ref{eq:N_ijkl_even_0}), we have, symmetrizing under the exchange $(k\leftrightarrow l)$, 
\begin{align}
    \left[i\varepsilon_{lml'} k_m T_{ijkl'} \right]_\mathrm{sym.} 
    \propto 
    N_{ij,kl}^{(1)}. 
    \label{eq:curl_N_ijkl_even_0}
\end{align}
Thus, the linear-in-$k$ contribution represents a preferential alignment of the white-noise tensor components in a specific rotational direction, i.e., handedness, which arises due to the breaking of statistical parity symmetry. 
Similarly, at the next-to-leading order, we find that the curl of $N_{ij,kl}^{(2,\alpha)}$ ($\alpha=1,2$) generates $N_{ij,kl}^{(3,\alpha)}$ for each $\alpha$. 
In this way, the parity-odd stochastic power spectrum naturally shows up as the curl of a parity-even stochastic term that begins at zeroth order in derivatives.

\subsection{Case study} 
\label{subsec:case_study} 
In this section, we assume the two parity-odd trispectrum models, squeezed-type and collapsed-type, introduced in Section~\ref{subsec:recap_trisp} and compute the expected signals of the parity-odd power spectrum of galaxy shapes. 
By substituting \eqs{eq:def_tau-}{eq:def_f-} into \eq{eq:P-_22}, we obtain an expression for each component of the parity-odd power spectrum in \eqs{eq:pop_spectrum_lambda1}{eq:pop_spectrum_lambda2}: 
\begin{widetext}
\begin{align}
    P_{XY}^{(\lambda,-)}(k) 
    = 
    &\int_{\bq_1,\cdots,\bq_4} (2\pi)^3\delD_{\bk-\bq_{12}} (2\pi)^3\delD_{\bk+\bq_{34}} 
    K_{XY}^{(\lambda,-)}(\bq_1,\bq_2,\bq_3,\bq_4) \nonumber\\
    &\times 
    \prod_{i=1}^4\left[\calM(q_i)\right] 
    \left[-i \bq_{12}\cdot\left(\bq_1\times\bq_3\right) \right] 
    \sum_{\sigma \in S_4} \mathrm{sgn}(\sigma) 
    f_-(q_{\sigma(1)},q_{\sigma(2)},q_{\sigma(3)},q_{\sigma(4)},q_{\sigma(1)\sigma(2)},q_{\sigma(1)\sigma(4)}), \label{eq:pop_spectrum_XY_explicit}
\end{align}
\end{widetext}
with $X,Y\in\{Q,R\}$ and we used the triangle condition: 
$\bq_1\cdot\left(\bq_2\times\bq_3\right)=-\bq_{12}\cdot\left(\bq_1\times\bq_3\right)$. 
\eq{eq:pop_spectrum_XY_explicit} is a key equation, as it allows us to compute all parity-odd power spectra of different helicities when a trispectrum template is provided. 
In particular, as shown in Fig.~\ref{fig:diagram_P22}, this corresponds to a two-loop integral, involving two free momenta.
In the following, we denote the two UV cut-off scales for the two-loop integral as $\Lambda$ and $\Lambda'$.

\subsubsection{Squeezed-type trispectrum} 
\begin{figure*}
    \centering
    \includegraphics[width=2.0\columnwidth]{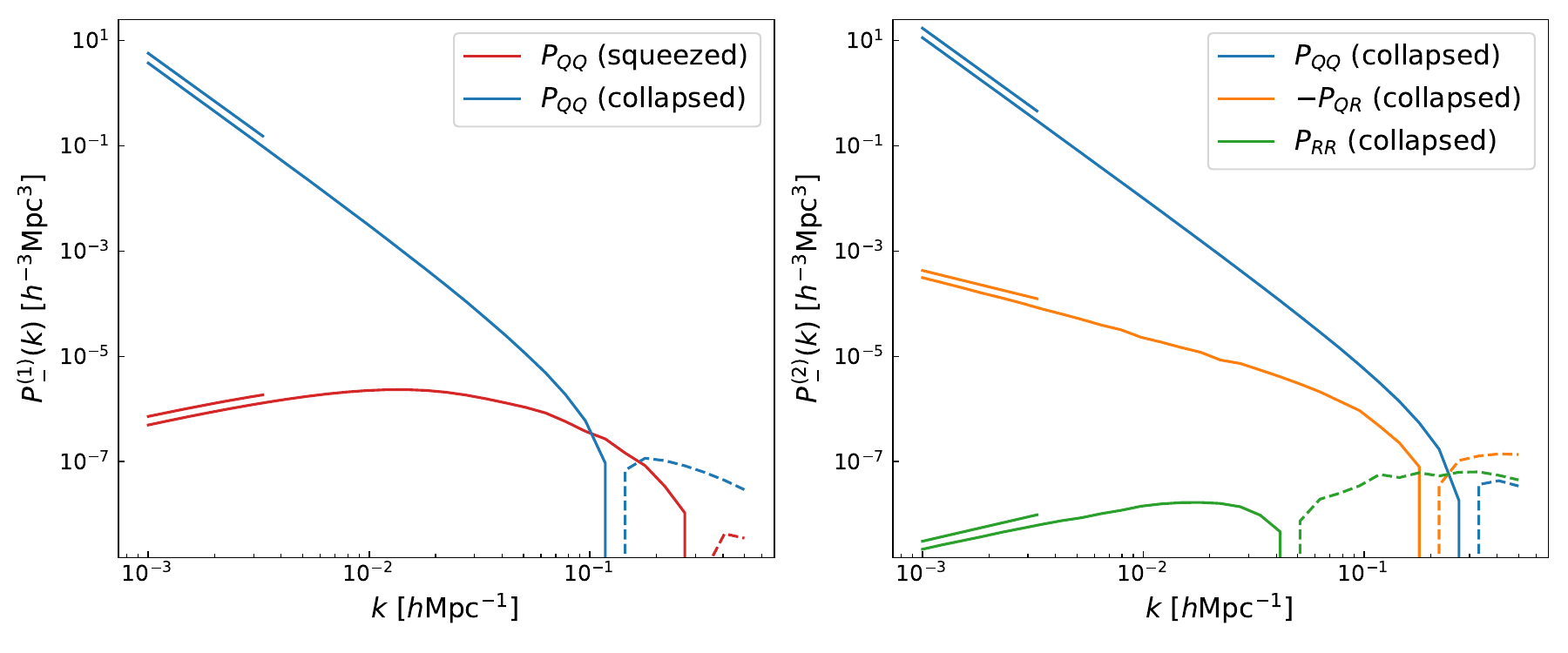}
    \caption{
    Parity-odd power spectra of helicity-1 (left panel) and helicity-2 (right panel), as defined in \eq{eq:pop_spectrum_XY_explicit}, sourced by the squeezed-type and collapsed-type parity-odd trispectra. 
    The helicity-2 spectra for the squeezed trispectrum are exactly zero due to the absence of diagonal dependence, and hence do not appear in the right panel. 
    The model amplitudes are set to $g_- = A_\mathrm{gauge} = 1$, and the linear matter power spectrum at $z=0$ is used. 
    For the loop integral, we set the IR and UV cutoff scales to $q_\mathrm{min} = 10^{-4}\hMpci$ and $\Lambda = \Lambda' = 1\hMpci$; however, the results are insensitive to $q_\mathrm{min}$ as long as $q_\mathrm{min} \ll k$. 
    To better visualize the behavior in the $k \to 0$ limit, we also show the analytically derived asymptotic scalings for each case as guides, plotted as short lines alongside the corresponding original curves in the low-$k$ region: 
    \eq{eq:Pk_m1_QQ_shlim_squeezed} for the squeezed trispectrum, and Eqs.~\eqref{eq:Pk_m1_QQ_dhlim_collapsed}--\eqref{eq:Pk_m2_RR_dhlim_collapsed} for the collapsed trispectrum. 
    Note that this is \textit{not} an expected physical signal, as the amplitude depends on the choice of the UV cut-off scale. 
    The key point here is the low-$k$ scaling, which originates from the trispectra in the soft limits and must be renormalized by introducing an additional tensor operator. 
    }
    \label{fig:pt}
\end{figure*}

The left panel of Fig.~\ref{fig:pt} presents the result of the two-loop calculation defined in \eq{eq:pop_spectrum_XY_explicit}, applied to the squeezed model introduced in \eq{eq:def_f-_squeezed} (red line).
The details of the numerical implementation of the loop integrals are provided in Appendix~\ref{subapp:implementation_of_loop_int_squeezed}. 
Interestingly, no helicity-2 parity-odd signals arise from this model at this order due to the absence of diagonal dependence. 
We provide a proof of this in Appendix~\ref{subapp:proof_no_helicity-2}. 
Thus, only the helicity-1 component $P_{QQ}^{(1,-)}$ appears, as shown in the left panel.
Since the overlap between the shapes of the gravitational kernels and the squeezed model is small, the result of the loop integral becomes small (particularly compared to the collapsed trispectrum case discussed later), as expected. 
Nevertheless, there is a non-negligible dependence on small-scale modes in the low-$k$ region, quantified by the dependence on the UV cut-off scale. 
Below, we examine the dependence on the high-$q$ modes in more detail.

First, let us consider the contribution from the collapsed (double-hard) limit, where both internal momenta are much larger than $k$ of interest: $k \ll q_1, q_3$. 
From the calculation presented in Appendix~\ref{subapp:uvlim_squeezed}, we find 
\begin{align}
    \left[P_{QQ}^{(1,-)}(k)\right]^{dh} = g_- \mathfrak{S}_{dh}(\Lambda,\Lambda') k^3,
    \label{eq:Pk_m1_QQ_dhlim_squeezed}
\end{align} 
showing a $k^3$ suppression with a cut-off dependent coefficient $\mathfrak{S}_{dh}(\Lambda,\Lambda')$ defined in \eq{eq:pop_spectrum_h1_QQ_squeezed_final_dhlim}. 
This suppression is consistent with the odd-power scaling of $k$ that naturally arises from the assumption of local stochasticity, as discussed in the previous section. 
In particular, at this order, the cut-off dependence is absorbed into the coefficient $a_{3,2}$ in \eq{eq:pop_spectrum_lambda1_stoc}. 
It is also in agreement with the fact that this trispectrum does not exhibit a large amplitude in the collapsed limit, where small-scale fluctuations are decoupled from large-scale modes. 

\begin{figure*}
    \centering
    \includegraphics[width=1.75\columnwidth]{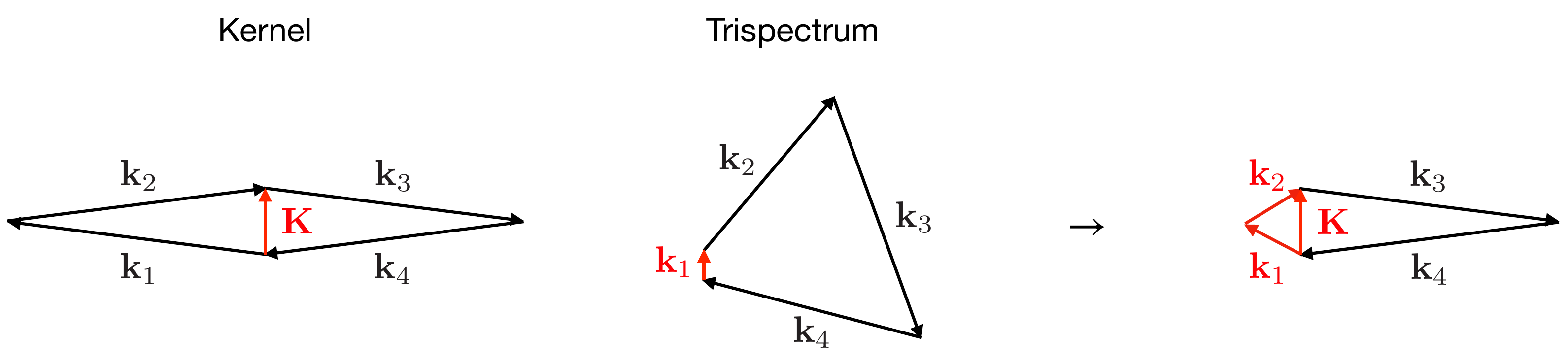}
    \caption{
    A conceptual diagram illustrating why the single-hard limit dominates in the case of a squeezed trispectrum. 
    The red (black) vectors represent soft (hard) momenta, respectively. 
    The (22)-type loop integral is sensitive to the collapsed limit (left). 
    However, if the trispectrum template has a shape that peaks in the squeezed limit (center), the dominant contribution after loop integration comes from the single-hard limit, where both effects overlap (right). 
    Note that for this specific case, where the diagonal $K=|\bk_{12}|$ is taken as the soft mode, by ``single-hard'' and ``double-hard'' we specifically refer to whether one or both of $k_1$ and $k_3$ are large. 
    }
    \label{fig:tetrahedral_product}
\end{figure*}
However, the actual behavior of the power spectrum at low $k$, as shown in the left panel of Fig.~\ref{fig:pt}, does not follow this $k^3$ scaling, indicating that this contribution is subdominant. 
As detailed in Appendix~\ref{subapp:uvlim_squeezed}, the dominant contribution actually comes from the \textit{single-hard} limit where only one of the internal momenta is much larger than $k$. 
This can be understood as follows: the given squeezed-type trispectrum, which has a sharp peak in the squeezed limit, is nearly orthogonal to the (22)-type loop integral that captures signals in the collapsed configuration.  
As a result, the dominant contribution instead arises from the ``product'' configuration of the two soft limits, specifically from the \textit{equilateral triangle-squeezed triangle configuration}, where, out of the two triangles glued together via $\bk$, only one is in the squeezed limit (see Fig.~\ref{fig:tetrahedral_product} for an illustration). 
This leads to a dominant single-hard contribution, which takes the form  
\begin{align}  
    \left[P_{QQ}^{(1,-)}(k)\right]^{sh} = g_- \mathfrak{S}_{sh}(\Lambda) 
    k^{2n_s+2+\alpha+\beta}, 
    \label{eq:Pk_m1_QQ_shlim_squeezed}
\end{align}  
where $\mathfrak{S}_{sh}(\Lambda)$ is a cut-off dependent coefficient defined in \eq{eq:pop_spectrum_h1_QQ_squeezed_final_shlim}. 
The asymptotic $k$-scaling explicitly depends on $\alpha$ and $\beta$ in the definition of the squeezed trispectrum template (Eq.~\ref{eq:def_f-_squeezed}).  
Substituting $(\alpha, \beta) = (-2, -1)$, as assumed in this work, into \eq{eq:Pk_m1_QQ_shlim_squeezed}, we find that it scales as $\propto k^{2n_s-1}$. 
As indicated by the red guide line in the left panel of Fig.~\ref{fig:pt}, this function correctly captures the low-$k$ scaling behavior. 
Thus, after the appropriate renormalization, the model of the parity-odd power spectrum of galaxy shapes under the squeezed trispectrum should take the following form: 
\begin{align} 
    P_-^{(1)}(k) = g_- b_-^2 k^{2n_s+2+\alpha+\beta}, 
\end{align} 
where $b_-$ is a renormalized bias parameter. 
However, the signal predicted by this model is intrinsically small due to the kernel-trispectrum mismatch, and it likely will be difficult to place strong constraints on the amplitude parameter with the galaxy shape power spectrum in this scenario.

\subsubsection{Collapsed-type trispectrum} 

In both panels of Fig.~\ref{fig:pt}, we show the results of the loop integrations for the $U(1)$-gauge model trispectrum defined in \eq{eq:def_f-_collapsed}.
We provide the details of the numerical implementation in Appendix~\ref{subapp:implementation_of_loop_int_collapsed}.
Unlike the squeezed-type trispectrum, this case predicts nonzero signals for both helicity-1 (left panel) and helicity-2 (right panel) modes.
In the low-$k$ regime, the behavior is dominated by the collapsed (double-hard) limit of the trispectrum, leading to the following $k$-dependencies for each spectrum (see Appendix~\ref{subapp:uvlim_collapsed} for derivations): 
For helicity $\lambda=1$: 
\begin{align}
    \left[P_{QQ}^{(1,-)}(k)\right]^{dh} = A_\mathrm{gauge} \mathfrak{S}_{QQ}^{(1,-)}(\Lambda,\Lambda') P_\phi(k),
    \label{eq:Pk_m1_QQ_dhlim_collapsed}
\end{align}
and for helicity $\lambda=2$: 
\begin{align}
    \left[P_{QQ}^{(2,-)}(k)\right]^{dh} &= A_\mathrm{gauge} \mathfrak{S}_{QQ}^{(2,-)}(\Lambda,\Lambda') P_\phi(k), 
    \label{eq:Pk_m2_QQ_dhlim_collapsed}\\
    \left[P_{(QR)}^{(2,-)}(k)\right]^{dh} &= A_\mathrm{gauge} \mathfrak{S}_{QR}^{(2,-)}(\Lambda,\Lambda') k^2P_\phi(k), 
    \label{eq:Pk_m2_QR_dhlim_collapsed}\\
    \left[P_{RR}^{(2,-)}(k)\right]^{dh} &= A_\mathrm{gauge} \mathfrak{S}_{RR}^{(2,-)}(\Lambda,\Lambda') k^4P_\phi(k), 
    \label{eq:Pk_m2_RR_dhlim_collapsed}
\end{align}
where $\mathfrak{S}_{XY}^{(\lambda,-)}(\Lambda,\Lambda')$ is defined in Eqs.~\eqref{eq:pop_spectrum_h1_QQ_collapsed_final_dhlim}--\eqref{eq:pop_spectrum_h2_RR_collapsed_final_dhlim}. 
As seen from Eqs.~\eqref{eq:prj_kernel_lambda1_QQ}--\eqref{eq:prj_kernel_lambda2_RR}, the relative suppressions of factor $k^2$ and $k^4$ for $P_{(QR)}$ and $P_{RR}$, respectively, exist at the kernel level, i.e., they are independent of the form of the trispectrum. 
Therefore, in the $k \to 0$ limit, the dominant contribution always comes from $P_{QQ}$ for both helicities, i.e., from the ``tidal torquing'' operator by definition of $Q_{ij}$ (Eqs.~\ref{eq:Oij_22} and \ref{eq:def_Qij_Rij}). 
In particular, in the case of the $U(1)$-gauge model, $P_{QQ}$ has a scale dependence proportional to $P_\phi(k) \sim k^{-3}$, leading to a significant enhancement at low $k$.\footnote{In general, the scale dependence is determined by the functional form of the diagonal dependence of the primordial trispectrum.}

Note that, in terms of the $d_n^\mathrm{odd}$-parametrization in \eq{eq:dn_odd_parametrization}, this scale-dependent enhancement comes only from the $n=1$ component, and the contributions from the $n=0$ component have an additional $k^2$-suppression due to cancellations (see Appendix~\ref{subapp:uvlim_collapsed}). 
Hence, $A_\mathrm{gauge}$ in Eqs.~\eqref{eq:Pk_m1_QQ_dhlim_collapsed}--\eqref{eq:Pk_m2_RR_dhlim_collapsed} can always be equivalently rewritten as $A_\mathrm{gauge} \leftrightarrow d_1^\mathrm{odd}/3$. 
In other words, since the scale-dependent enhancement of the galaxy shape power spectrum is primarily sensitive to $d_1^\mathrm{odd}$, it is consequently also sensitive to $A_\mathrm{gauge}$. 

We hereafter focus on the leading term, $P_{QQ}$, and carry out its renormalization. 
(The same discussion applies to the other components, $P_{QR}$ and $P_{RR}$.) 
To renormalize the UV cut-off dependence of $P_{QQ}$ in \eqs{eq:Pk_m1_QQ_dhlim_collapsed}{eq:Pk_m2_QQ_dhlim_collapsed}, we introduce additional tensor operators in the bias expansion of the shape field (Eq.~\ref{eq:def_eft_expansion}): 
\begin{align}
    S_{ij} \supset \bar{\psi}_{ij}^{(\lambda)}, 
\end{align}
for $\lambda=1,2$, satisfying 
$\langle\calO_{ij} \bar{\psi}_{kl}^{(\lambda)}\rangle = 0$ for any of the previously introduced ``Gaussian'' operators $\calO_{ij}$, and 
\begin{align}
    P_{\bar{\psi}\bar{\psi}}^{(\lambda,-)}(k) 
    &\equiv 
    \left[\Lambda^{(\lambda,-)}_{ij,kl}(\hbk)\right]^* 
    \avrg{\bar{\psi}_{ij}^{(\lambda)}(\hbk)\bar{\psi}_{kl}^{(\lambda)}(\hbk')}' \nonumber\\
    &= P_\phi(k). 
    \label{eq:def_power_psi}
\end{align}
Thus, $\bar{\psi}_{ij}^{(\lambda)}$ is a purely helical tensor field with helicity $\lambda$, whose power spectrum matches (by our choice of definition) the primordial scalar power spectrum $P_\phi$. 
Therefore, the model of the parity-odd power spectrum for galaxy shapes under the collapsed trispectrum takes the following simple form for each helicity: 
\begin{align}
    P_-^{(\lambda)}(k) &= A_\mathrm{gauge}\left(b_-^{(\lambda)}\right)^2 P_\phi(k), 
    \label{eq:eft_model_Agauge}
\end{align}
where $b_-^{(\lambda)}$ ($\lambda=1,2$) is a renormalized bias parameter. 
Note that this is the only contribution to the parity-odd shape statistics, since gravitational evolution from Gaussian (i.e., parity-conserving) initial conditions does not lead to odd-parity contributions at any order. 

Lastly, we provide a physical interpretation for the additional bias operator in the presence of the parity-odd collapsed trispectrum. 
We start with an expression for the matter trispectrum in the collapsed limit, derived in Appendix~\ref{subapp:uvlim_collapsed}.
For simplicity of notation, we relabel the UV momenta as $\bq \equiv \bq_1$ and $\bq' \equiv \bq_3$, and take the limit of $k\ll q,q'$. 
From \eq{eq:def_T-_s-channel_vectorized_dhlim}, we obtain the following expression: 
\begin{align}
    T_\delta^{(-)} 
    \xrightarrow{dh} 
    \left[-i \hbk\cdot\left(\hbq\times\hbq'\right) \right] 
    \left(\hbq\cdot\hbq'\right)
    P_\phi(k) 
    P(q)
    P(q'), \label{eq:T_delta_dhlim_main}
\end{align}
where $P(q)$ is the linear matter power spectrum, and a constant prefactor is omitted for simplicity. 
We find that this particular angular dependence can be rewritten as a tensor contraction using the parity-odd projectors defined in \eqs{eq:Lambda1-_def}{eq:Lambda2-_def}: 
\begin{align}
    \left[-i \hbk\cdot\left(\hbq\times\hbq'\right) \right] 
    \left(\hbq\cdot\hbq'\right)
    =
    \left(\Lambda_{ij,kl}^{(1,-)} + 2\Lambda_{ij,kl}^{(2,-)}\right) \hq_i\hq_j \hq'_k\hq'_l. 
\end{align}
Substituting this into \eq{eq:T_delta_dhlim_main}, we obtain an explicit form of the parity-odd contribution from the UV modes for each helicity: 
\begin{align}
    T_\delta^{(-)} 
    \supset 
    \Lambda_{ij,kl}^{(\lambda,-)}(\bk) P_\phi(k) 
    \hq_i\hq_j P(q) \hq'_k\hq'_l P(q'). 
    \label{eq:T_delta_dhlim_main_helicity_decomp}
\end{align}
To develop a physical understanding of this, we consider a case where the local (position-dependent) power spectrum is modulated by the large-scale modes of $\bar{\psi}_{ij}^{(\lambda)}$: 
\begin{align}
    P(q;\bx) \equiv \left(1 + \hq_i\hq_j \bar{\psi}_{ij}^{(\lambda)}(\bx) \right)P(q),
\end{align} 
This corresponds to fixing a long-wavelength realization and taking an ensemble average over the short modes.
Considering the large-scale correlation of the quadrupolar modulation, defined as
$\Delta P(q;\bx) \equiv P(q;\bx) - P(q)$, we obtain:
\begin{align}
    &
    \int \rmd\bx e^{-i\bk\cdot\bx} 
    \int \rmd\bx' e^{-i\bk'\cdot\bx'} 
    \avrg{
    \Delta P(q;\bx)
    \Delta P(q';\bx')}' \nonumber\\
    &= 
    \int \rmd\br e^{-i\bk\cdot\br} 
    \avrg{\bar{\psi}_{ij}^{(\lambda)}(\bx) \bar{\psi}_{ij}^{(\lambda)}(\bx')} 
    \hq_i\hq_j P(q) \hq'_k\hq'_l P(q') \nonumber\\
    &= 
    \Lambda_{ij,kl}^{(\lambda,-)}(\bk) P_\phi(k) 
    \hq_i\hq_j P(q) \hq'_k\hq'_l P(q'), 
\end{align} 
which is identical to \eq{eq:T_delta_dhlim_main_helicity_decomp}. 
Note that we used the definition of the power spectrum of $\bar{\psi}_{ij}^{(\lambda)}$, given in \eq{eq:def_power_psi} in the third line. 
Thus, the collapsed limit of the parity-odd trispectrum quantifies the large-scale correlation of the \textit{anisotropic (quadrupolar) modulation} in the small-scale power spectrum, mediated by a \textit{helical} tensor field, $\bar{\psi}_{ij}^{(\lambda)}$, that is uncorrelated with large-scale scalar perturbations\footnote{For the parity-even collapsed trispectrum, i.e., the $\tau_\mathrm{NL}$-trispectrum, it has already been shown that an additional scalar field (denoted here as $\psi$) appears in the bias expansion of the galaxy density field.
Physically, this is interpreted as the effect where the local power spectrum is modulated by large-scale modes of an uncorrelated scalar field $\psi$, as shown in \citep{Tseliakhovich+2010:Stoch_PNG,Smith&LoVerde2011:Stoch_NbodySim,Baumann+2013:Stoch_Bias_PNG,Assassi+2015:PNG}:
\begin{align} P(q;\bx) = \left(1 + \psi(\bx) \right)P(q).
\end{align}
Thus, our analysis provides a natural generalization to the case of a parity-odd trispectrum.}. 
This new degree of freedom subsequently appears in the bias expansion of the shape field.

\section{\texorpdfstring{$N$}{N}-body simulation}
\label{sec:nbody_sim}

So far, we have derived analytical expressions for the parity-odd power spectrum of galaxy shapes in the EFT framework. 
In this section, in order to test whether these predictions are actually realized and to estimate the amplitude of the response of shapes to parity-violating PNG, which remains undetermined in the EFT expansion, we perform a set of numerical simulations.
In this study, we use the shapes of dark matter halos as a proxy for galaxy shapes, and analyze the signal by running $N$-body simulations.
In the previous section, we showed that the shape power spectrum is expected to be particularly sensitive to the collapsed-type parity-odd trispectrum. 
From now on, we therefore focus only on the collapsed-type model, rather than the squeezed-type model. 
We first propose an efficient method to implement a specific class of parity-odd collapsed-type trispectra into the initial conditions.
We then demonstrate that our new method is sufficiently flexible to imprint the parity-odd trispectrum predicted by the $U(1)$-gauge model into the initial conditions, and we validate this implementation.
Subsequently, we run $N$-body simulations using these initial conditions and show that the IA power spectrum at late times is indeed sensitive to the parity-violating initial conditions. 
We also confirm that the resulting signal matches the functional form predicted by EFT.
Finally, by performing a fit to the measured power spectrum, we estimate the amplitude of the response of halo shapes to the parity-violating PNG.

\subsection{Parity-violating initial conditions} 
\label{subsec:parity-odd_ic}

\subsubsection{Idea} 
Let us start with the simplest method that produces a (parity-even) collapsed-type trispectrum, known as the $\tau_\mathrm{NL}$-trispectrum. 
We assume that the Bardeen potential includes a quadratic non-Gaussian correction:  
\begin{align}
    \Phi(\bx) = \phi(\bx) + A\phi^{(2)}(\bx;\sigma),
\end{align}
where  
\begin{align}
    \phi^{(2)}(\bx;\sigma) \equiv \phi(\bx) \sigma(\bx),  
    \label{eq:phi2_tauNL}
\end{align}
with $A$ being the amplitude parameter and $\phi$ and $\sigma$ being uncorrelated Gaussian random fields \citep{Smith&LoVerde2011:Stoch_NbodySim,Smith+2015:OptAnaCMBTriSpec}. 
In Fourier space, we equivalently have 
\begin{align}
    \phi^{(2)}(\bk;\sigma)
    = 
    \int_{\bq_1,\bq_2} (2\pi)^3\delD_{\bk-\bq_{12}} 
    \phi(\bq_1)
    \sigma(\bq_2). 
    \label{eq:phi2_tauNL_FS}
\end{align} 
We will suppress the argument $\sigma$ in $\phi^{(2)}$ to condense the notation. 
In this model, polyspectra involving an odd number of $\Phi$ vanish at any order, and the leading-order trispectrum is given by  
\begin{align}
    &T_\Phi(\bk_1,\bk_2,\bk_3,\bk_4) \nonumber\\
    &= A^2 \avrg{\phi_{\bk_1} \phi_{\bk_2}^{(2)} \phi_{\bk_3} \phi_{\bk_4}^{(2)}}' + 5~\mathrm{perms} + \calO(A^4) \nonumber\\
    &= A^2 P_\phi(k_1) P_\phi(k_3) P_\sigma(k_{12}) + 11~\mathrm{perms} + \calO(A^4). 
    \label{eq:trispectrum_tauNL}
\end{align} 
Although $P_\sigma$, determining the scale dependence of the diagonal, can be taken to be different from $P_\phi$ in general, we assume $P_\sigma=P_\phi$ hereafter.
Thus, $A$ is related to the well-known amplitude parameter $\tau_\mathrm{NL}$ as $A^2 = 25\tau_\mathrm{NL}/9$. 
Regarding the structure of the collapsed-type trispectrum, we note that since $\sigma$ appears only in the quadratic correction and not in the Gaussian part, it always carries the diagonal dependence of the trispectrum at the tree level.
Thus, $\bq_2$ in \eq{eq:phi2_tauNL_FS} corresponds to one of the diagonal momenta, for example, as $\bk_{12}$ in \eq{eq:trispectrum_tauNL}. 
On the other hand, the momentum $\bq_1$ of $\phi$ corresponds to one of the edge momenta $k_i$. 

\begin{figure}
    \centering
    \includegraphics[width=0.6\columnwidth]{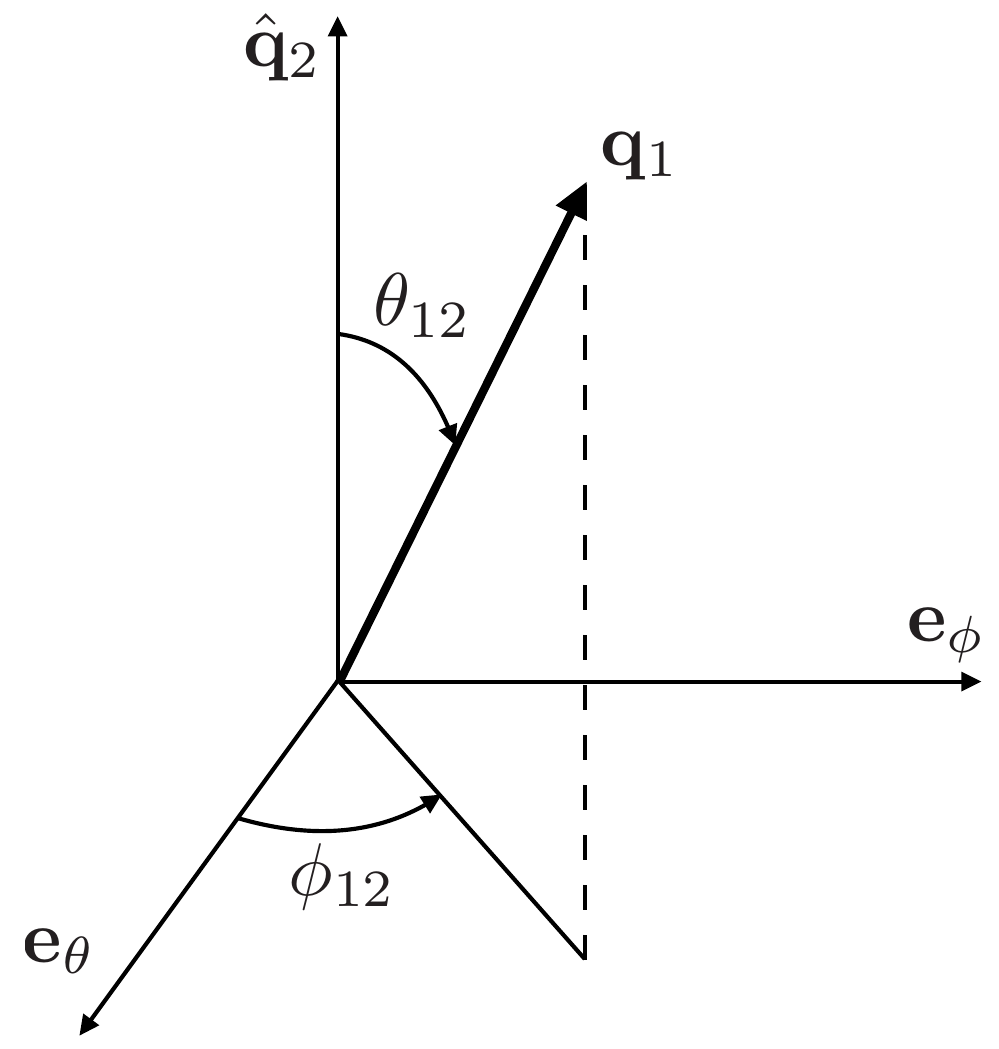}
    \caption{Definitions of the polar and azimuthal angles $(\theta_{12},\phi_{12})$ in \eq{eq:phi2_Ylm_FS}. 
    }
    \label{fig:coordinate_system_for_quad_corr}
\end{figure}

Now, interpreting the convolution kernel in \eq{eq:phi2_tauNL_FS} as unity or ``monopole,'' it can be generalized using spherical harmonics as 
\begin{align}
    \phi_{\ell m}^{(2)}\left(\bk; \sigma_{\ell m}\right)
    \equiv 
    &\int_{\bq_1,\bq_2} 
    (2\pi)^3\delD_{\bk-\bq_{12}} \nonumber\\ 
    &\times
    i^m
    Y_{\ell m} (\hbq_1;\hbq_2)
    \phi(\bq_1)
    \sigma_{\ell m}(\bq_2), 
    \label{eq:phi2_Ylm_FS}
\end{align}
where we define the polar and azimuthal angles for the spherical harmonics, 
$Y_{\ell m} (\hbq_1;\hbq_2) \equiv Y_{\ell m} (\theta_{12},\phi_{12})$, 
as the angles for $\bq_1$ measured in the Cartesian coordinate system, where the polar basis associated with $\bq_2$, $\{\be_\theta(\hbq_2), \be_\phi(\hbq_2), \hbq_2\}$, is regarded as the coordinate axes in the $x$, $y$, and $z$ directions, respectively (see Fig.~\ref{fig:coordinate_system_for_quad_corr}). 
As mentioned above, since $\bq_1$ and $\bq_2$ correspond to the edge and diagonal momenta for the trispectrum, respectively, the spherical harmonics are designed to encode the angular dependence associated with the orientation of the triangle ($\bq_1$) around the diagonal momentum ($\bq_2$). 
The phase factor $i^m$ is needed to satisfy the reality condition: 
$[\phi_{\ell m}^{(2)}(\bk)]^* = \phi_{\ell m}^{(2)}(-\bk)$ (see Appendix~\ref{subapp:reality_condition} for details). 
We introduced distinct, uncorrelated Gaussian random fields $\sigma_{\ell m}$ for each $m$-state ($m=0,\pm1,\cdots,\pm\ell$). 
These fields are constructed to satisfy
$\avrg{\sigma_{\ell m}\sigma_{\ell' m'}}' = \delK_{mm'} \rho_{\ell\ell'} P_\sigma$,
ensuring statistical isotropy, where $|\rho_{\ell\ell'}| \leq 1$ denotes the correlation matrix. 
By expanding the quadratic kernel of $\phi^{(2)}$ using spherical harmonics: 
\begin{align}
    \Phi(\bk) = \phi(\bk) 
    + 
    \sum_{\ell m} \tilde{A}_{\ell m} \phi_{\ell m}^{(2)}\left(\bk;\sigma_{\ell m}\right), 
    \label{eq:def_phi2_sum_phi2lm}
\end{align}
where $\tilde{A}_{\ell m}$ are real coefficients, one can express an arbitrary relative angular dependence between the pair $(\bq_1, \bq_2)$. 
The tilde notation $\tilde{A}_{\ell m}$ represents a temporary parameter introduced before redefining the final amplitude parameter $A_{\ell m}$ (without the tilde) in \eq{eq:def_Phix_tensor_prod}. 
Note that \eq{eq:phi2_tauNL_FS} corresponds to $(\ell,m)=(0,0)$. 

\begin{figure}
    \centering
    \includegraphics[width=0.6\columnwidth]{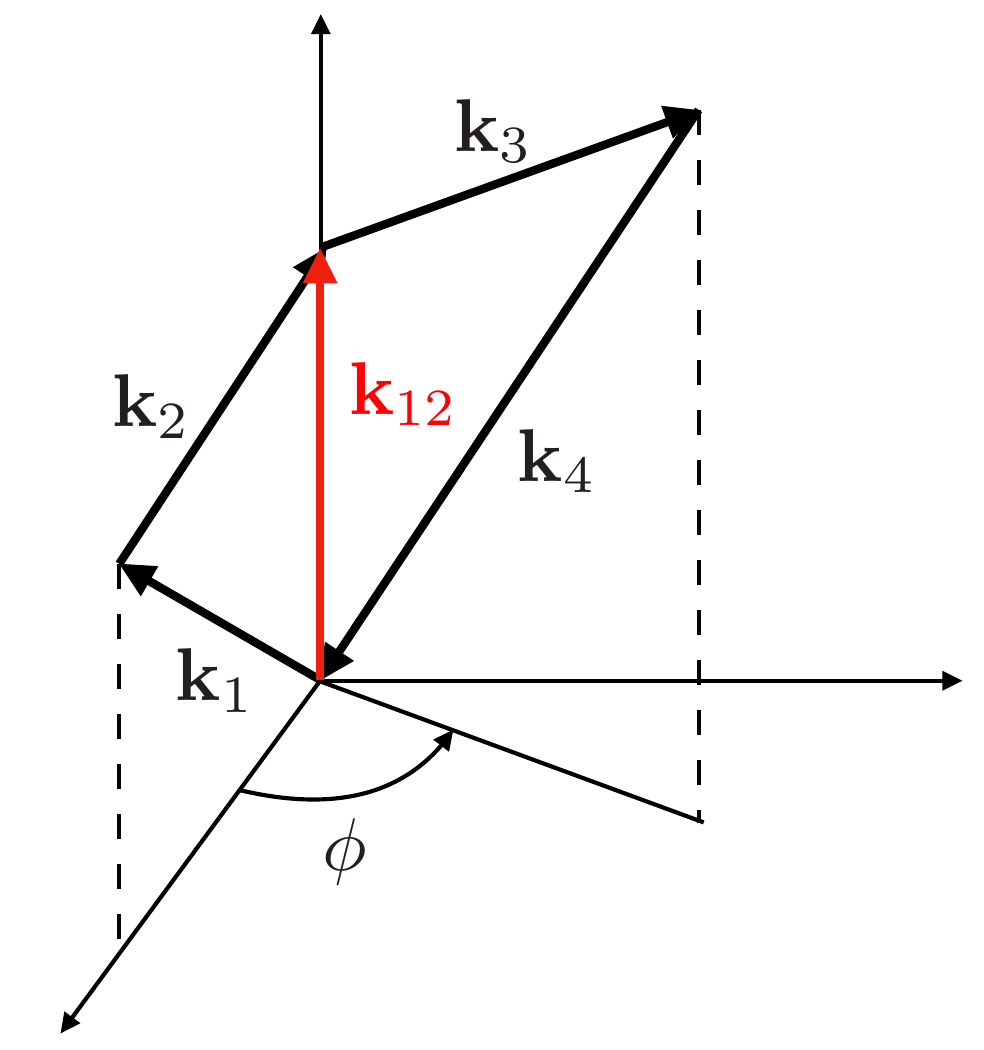}
    \caption{The shape of a tetrahedron is determined by two triangles that share a diagonal momentum $\bk_{12}$: one defined by $\bk_1$ and the other by $\bk_3$, along with the angle $\phi$ between their planes. 
    This angle $\phi$ is given by the dot and cross products of $\hbk_1$, $\hbk_3$, and $\hbk_{12}$ (see Eqs.~\ref{eq:triangles_cos_phi} and \ref{eq:triangles_sin_phi}). 
    }
    \label{fig:tetrahedron_triangles_phi}
\end{figure}

With the quadratic correction given in \eq{eq:def_phi2_sum_phi2lm}, we obtain the leading-order trispectrum (see Appendix~\ref{subapp:ini_trispectrum} for details): 
\begin{align}
    &T_\Phi(\bk_1,\bk_2,\bk_3,\bk_4) \nonumber\\
    &= 
    \sum_{\ell\ell'm} 
    \rho_{\ell\ell'}
    \tilde{A}_{\ell m} \tilde{A}_{\ell' m}
    (-1)^\ell
    Y_{\ell m} (\hbk_1;\hbk_{12}) 
    Y_{\ell' m}^* (\hbk_3;\hbk_{12}) \nonumber\\
    &\quad\quad\quad\times 
    P_\phi(k_1) P_\phi(k_3) P_\sigma(k_{12}) + 11~\mathrm{perms}, 
    \label{eq:trispectrum_YlmYlm}
\end{align} 
which includes the typical collapsed-type scale dependence $P_\phi(k_1) P_\phi(k_3) P_\sigma(k_{12})$, along with the angular dependence expressed as the product of two spherical harmonics that capture the orientations of the two triangles sharing the diagonal $\hbk_{12}$ (see Fig.~\ref{fig:tetrahedron_triangles_phi} for an illustration). 
As shown in Appendix~\ref{subapp:ini_trispectrum}, 
this trispectrum template has both nonzero parity-even (real) and parity-odd (imaginary) parts in general. 
Furthermore, the angular dependence of each component is represented as a polynomial composed of three cosines,  
$\hbk_{12} \cdot \hbk_1$,  
$\hbk_{12} \cdot \hbk_3$, and 
$\hbk_1 \cdot \hbk_3$ (and their permutations), 
with a single factor of the scalar triple product for the parity-odd component: 
$i \bk_{12} \cdot \left(\bk_1 \times \bk_3\right)$, which ensures statistical isotropy (see Eq.~\ref{eq:two_Ylm_in_polys} for the explicit expression). 
Note that, for the pair of $\pm |m|$ states, when the two amplitudes $\tilde{A}_{\ell \pm |m|}$ are equal, the imaginary parts in \eq{eq:trispectrum_YlmYlm} cancel out, making the trispectrum purely parity-even. 
In other words, the difference between $\pm |m|$ states generates the parity-odd component. 
We will provide a more intuitive explanation for the generation of nonzero parity-odd components in this model in the following implementation section. 

\subsubsection{Implementation} 
To implement the correction given in \eq{eq:phi2_Ylm_FS} by utilizing FFTs, we recall an expression for spherical harmonics in terms of the rank-$\ell$ trace-free tensor contraction \citep[e.g.,][]{Thorne1980:Multipole_GW}: 
\begin{align}
    &Y_{\ell m}(\theta,\phi) 
    = 
    c_{\ell m} 
    \mathrm{TF}\left[
    \hn_{i_1} \cdots \hn_{i_\ell}
    \right] \nonumber\\
    &\quad\quad\quad\quad\times
    \left( 
    e_{s,i_1}(\hbz) \cdots e_{s,i_{|m|}}(\hbz)
    \hz_{i_{|m|+1}} \cdots \hz_{i_\ell} 
    \right),
    \label{eq:Ylm_TF}
\end{align}
where 
$\hn_i(\theta,\phi)$ is a unit vector, and 
$c_{\ell m}$ is the normalization constant:
\begin{align}
    c_{\ell m} = (2\ell-1)!! 
    \sqrt{\frac{2\ell+1}{4\pi} \frac{2^{|m|}}{(\ell+m)!(\ell-m)!}}, 
    \label{eq:def_clm}
\end{align}
$s \equiv \mathrm{sgn}(m) \in\{\pm\}$ is the sign of $m$, 
$e_{s,i}(\hbz)$ is the complex unit vector defined in \eq{eq:def_epm}, and $\mathrm{TF}\left[X_{i_1,\cdots,i_\ell}\right]$ denotes the trace-free part of $X_{i_1,\cdots,i_\ell}$\footnote{Due to the property of the trace-free part, we have 
\begin{align}
    &\mathrm{TF}\left[
    \hn_{i_1} \cdots \hn_{i_\ell} 
    \right] 
    \left( 
    e_{\pm,i_1}(\hbz) \cdots e_{\pm,i_{|m|}}(\hbz)
    \hz_{i_{|m|+1}} \cdots \hz_{i_\ell} 
    \right) \nonumber\\
    &= 
    \hn_{i_1} \cdots \hn_{i_\ell} 
    \mathrm{TF}
    \left[
    e_{\pm,(i_1}(\hbz) \cdots e_{\pm,i_{|m|}}(\hbz)
    \hz_{i_{|m|+1}} \cdots \hz_{i_\ell)} 
    \right]. 
    \label{eq:Ylm_TF_footnote}
\end{align}
Note that the trace-free operation in the second line is applied after the symmetrization for the indices, $i_1, \cdots, i_\ell$.}. 
By replacing $\hbn$ with $\hbq_1$ and $\hbz$ with $\hbq_2$ in \eq{eq:Ylm_TF}, we obtain an expression for the spherical harmonics defined in \eq{eq:phi2_Ylm_FS}: 
\begin{align}
    &Y_{\ell m}(\hbq_1;\hbq_2) 
    = 
    c_{\ell m} 
    \mathrm{TF}\left[
    \hq_{1,i_1} \cdots \hq_{1,i_\ell}
    \right] \nonumber\\
    &\quad\quad\times
    \left( 
    e_{s,i_1}(\hbq_2) \cdots e_{s,i_{|m|}}(\hbq_2)
    \hq_{2,i_{|m|+1}} \cdots \hq_{2,i_\ell} 
    \right).
    \label{eq:Ylm_TF_12}
\end{align}
Notably, the spherical harmonics are now written in a sum-separable form with respect to $\bq_1$ and $\bq_2$. 
This allows the quadratic convolution in \eq{eq:phi2_Ylm_FS} to be implemented just as a product of two tensor fields in real space. 
By substituting \eq{eq:Ylm_TF_12} into \eq{eq:phi2_Ylm_FS}, we can finally express the quadratic correction as follows: 
\begin{align} 
    \Phi(\bx) 
    &\equiv \phi(\bx) + \sum_{\ell m} \tilde{A}_{\ell m} \phi_{\ell m}^{(2)}\left(\bx;\sigma_{\ell m}\right) \\
    &\equiv \phi(\bx) + \sum_{\ell m} A_{\ell m} 
    \phi_{i_1\cdots i_\ell}^{(\ell)}(\bx) 
    \sigma_{i_1\cdots i_\ell}^{(\ell m)}(\bx), \label{eq:def_Phix_tensor_prod}
\end{align} 
where the amplitude parameters are redefined as 
\begin{align}
    A_{\ell m} \equiv (-1)^\ell\tilde{A}_{\ell m} c_{\ell m}, 
    \label{eq:redef_Alm}
\end{align} 
in the second line, 
and the two rank-$\ell$ tensors are defined in Fourier space as
\begin{align}
    \phi_{i_1\cdots i_\ell}^{(\ell)}(\bk) 
    &\equiv 
    \mathrm{TF}\left[ i\hk_{i_1} \cdots i\hk_{i_\ell}\right] \phi(\bk), \label{eq:def_phi_ell}\\
    \sigma_{i_1\cdots i_\ell}^{(\ell m)}(\bk) 
    &\equiv 
    e_{i_1}^{-s}(\hbk) \cdots e_{i_{|m|}}^{-s}(\hbk) 
    i\hk_{i_{|m|+1}} \cdots i\hk_{i_\ell} \sigma_{\ell m}(\bk), \label{eq:def_sigma_ell_m}
\end{align} 
with $e_i^{s} \equiv e_{s,i}^*$ being the dual basis vector satisfying: 
$e^s \cdot e_{s'} =\delK_{ss'}$ with $s\in\{\pm\}$. 
Note that both $\phi_{i_1\cdots i_\ell}^{(\ell)}$ and $\sigma_{i_1\cdots i_\ell}^{(\ell m)}$ satisfy the reality condition, and it does not affect the result whether the indices of $\sigma_{i_1\cdots i_\ell}^{(\ell m)}$ are symmetrized in the definition. 
$\phi_{i_1\cdots i_\ell}^{(\ell)}$ is a pure longitudinal tensor, while $\sigma_{i_1\cdots i_\ell}^{(\ell m)}$ is a helical tensor with helicity $m$. 
Therefore, in this model, the quadratic mode coupling between the helical and non-helical tensor field sources both parity-odd and parity-even trispectra.

The simplest case where a parity-odd component arises is when only the $m=+1$ or $m=-1$ state for $\ell=1$ is nonzero, while all others vanish: 
\begin{align} 
    \phi_\pm^{(2)}(\bx;\sigma_{1\pm1}) \equiv A \phi_i(\bx) \sigma_i^{(1\pm1)}(\bx), \label{eq:phi2_d0_odd} 
\end{align}
where $A \equiv A_{1\pm1}$, and
\begin{align}
    \phi_i^{(1)}(\bk) 
    &=
    i \hk_i \phi(\bk), \\
    \sigma_i^{(1\pm1)}(\bk) 
    &=
    e_i^\mp(\hbk) \sigma_{1\pm1}(\bk). 
\end{align}  
Here, $\sigma_i^{(1m)}$ is interpreted as a purely helical Gaussian vector field whose power spectrum is given by $P_\sigma$. 
Computing the trispectrum with this non-Gaussian correction (see Appendix~\ref{subapp:ini_trispectrum} for detailed derivation), we find that the parity-odd component corresponds to the ``$d_0^\mathrm{odd}$-only model'' in the parity-odd template of \eq{eq:def_dn_odd_template}, with $d_0^\mathrm{odd} = \mp3A^2/100$. 

Furthermore, by extending the correction to $\ell=2$, we consider the following quadratic term: 
\begin{align}
    &\phi_\pm^{(2)}(\bx;\sigma_{1\pm1},\sigma_{2\pm1},\sigma_{2\pm2}) \nonumber\\
    &\equiv 
    A 
    \Bigl[
    \phi_i^{(1)}(\bx) \sigma_i^{(1\pm1)}(\bx) \nonumber\\
    &\quad\quad+
    \phi_{ij}^{(2)}(\bx) \sigma_{ij}^{(2\pm1)}(\bx)
    +
    \phi_{ij}^{(2)}(\bx) \sigma_{ij}^{(2\pm2)}(\bx)
    \Bigr], 
    \label{eq:phi2_U1gauge}
\end{align}
where $A \equiv A_{1\pm1}=A_{2\pm1}=A_{2\pm2}$, and 
\begin{align}
    \phi_{ij}^{(2)}(\bk) 
    &= - \hk_i \hk_j \phi(\bk), \\
    \sigma_{ij}^{(2\pm1)}(\bk) 
    &= i \hk_{i} e_{j}^\mp(\hbk) \sigma_{2\pm1}(\bk), \\
    \sigma_{ij}^{(2\pm2)}(\bk) 
    &= e_i^\mp(\hbk) e_j^\mp(\hbk) \sigma_{2\pm2}(\bk).  
\end{align}
Here, $\sigma_{ij}^{(2m)}$ are helical rank-two tensor fields with helicity $m$. 
Following the same computation as before (see Appendix~\ref{subapp:ini_trispectrum}), we can show that the resulting parity-odd trispectrum takes the form of the $U(1)$-gauge model (Eq.~\ref{eq:def_f-_collapsed}) with $A_\mathrm{gauge} = \pm3A^2/100$, assuming the correlation matrix $\rho_{\ell\ell'}=1$ for any $\ell,\ell'=1,2$. 
Equivalently, this can be expressed in terms of \eq{eq:dn_odd_parametrization} as $d^{\mathrm{odd}}_{0} = \mp 3A^2/100$ and $d^{\mathrm{odd}}_{1} = \pm 9A^2/100$.

We implement the quadratic non-Gaussian correction defined in Eq.~\eqref{eq:phi2_U1gauge} into the initial conditions to realize the $U(1)$-gauge parity-odd trispectrum, by modifying the public initial condition generator \texttt{MUSIC2-monofonIC}~\citep{Michaux+2021:monofonIC,Hahn+2021:monofonIC}.
The procedure is briefly summarized as follows. 
First, we generate three Gaussian random fields, $\phi$, $\sigma_{1+1}$, and $\sigma_{2+2}$, from the assumed $P_\phi$. 
For opposite helicities ($m < 0$), we use the same random seed, i.e., we set $\sigma_{\ell m} = \sigma_{\ell -m}$. 
Using these fields, we construct the relevant vector and tensor fields in Fourier space: 
$\phi_i^{(1)}$, $\phi_{ij}^{(2)}$, $\sigma_i^{(1+1)}$, $\sigma_{ij}^{(2+1)}$, and $\sigma_{ij}^{(2+2)}$.
These are then transformed to real space via inverse FFTs, and the tensor contractions in Eq.~\eqref{eq:phi2_U1gauge} are evaluated to obtain the final non-Gaussian field $\Phi$.
Based on this $\Phi$, we solve the Lagrangian dynamics up to second order to generate the initial particle distribution \citep{Crocce+2006:IC,Scoccimarro+2012:2LPT}. 

\begin{figure*}
    \centering
    \includegraphics[width=2.0\columnwidth]{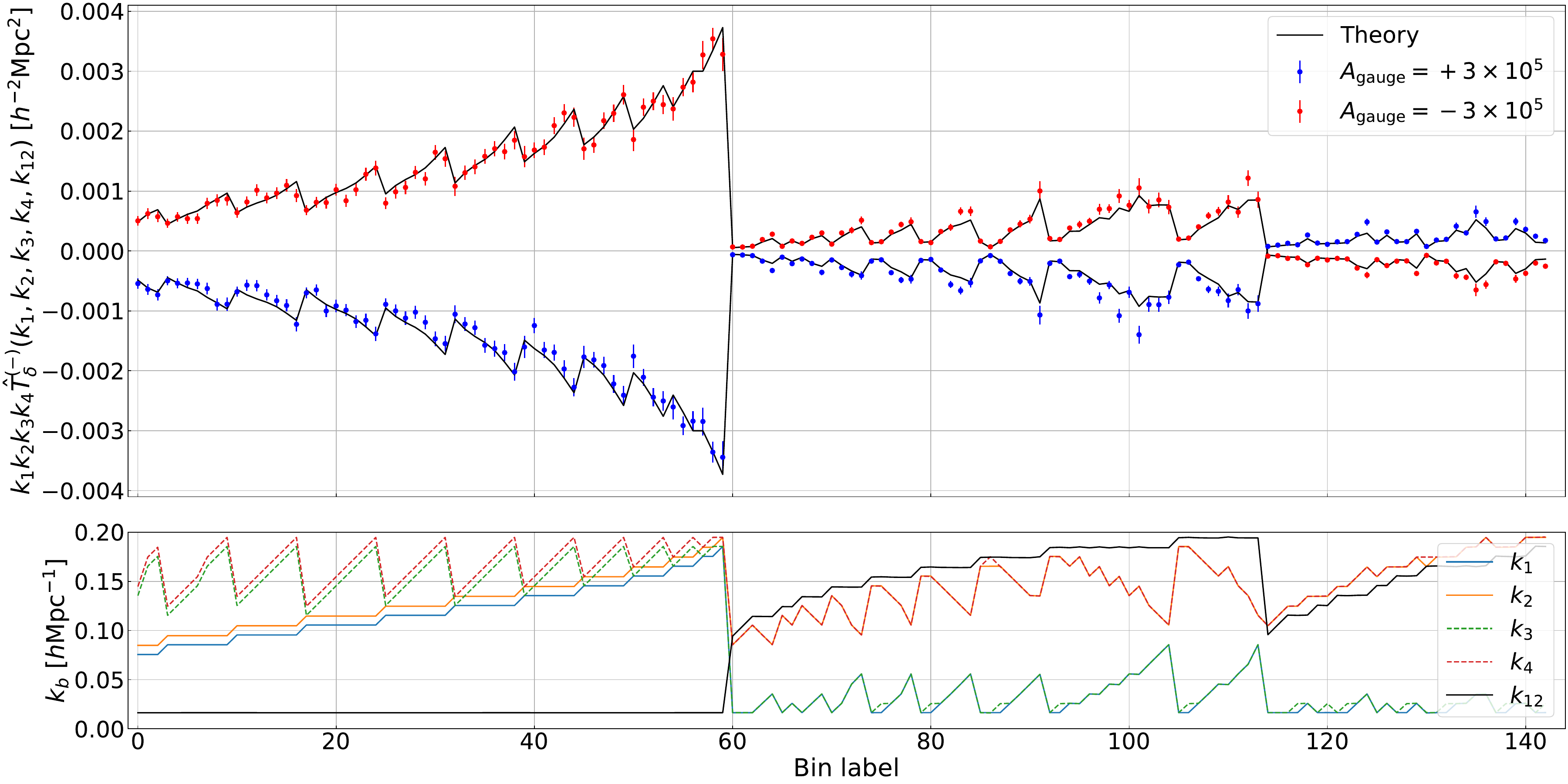}
    \caption{A validation test for our initial conditions, comparing the measurements of the parity-odd trispectrum in the initial conditions with its expectation values calculated from the analytic expression (see Appendix~\ref{app:validation_tests_ic} for details). 
    The top panel shows the parity-odd trispectrum measured from 1000 realizations of the initial conditions. 
    Blue (red) data points correspond to the positive (negative) amplitude $A_\mathrm{gauge}$. 
    To suppress noise, we plot the difference relative to the corresponding Gaussian realization. 
    The box size is $1\,h^{-1}\mathrm{Gpc}$, and the density field is constructed on a $64^3$ grid based on the initial particle distribution. 
    For the trispectrum measurement, we use binning with $k_\mathrm{min}=0.0$, $k_\mathrm{max}=0.2~\hMpci$, and $N_\mathrm{bin}=20$, applied to each of the five wavenumbers $k_b$ ($b=1, \cdots, 4, 12$).
    The bottom panel shows the corresponding configurations $(k_1, k_2, k_3, k_4, k_{12})$. 
    We plot the configurations with the highest signal-to-noise ratios among the independent data points. 
    As expected, the configurations with the strongest trispectrum signals are those in the collapsed limit, where one of the diagonal momenta is significantly smaller than the edge momenta. 
    The black line in the top panel represents the theoretical expectation (computed by using a sub-gridding procedure), which is in good agreement with the measurements. 
    }
    \label{fig:trisp_u1}
\end{figure*}

Fig.~\ref{fig:trisp_u1} shows the results of the validation test based on the parity-odd trispectrum measured from the initial conditions. 
Details of the parity-odd trispectrum estimator and the analytical expression for the angle-averaged trispectrum corresponding to the measurement are provided in Appendix~\ref{app:validation_tests_ic}. 
The results are consistent with theoretical expectations, which confirms the validity of our method.  

Here, we comment on the relation to existing methods for generating parity-violating initial conditions. Ref.~\cite{Coulton+2024:Quijote-Odd} introduces a cubic correction to Gaussian initial conditions to realize a parity-odd trispectrum enhanced in the squeezed limit. 
Although both our approach and theirs introduce parity violation through scalar perturbations, we focus on collapsed-type parity-odd trispectra and thus on a different trispectrum shape. 
Other works implement parity-violating signals from helical vector and tensor fossil effects \citep{Shim+2025:PVIC_VectorFossil,Zhu&Pen2025:PVIC_VectorTensorFossil}, whereas we consider scalar-sector primordial non-Gaussianity and therefore probe a different physical scenario.

\subsection{Data} 
We generate the initial particle distribution at $z = 49$ with $N_\mathrm{p} = 1024^3$ particles in a comoving simulation box of size $1\,h^{-1}\mathrm{Gpc}$. 
We evolve the particle distribution by using \texttt{Gadget-4}~\citep{Springel+2021:Gadget-4}. 
The particle mass corresponds to $m_\mathrm{p} \simeq 8\times10^{10}\,\hiMsun$. 
We perform a total of 40 $N$-body simulations, consisting of eight realizations for each of five different initial conditions: one Gaussian case and four non-Gaussian collapsed-type cases with different amplitudes of $A_\mathrm{gauge}$, specifically 
$A_\mathrm{gauge} = \pm6 \times 10^4$ and $\pm3 \times 10^5$ (corresponding to $A^2 = 2 \times 10^6$ and $1 \times 10^7$, respectively). 
For each realization, we use the same random seed for the Gaussian part $\phi$ to reduce sample variance when comparing different models. 
Note that for each nonzero amplitude, the nonlinear correction to the initial power spectrum increases monotonically with $k$ and peaks at the Nyquist frequency, where the relative correction is $P_\Phi / P_\phi -1 \lesssim 0.01$ and $0.04$, respectively (see Appendix~\ref{subapp:correction_initial_power} for the analytical expression). 
Since these corrections are sufficiently small for the purposes of this study, we do not apply any rescaling to the initial power spectrum in our simulations.

At low-redshift snapshots, we identify dark matter halos using the friends-of-friends (FoF) algorithm implemented in \texttt{Gadget-4}. 
For each halo, we define its shape by computing the inertia tensor from the relative positions of its member particles with respect to the halo center. 
We adopt the ``reduced'' inertia tensor (with $w(r) = 1/r^2$ in Eq.~\ref{eq:def_Iij}) for our main results, rather than the standard inertia tensor (with $w(r) = 1$), in order to better capture the shape of the inner halo region, which is often associated with the formation of central galaxies \cite[see e.g.,][]{Bett2012:HaloShape,Tenneti+2015a:IA_hydro,Osato+2018:IA_nbody,Kurita+2020:IA_nbody}. 
The results using the standard inertia tensor are presented in Appendix~\ref{app:shape_diff}. 
They are in overall qualitative agreement, with minor differences discussed later. 
We then assign these halo shapes to a regular grid using the Cloud-In-Cell (CIC) scheme to construct the three-dimensional shape field $\hat{S}_{ij}(\bx)$. 
In Fourier space, we apply the interlacing technique \citep{Sefusatti+2016:FFT_aliasing_interlacing} to suppress aliasing effects, in addition to deconvolving the CIC kernel.

We measure the parity-odd power spectra for helicities $\lambda = 1$ and $2$ using the standard binned power spectrum estimator with a parity-odd projection:
\begin{align}
    \hat{P}_-^{(\lambda)}(k_b) 
    = \frac{1}{N_b} \sum_{\bk\in \mathrm{bin}\,b} 
    \left[\Lambda^{(\lambda,-)}_{ij,kl}(\hbk)\right]^* 
    \hat{S}_{ij}(\bk) \hat{S}_{kl}(-\bk),
\end{align}
where $N_b$ is the number of Fourier modes within the $b$-th spherical shell, and $k_b$ is the bin-averaged wavenumber defined by $k_b = \sum_{\bk \in \mathrm{bin}\,b} |\bk| / N_b$. 
We use logarithmically spaced bins with width $\Delta \log_{10} k = 0.1$.

\subsection{Results} 
\subsubsection{Comparison of EFT and simulation} 
\begin{figure*}
    \centering
    \includegraphics[width=2.0\columnwidth]{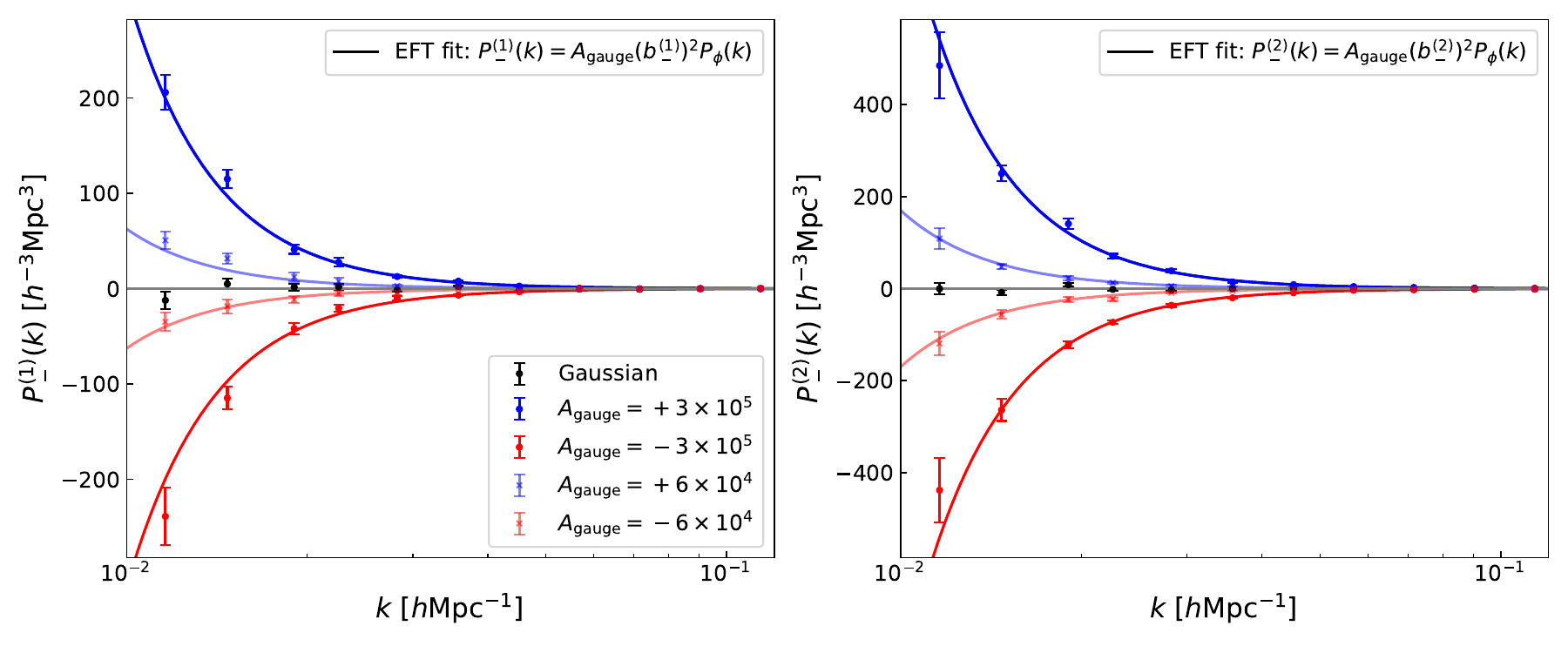}
    \caption{Comparison between the measured parity-odd power spectra of halo IA for different initial conditions and the corresponding EFT best-fit models (solid lines) defined in \eq{eq:eft_model_Agauge}: 
    $(b_-^{(1)})^2 = 0.065$ and $(b_-^{(2)})^2 = 0.177$. 
    The left (right) panel shows the helicity-1 (helicity-2) component of the power spectrum. 
    The power spectra are measured at $z = 1$ from halo samples with 
    $1 \times 10^{13} < M_\mathrm{h} < 4 \times 10^{13}~h^{-1}M_\odot$. 
    }
    \label{fig:EFT_fit}
\end{figure*}
Fig.~\ref{fig:EFT_fit} shows the parity-odd power spectra measured from dark matter halo shapes for both helicities $\lambda = 1, 2$. 
Clear nonzero signals at large scales are observed in the cases with parity-violating non-Gaussian initial conditions, in contrast to the null results obtained from Gaussian initial conditions.  
We also show the best-fit EFT curves as defined in Eq.~\eqref{eq:eft_model_Agauge}.  
For the fitting, we restrict the analysis to linear scales, using data up to $k_\mathrm{max} = 0.06\,\hMpci$. 
To suppress noise, we define the signal as the half difference between the simulations with opposite parity-violating amplitudes, $\pm A_\mathrm{gauge}$.
The fact that the amplitudes of the measured signals linearly scale with the input parameter $A_\mathrm{gauge}$, both in magnitude and in sign, and that their scale-dependent enhancement matches the shape of $P_\phi(k)$ is in excellent agreement with the EFT prediction. 

\begin{figure*}
    \centering
    \includegraphics[width=2.0\columnwidth]{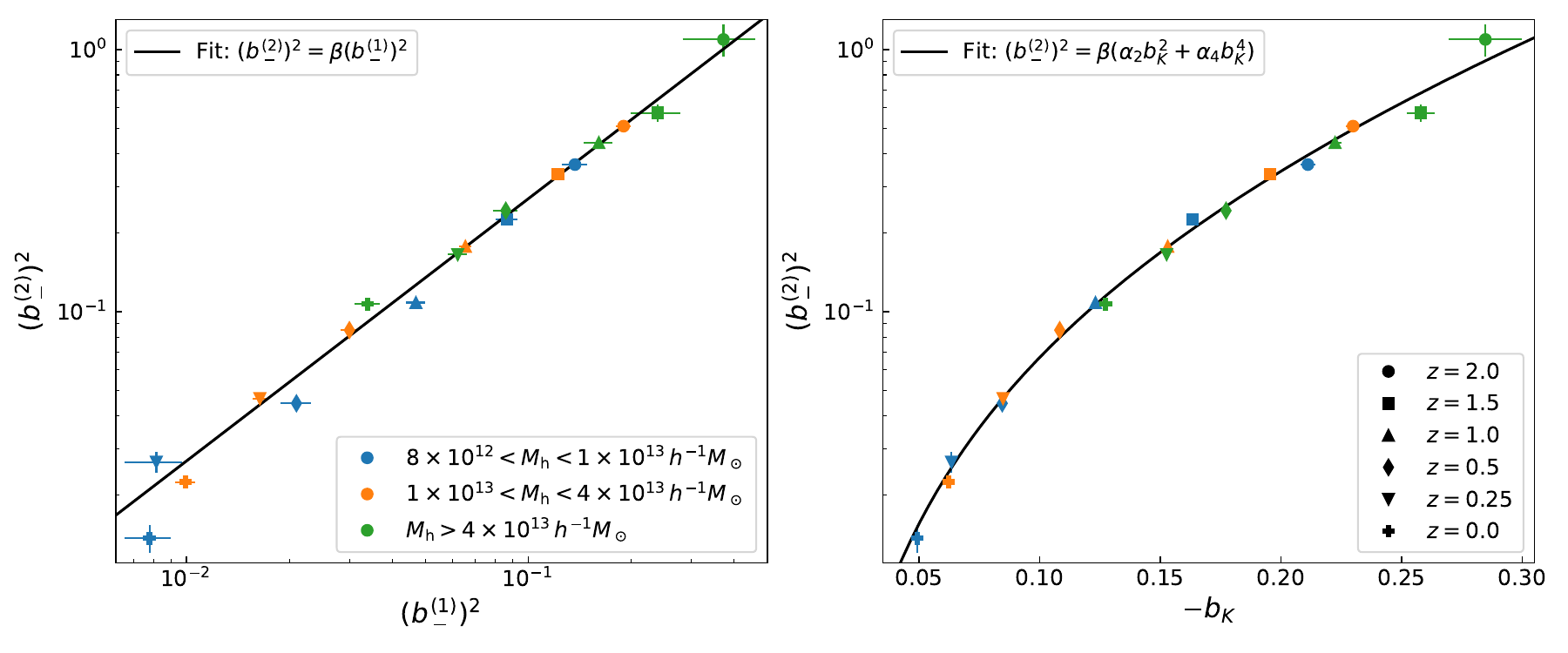}
    \caption{Relations among the measured bias parameters and results of polynomial fitting.  
    The left panel shows the relation between the two parity-violating biases for helicities $\lambda = 1, 2$: $b_-^{(1)}$ and $b_-^{(2)}$. 
    The black line indicates the best-fit linear model with slope parameter in \eq{eq:bias_best-fit_beta}.  
    The right panel shows the relation between $b_-^{(2)}$ and the standard Gaussian linear bias $b_K$. 
    The black curve shows the best-fit polynomial with parameters in \eq{eq:bias_best-fit_alpha}. 
    We omit the plot of the relation between $b_-^{(1)}$ and $b_K$ as it is redundant given the two panels above.
    }
    \label{fig:bias_fit}
\end{figure*}

\subsubsection{PNG bias: Different helicities} 
Since, as shown in Eq.~\eqref{eq:eft_model_Agauge}, the parity-odd power spectrum is proportional to the product of the underlying parity-violating amplitude $A_\mathrm{gauge}$ and the PNG-induced bias $(b_-^{(\lambda)})^2$, prior knowledge of the PNG-induced bias is required to directly constrain $A_\mathrm{gauge}$ from measurements. 
This situation is analogous to the well-known degeneracy between the local-type PNG parameter $f_\mathrm{NL}$ and the PNG-induced bias $b_\phi$ in the scale-dependent bias of the galaxy clustering power spectrum \citep{Dalal+2008:PNG_halo}. 
Therefore, we here derive fitting formulae that relate the PNG-induced biases $b_-^{(\lambda)}$ to the linear (Gaussian) shape bias $b_K$.

First, the left panel of Fig.~\ref{fig:bias_fit} shows the relation between the parity-violating PNG-induced biases for the two helicities $\lambda = 1, 2$. 
They show a clear linear relation, which is well described by 
\begin{align}
    \left(b_-^{(2)}\right)^2 
    &= \beta \left(b_-^{(1)}\right)^2, 
    \label{eq:bias_relations:bm1_bm2} 
\end{align}
with the best-fit value
\begin{align}
    \beta = 2.70 \pm 0.08. 
    \label{eq:bias_best-fit_beta}
\end{align}
Note that $b_-^{(\lambda)}$ are only defined up to a sign, since they always enter quadratically in the statistics. 
We found that adding a constant or quadratic term does not improve the quality of the fit. 
The linear relation between the biases for different helicities indicates that the response of the halo shape to anisotropic modulations of local power spectrum is essentially the same for both helicities, apart from an overall multiplicative factor (i.e., the slope), regardless of the specific halo sample. 

On the other hand, the fact that the slope deviates from unity is nontrivial and is generally expected to be model-dependent. 
In particular, for the $U(1)$-gauge model considered here, a renormalization calculation presented in Appendix~\ref{app:calc_detail_collapsed} shows that the cutoff-scale dependence differs between helicities exactly by a factor two (see Eqs.~\ref{eq:pop_spectrum_h1_QQ_collapsed_final_dhlim} and \ref{eq:pop_spectrum_h2_QQ_collapsed_final_dhlim}): 
\begin{align} 
    \mathfrak{S}_{QQ}^{(2,-)}(\Lambda,\Lambda') = 2\, \mathfrak{S}_{QQ}^{(1,-)}(\Lambda,\Lambda'). 
    \label{eq:relation_Sigma}
\end{align} 
Thus, the double-hard limits for both helicities are proportional, and this proportionality is also confirmed by the halo bias measurements, although with a different coefficient. 

We note that this result is based on halo shapes defined using the reduced inertia tensor, which gives more weight to particles in the inner regions of the halo. 
Interestingly, even when we use the unweighted inertia tensor instead, the slope parameter $\beta$ remains nearly unchanged (consistent within the $1\sigma$ level; see Appendix~\ref{app:shape_diff}). 
This suggests that both the inner and outer regions of the halo respond similarly to modulations of the local power spectrum.

\subsubsection{PNG bias: Relation to Gaussian bias} 
Next, we investigate the relation between these PNG-induced biases and the Gaussian linear shape bias $b_K$. 
To estimate $b_K$, we used the ratio between the matter-shape cross power spectrum to the matter power spectrum at linear scales in the Gaussian simulations \citep{Akitsu+2023:QuadShapeBias}. 
The right panel of Fig.~\ref{fig:bias_fit} shows a scatter plot comparing the PNG-induced bias and the Gaussian bias, and we fit the data with a polynomial function: 
\begin{align}
    \left(b_-^{(1)}\right)^2 
    &= \alpha_2 b_K^2 + \alpha_4 b_K^4, 
    \label{eq:bias_relations:bm1_bK} 
\end{align}
with the best-fit values: 
\begin{align}
    \alpha_2 = 2.25 \pm 0.11,~
    \alpha_4 = 23.38 \pm 4.15. 
    \label{eq:bias_best-fit_alpha}
\end{align}
The relation between $b_-^{(2)}$ and $b_K$ can then be obtained by substituting \eq{eq:bias_relations:bm1_bK} into \eq{eq:bias_relations:bm1_bm2} as shown in Fig.~\ref{fig:bias_fit}. 
We found that including additional lower- or higher-order polynomial terms (from zeroth to sixth order) does not improve the quality of the fit, as all coefficients other than the quadratic and quartic terms are consistent with zero. 

When using the unweighted inertia tensor instead of the reduced one, we find that the fitting parameters $\alpha_4$ differ significantly (see Appendix~\ref{app:shape_diff}), in contrast to the relation between the helicities discussed above, which remains nearly unchanged.  
This discrepancy reflects the fact that the response to the large-scale tidal field, encoded in $b_K$, differs between the inner and outer regions of the halo. 
In the forecast analysis in the next section, we employ the fitting results from the reduced inertia tensor (Eq.~\ref{eq:bias_best-fit_alpha}), which better captures the inner halo shape and is often considered more relevant for describing the region associated with central galaxy formation.

As a caveat, we have not investigated possible assembly bias or other secondary dependencies of $b_-^{(\lambda)}$. 
Such effects are known to be significant for the local PNG-induced bias of galaxy density \citep[e.g.][]{Barreira+2020:PNG_bias_TNG}, and may also affect $b_-^{(\lambda)}$. 
A detailed study of these effects is beyond the scope of this work and left for future investigation. 

\section{Projection and angular statistics} 
\label{sec:projection_effects}

In this section, we forecast how well IA can constrain the $U(1)$-gauge model, particularly its amplitude parameter $A_\mathrm{gauge}$, with the parity-violating signals. 
In the previous sections, we focused on three-dimensional galaxy and halo shapes, as they naturally form in three-dimensional space. 
In actual imaging galaxy surveys, however, we do not have direct access to the full three-dimensional shapes. 
Instead, only the two-dimensional shapes projected onto the sky are observable. 
If spectroscopic redshifts are available from a spectroscopic survey, we can infer the three-dimensional positions of galaxies. 
In that case, the three-dimensional power spectrum of projected shapes becomes the observable quantity. 
On the other hand, in imaging surveys, where only photometric redshifts are generally available, the galaxy positions are effectively projected onto the sky, and the angular power spectrum becomes the primary observable.
We demonstrate that in both cases, the parity-odd power spectra found in the three-dimensional formalism show up as nonzero $EB$ cross-power spectra after the projection of galaxy shapes: $P_{EB}(k,\mu)$ for spectroscopic surveys and $C_{EB}(\ell)$ for photometric surveys, respectively\footnote{Our framework is not restricted to the inflationary model considered here and can be applied to other parity-odd primordial trispectra \citep[e.g.,][]{Creque-Sarbinowski+2023:PVTrisp_CSGrav,Yura+2025:PVTrisp_PMF}. 
Different collapsed-type trispectra generally show distinct scale dependences and helicity structures. 
These differences can in principle be distinguished through the multipole dependence of $P_{EB}(k,\mu)$ or the $\ell$-dependence of $C_{EB}(\ell)$ (see below). }.
Finally, we perform Fisher analyses to estimate how tightly current and future surveys can constrain the parameter $A_\mathrm{gauge}$.

\subsection{Spectroscopic sample} 
\label{subsec:specz_sample}
\subsubsection{3D power spectrum: \texorpdfstring{$P_{EB}(k,\mu)$}{PEB(k,mu)}}
The projected traceless shape tensor at position $\bx$ along the line-of-sight direction $\hbn$, denoted by $\gamma_{ij}(\bx;\hbn)$, is defined as \citep{Schmidt+2015:IA_PNG} 
\begin{align}
    \gamma_{ij}(\bx;\hbn) \equiv \calP_{ijkl}(\hbn) S_{kl}(\bx), 
    \label{eq:projection_gamma_x} 
\end{align}
where $S_{ij}$ is the original three-dimensional shape tensor, and $\mathcal{P}_{ijkl}$ is a rank-four projection operator given by
\begin{align}
    \calP_{ijkl}(\hbn) 
    &\equiv 
    \frac{1}{2} \left( \mathcal{P}_{ik}\mathcal{P}_{jl} + \mathcal{P}_{il}\mathcal{P}_{jk} - \mathcal{P}_{ij}\mathcal{P}_{kl} \right) \\
    &= 
    2 \Lambda^{(2,+)}_{ij,kl}(\hbn), 
\end{align}
which corresponds to the helicity-2, parity-even projection tensor with respect to the direction $\hbn$ (see Eq.~\ref{eq:Lambda2+_def}).
Here, we adopt the global plane-parallel (or distant-observer) approximation, where the line-of-sight direction for all galaxies in the survey is taken to be a fixed unit vector $\hbn$. 
Note, however, that when we later consider the angular power spectrum, we will adopt the full-sky formalism. 
Under this approximation, the projection in Fourier space can also be written in the same multiplicative form: 
\begin{align}
    \gamma_{ij}(\bk;\hbn) = \calP_{ijkl}(\hbn) S_{kl}(\bk). 
    \label{eq:projection_gamma_k} 
\end{align}

The two degrees of freedom in $\gamma_{ij}$ can be decomposed into the rotationally invariant $E$- and $B$-modes in Fourier space. 
Following Ref.~\cite{Kurita&Takada2022:AnalysisIAPS}, we extend the $E/B$ decomposition to include possible nonzero parity-violating signals, and derive the expression for the $EB$ cross-power spectrum (see Appendix~\ref{app:3d_EB_power_spectrum} for details):
\begin{align}
    P_{EB}(k,\mu) = -\frac{i}{2}\mu(1-\mu^2)P_-^{(1)}(k) -\frac{i}{4}\mu(1+\mu^2)P_-^{(2)}(k), 
    \label{eq:EB_power_k_mu} 
\end{align}
where $\mu \equiv \hbk \cdot \hbn$ is the cosine of the angle between the Fourier mode $\bk$ and the line-of-sight direction. 
Note that $P_{EB}$ is purely imaginary, and, hence, odd in $\mu$ due to the reality condition of the shape field. 
$P_{EB}$ becomes nonzero only in the presence of intrinsic parity violation, i.e., $P_-^{(\lambda)}\ne0$, and thus deviations of $P_{EB}$ from zero serve as a direct probe of parity-violating physics. 
We expand $P_{EB}$ in terms of Legendre polynomials: 
\begin{align}
    P_{EB}(k,\mu) = \sum_{\ell=1,3,\cdots} P_{EB}^{(\ell)}(k) \calL_\ell(\mu), 
    \label{eq:PEB_multipole_expansion}
\end{align}
and obtain the following expressions for the non-zero multipoles: 

\begin{align}
    P_{EB}^{(1)}(k) 
    &= 
    -\frac{i}{5}\left(P_-^{(1)}(k) + 2P_-^{(2)}(k)\right), \label{eq:PEB_L1}\\
    P_{EB}^{(3)}(k) 
    &= 
    -\frac{i}{10}\left(2P_-^{(1)}(k) + P_-^{(2)}(k)\right). \label{eq:PEB_L3}
\end{align} 
Although the projection reduces the number of independent degrees of freedom from five (helicities) to two ($E/B$-modes), which may appear to limit access to the underlying 3D shape information, it is worth noting that the helicity-$1$ and $2$ components have distinct $\mu$-dependence in \eq{eq:EB_power_k_mu}.
These lead to different contributions to the multipoles in \eq{eq:PEB_L1}-\eq{eq:PEB_L3}.
Therefore, in principle, the underlying helicity components can still be disentangled, as similarly demonstrated for parity-even signals in Ref.~\cite{Kurita&Takada2022:AnalysisIAPS}. 

\begin{figure}
    \centering
    \includegraphics[width=1.0\columnwidth]{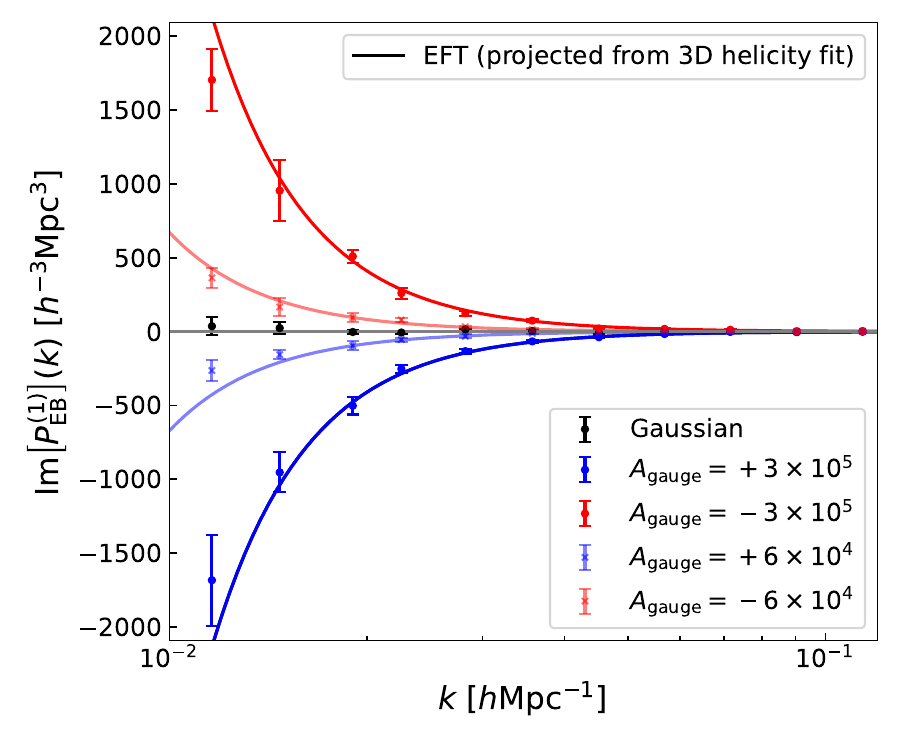}
    \caption{Dipole moment ($\ell=1$) of the $E/B$ power spectrum measured from the same halo sample as in Fig.~\ref{fig:EFT_fit} after the projection of their shapes. 
    The solid curve shows the EFT prediction given in Eq.~\eqref{eq:PEB_L1}, computed by inserting the best-fit bias parameters obtained from the three-dimensional helicity analysis in the previous section. 
    Thus, this is not a direct fit to $P_{EB}^{(1)}$ itself. 
    See the main text for further details. 
    }
    \label{fig:PEB_L1}
\end{figure}

Fig.~\ref{fig:PEB_L1} shows the dipole moment ($\ell=1$) of the $EB$ cross-power spectrum measured from the projected shapes of dark matter halos, using the method introduced in Ref.~\cite{Kurita+2020:IA_nbody}. 
We omit the octopole moment due to the low signal-to-noise ratio. 
The solid curve represents the EFT prediction, which is not obtained by directly fitting $P_{EB}$, but instead derived by projecting the theoretical model from the previous section. 
Specifically, we substitute the best-fit bias parameters from the three-dimensional helicity-spectrum analysis into Eq.~\eqref{eq:eft_model_Agauge}, and then evaluate the projected $EB$ spectrum using the analytic expression in Eq.~\eqref{eq:PEB_L1}. 
The good agreement between the simulation measurement and the EFT prediction confirms the consistency between the projection procedure applied to the simulation data and the corresponding theoretical modeling.

The multipole moments in real space are nonzero only for $\ell = 1$ and $3$, and this structure arises purely from the geometry of the projection. 
In particular, it is independent of the underlying parity-violating model. 
In actual galaxy surveys, redshift-space distortions (RSD) can break this structure by introducing additional angular dependencies. 
However, for IA signals, the impact of RSD is a higher-order effect \citep{Singh+2015:IA_measurement}. 
Moreover, since our analysis focuses on the large-scale limit, where such effects are suppressed, this approximation should remain valid for our purposes.
Therefore, higher multipoles $\ell \geq 5$ can be used as consistency checks or to identify potential systematics in the data. 
In the following analysis, we will use the multipoles $\ell=1$ and 3 as the observable signals.

\subsubsection{Fisher analysis: DESI Y5} 
Here, we perform a Fisher analysis to estimate the constraining power of the IA power spectrum on the amplitude parameter $A_\mathrm{gauge}$. 
The observable signals are the $EB$ cross-power spectra, and their amplitudes, i.e., the PNG-induced bias parameters $b_-^{(\lambda)}$, are determined as follows. 
We consider luminous red galaxies as a typical galaxy sample, since for these galaxies, the standard Gaussian (parity-even) IA signal has already been measured \citep[see e.g.,][]{Singh+2015:IA_measurement}. 
For example, Ref.~\cite{Kurita&Takada2023:AniPNGfromIA} measured the cross-power spectrum between the galaxy density field and the $E$-mode shape field, $P_{\mathrm{g}E}$, using the SDSS-III BOSS LOWZ and CMASS samples \citep{Reid+2016:LSScatalog}, and found 
$b_K^\mathrm{2D} \simeq -0.05$ at $z \simeq 0.5$.
We adopt this value as the fiducial value in our forecast and assume that it is well constrained by parity-even signals such as $P_{\mathrm{g}E}$. 
Therefore, we treat it as a fixed parameter rather than marginalizing over it as a free parameter.
Based on this fiducial value, we assume the redshift dependence to be
\begin{align}
    b_K^\mathrm{2D}(z) = -0.05 \times \frac{\tilde{D}(0.5)}{\tilde{D}(z)}, 
    \label{eq:bK_2D_assumption}
\end{align}
where $\tilde{D}$ is the linear growth factor normalized to unity at $z = 0$.
This redshift dependence is the same as that adopted in Ref.~\cite{Schmidt+2015:IA_PNG}, although the overall amplitude has been updated based on more recent observational results. 
It corresponds to assuming that 
$A_\mathrm{IA}(z) \propto b_K^\mathrm{2D}(z)\tilde{D}(z)$, 
a commonly used form of the IA amplitude in the weak lensing literature, is constant. 
This assumption is sometimes referred to as the ``primordial alignment'' scenario \citep{Hirata&Seljak2004:IA_LA}.

We explicitly write ``2D'' here because in actual observations, the galaxy ellipticity is defined based on the projected shape, which is normalized by the two-dimensional trace, $I_{11}+I_{22}$. 
This is different from our three-dimensional definition, which uses the full trace $I_{11}+I_{22}+I_{33}$, and thus results in a different normalization of $b_K$. 
As noted by Refs.~\cite{Bakx+2023:EFTofIAvsSims} and \cite{Maion+2024:HYMALAIA}, the two are related at leading order by 
\begin{align}
    b_K^\mathrm{2D} = 3 b_K. 
\end{align}
Taking this into account, we convert the fiducial $b_K^\mathrm{2D}$ into the corresponding three-dimensional $b_K$, and then use the fitting formula in Eq.~\eqref{eq:bias_best-fit_alpha} to determine the values of $b_-^{(\lambda)}$. 
Finally, we rescale the result as $[b_-^{(\lambda)}]^\mathrm{2D} = 3b_-^{(\lambda)}$ to properly match the two-dimensional normalization used in actual observations. 

Note that \eq{eq:bK_2D_assumption} corresponds to the commonly used IA amplitude parameter \citep{Joachimi+2011:AIA_C1} 
$A_{\rm IA} \simeq 4 \sim 5$, 
chosen to reproduce an IA amplitude of elliptical galaxies consistent with current observations \citep{Joachimi+2011:AIA_C1,Singh+2015:IA_measurement,Samuroff+2019:DESY1_IA,Johnston+2019:IA_measurement_GAMA,Fortuna+2021:IA_KiDS_LRGs,Samuroff+2023:IA_measurement_DES&eBOSS,Georgiou+2025:IA_KiDS_Bright,HervasPeters+2025:IA_UNIONS,Navarro-Girones+2026:IA_PAUS,Siegel+2026:IA_DESI_Y1} and measurements from hydrodynamical simulations \cite{Chisari+2016:IA_hydro,Shi+2021:IAPS_TNG,Samuroff+2021:IA_hydro,Delgado+2023IA_MTNG,Ferlito+2025:WL_IA_MTNG}. 
Hence, we do not adopt the $b_K$ value of dark matter halos. 
The IA amplitude entering the Fisher analysis is intended to be realistic for galaxies, while the halo calibration is used only to map this amplitude into an expected parity-odd response via the $b_-$--$b_K$ relation in \eq{eq:bias_relations:bm1_bK}. 
This procedure is directly analogous to constraints on local-type PNG $f_\mathrm{NL}$ from the scale-dependent bias in galaxy clustering, where a calibrated relation $b_\phi(b_1)$ is used to infer $b_\phi$ from the measured $b_1$. 

Given the model signal, and assuming that all bias and cosmological parameters other than the parameter of interest $A_\mathrm{gauge}$ are fixed, we obtain the expected $1\sigma$ error on $A_\mathrm{gauge}$ as 
\begin{widetext}
\begin{align}
    \sigma^{-2}\left(A_\mathrm{gauge}\right) 
    = 
    V_\mathrm{S} 
    \int_{k_\mathrm{min}}^{k_\mathrm{max}} 
    \frac{k^2\rmd k}{2\pi^2} 
    \sum_{\ell,\ell'} 
    \frac{\partial P_{EB}^{(\ell)}(k)}{\partial A_\mathrm{gauge}} 
    \left[\mathrm{Cov}\left[ P_{EB}^{(\ell)}, P_{EB}^{(\ell')}\right](k) \right]^{-1}
    \frac{\partial P_{EB}^{(\ell')}(k)}{\partial A_\mathrm{gauge}}, 
    \label{eq:fisher_mat_Agauge}
\end{align}
with $V_\mathrm{S}$ being the survey volume and the Gaussian covariance per mode $k$ 
\begin{align}
    \mathrm{Cov}\left[ P_{EB}^{(\ell)}, P_{EB}^{(\ell')}\right](k) 
    = 
    (2\ell+1) (2\ell'+1) 
    \int_{-1}^1\frac{\rmd \mu}{2} 
    \mathcal{L}_\ell(\mu) 
    \mathcal{L}_{\ell'}(\mu) \left[ \hat{P}_{EE}(k,\mu)\hat{P}_{BB}(k,\mu) + \hat{P}_{EB}^2(k,\mu) \right].   
    \label{eq:Cov_PEB}
\end{align}
\end{widetext}
We employ the linear theory for the sample variance with the Poisson shape noise under the fiducial (Gaussian) cosmology: 
\begin{align}
    \hat{P}_{EE}(k,\mu) 
    &= \frac{(1-\mu^2)^2}{4} \left(b_K^\mathrm{2D}\right)^2P(k) + \frac{\sigma_\gamma^2}{\bar{n}_\rmg}, \\
    \hat{P}_{BB}(k,\mu) 
    &= \frac{\sigma_\gamma^2}{\bar{n}_\rmg}, \\
    \hat{P}_{EB}(k,\mu) 
    &= 0, 
\end{align}
where $\sigma_\gamma$ is the intrinsic shape noise, and $\bar{n}_\rmg$ is the mean number density of the galaxy sample. 
We set $\sigma_\gamma = 0.17$ based on Ref.~\cite{Kurita&Takada2023:AniPNGfromIA} and neglect the redshift dependence of $\sigma_\gamma$.\footnote{Here, $\gamma$ refers to the shear, which is defined as the measured ellipticity $\epsilon$ divided by the shear responsivity factor $2\calR \equiv 2(1-\sigma_\epsilon^2)$, i.e., $\gamma = \epsilon/(2\calR)$ \citep{Bernstein&Jarvis2002:Responsivity}. 
The amplitude parameter shown in \eq{eq:bK_2D_assumption} is also defined with this normalization. 
For the sample used in Ref.~\cite{Kurita&Takada2023:AniPNGfromIA}, $\calR \simeq 0.93$, which corresponds to $\sigma_\epsilon \simeq 0.27$ \citep{Singh&Mandelbaum2016:IA_measurement}.} 
Note that since there is no fiducial signal in $P_{EB}$ and no contributions other than shape noise in $P_{BB}$, the covariance does not contain terms proportional to $P^2(k)$. 
Additionally, in practice, since $(b_K^\mathrm{2D})^2P(k) < \sigma_\gamma^2/\bar{n}_\rmg$, the covariance is dominated by shape noise. 
As a result, the covariance matrix for the multipoles per mode is approximately diagonal. 

We adopt survey parameters motivated by the final year dataset of the Dark Energy Spectroscopic Instrument (DESI) survey \citep{DESI_Collab2016:DESI_PartI}, hereafter DESI Y5. 
We assume sky coverages of $f_\mathrm{sky} = 0.33$. 
As our target galaxy samples, we first consider the DESI Luminous Red Galaxy (LRG) sample covering the redshift range $0.4 < z < 1.0$ \citep{Zhou+2023:DESI_LRG}, and assume 
a redshift-dependent galaxy number density, following Ref.~\citep{DESI_Collab2025:DR2_BAO}, with 
$\bar{n}_\rmg = 5 \times 10^{-4}~h^3\,\mathrm{Mpc}^{-3}$ for $0.4 < z < 0.8$, 
decreasing to $4 \times 10^{-4}$ at $0.8 < z < 0.9$, 
and $2 \times 10^{-4}$ at $0.9 < z < 1.0$. 
The effective number density and redshift for the sample are 
$\bar{n}_{\rmg,\mathrm{eff}} = 4.5 \times 10^{-4}~h^3\,\mathrm{Mpc}^{-3}$ and $z_\mathrm{eff} = 0.7$, respectively. 
In addition, we include the DESI Bright Galaxy Sample (BGS) covering $0.1 < z < 0.4$ \citep{Hahn+2023:DESI_BGS}. 
Since the BGS is selected based on an $r$-band magnitude cut, it includes both red and blue galaxies. 
Among them, the BGS red galaxies are expected to exhibit IA signals similar to those of the BOSS LOWZ sample, which consists of red, early-type galaxies in the same redshift range. 
Ref.~\cite{Siudek+2024:DESI_EDR_catalog} reported that approximately 50\% of the flux-limited BGS sample are red galaxies, based on an analysis using the DESI Early Data Release (EDR) galaxies. 
Based on this, we adopt an effective number density for IA-contributing red galaxies in the BGS sample as 
$\bar{n}_\rmg = 5 \times 10^{-4}~h^3\mathrm{Mpc}^{-3}$, 
corresponding to 50\% of the total BGS sample \citep{DESI_Collab2025:DR2_BAO}. 
The combined BGS (red) and LRG samples cover the redshift range $0.1 < z < 1.0$.  
Accounting for sky coverage, the corresponding comoving survey volume is 
$V_\mathrm{S} \simeq 17~h^{-3}\mathrm{Gpc}^3$ for DESI Y5. 

\begin{figure*}
    \centering
    \includegraphics[width=2.0\columnwidth]{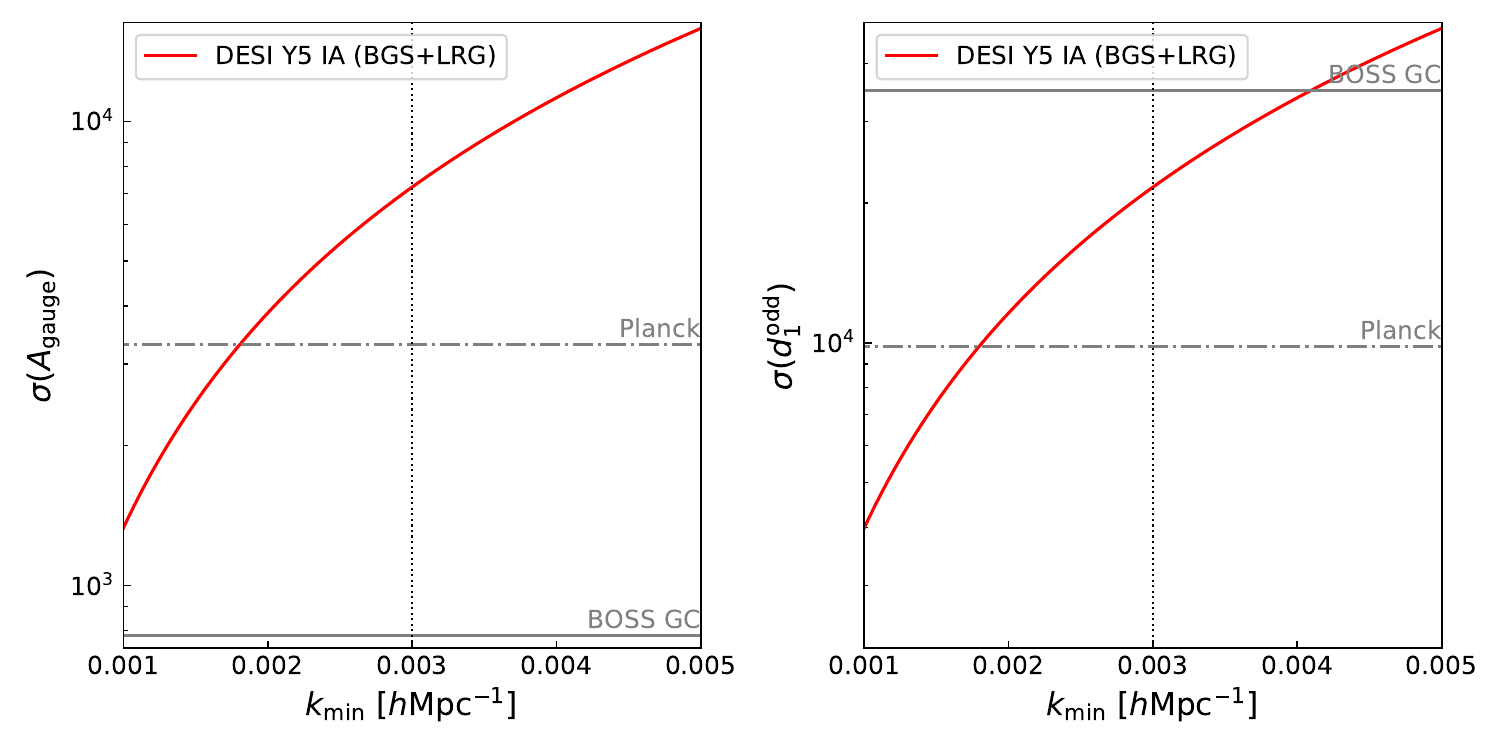}
    \caption{
    Forecasted $1\sigma$ constraints on the amplitude parameter of the $U(1)$-gauge model, $A_\mathrm{gauge}$ (left), and the $n=1$ component of the $d_n^\mathrm{odd}$ template, $d_1^\mathrm{odd}$ (right), using the three-dimensional IA power spectrum from BGS and LRG samples in DESI Y5, shown as functions of the minimum wavenumber $k_\mathrm{min}$ (red curves). 
    For comparison, current constraints from the galaxy 4PCF analysis of BOSS DR12 \citep{Philcox2022:PV4PCF_BOSS} and from the CMB trispectrum analysis using Planck $T$ and $E$ modes \citep{Philcox&Shiraishi2024:Constraints_PV_CMBTE} are shown as horizontal solid and dot-dashed lines, respectively. 
    The vertical dotted line indicates our fiducial scale cut, $k_\mathrm{min} = 0.003$, which matches the conservative choice adopted in the $f_\mathrm{NL}$ analysis of DESI DR1 \citep{Chaussidon+2024:DESI_2024_fnl}. 
    The specific values are also summarized in Table~\ref{tab:constraints_params}. 
    }
    \label{fig:fisher_1d_DESI}
\end{figure*}
\begin{table}
    \centering
    \begin{tabular}{cccc}
    \toprule\midrule
     & $10^{-2}\sigma(A_\mathrm{gauge})$ & $10^{-2}\sigma(d_0^\mathrm{odd})$ & $10^{-2}\sigma(d_1^\mathrm{odd})$\\
    \midrule
    BOSS GC & $ 7.8 $ & $ 7.8 $ & $ 350 $\\
    Planck & $ 33 $ & $ 1.1\times10^7 $ & $ 98 $\\
    \midrule
    DESI IA (fid) & 73 & - & 220 \\
    DESI IA (opt) & 14 & - & 41 \\
    LSST IA (fid) & 18 & - & 53 \\
    LSST IA (opt) & 2.7 & - & 8.0\\
    \midrule\bottomrule
    \end{tabular}
    \caption{Summary of $1\sigma$ constraints on the amplitude parameters of the parity-odd primordial trispectrum from the $U(1)$-gauge model: 
    $A_\mathrm{gauge}$, $d_0^\mathrm{odd}$, and $d_1^\mathrm{odd}$. 
    Top rows show current constraints from the galaxy clustering 4PCF analysis of BOSS DR12 \citep{Philcox2022:PV4PCF_BOSS} and the Planck CMB trispectrum analysis ($T$ and $E$) \citep{Philcox&Shiraishi2024:Constraints_PV_CMBTE}. 
    Bottom rows show forecasted constraints from IA measurements using DESI Y5 (BGS + LRG) and LSST Y10 (red galaxies), under both fiducial and optimistic scale cuts. 
    For DESI, we assume $k_\mathrm{min} = 0.003$ (fiducial) or $0.001$ (optimistic), and for LSST, we assume $\ell_\mathrm{min} = 10$ (fiducial) or $4$ (optimistic) with $f_\mathrm{red} = 0.1$. 
    Constraints from IA are not shown for $d_0^\mathrm{odd}$, as the signal is insensitive to this component. 
    The relation $A_\mathrm{gauge} = d_1^\mathrm{odd} / 3$ is used to translate between the parameterizations. 
    See also Figs.~\ref{fig:fisher_1d_DESI} and \ref{fig:fisher_1d_LSST} for the detail dependences on $k_\mathrm{min}$ and $\ell_\mathrm{min}$, respectively. 
    }
    \label{tab:constraints_params}
\end{table}
The left panel of Fig.~\ref{fig:fisher_1d_DESI} shows the $1\sigma$ constraint on the amplitude parameter of the $U(1)$-gauge model, $A_\mathrm{gauge}$, as a function of the minimum wavevector used in the analysis, $k_\mathrm{min}$. 
Throughout the analysis, we fix the maximum wavevector to $k_\mathrm{max} = 0.1$, which we have confirmed to have little impact on the results. 
Since the $P_{EB}$ signal is enhanced as $P_\phi(k) \propto k^{-3}$ in the large-scale limit, the constraints from IA improve significantly as $k_\mathrm{min}$ is pushed to smaller values. 
For the choice of $k_\mathrm{min}$, we follow the recent $f_\mathrm{NL}$ analysis in Ref.~\cite{Chaussidon+2024:DESI_2024_fnl}, which carefully investigated imaging systematics mitigation in large-scale galaxy power spectrum measurements from DESI DR1 data. 
We adopt $k_\mathrm{min}=0.003$ as the fiducial scale cut and consider $k_\mathrm{min}=0.001$ as the optimistic limit, based on the scales where their method is considered reliable for correcting geometrical systematics in DESI DR1 data\footnote{For simplicity, we use the (local) plane-parallel approximation for the Fisher forecast with the three-dimensional IA power spectrum. 
While standard in IA and galaxy power spectrum estimators \citep{Yamamoto+2006:estimator,Kurita&Takada2022:AnalysisIAPS}, wide-angle effects may become relevant on the very large scales considered here, and should be mitigated or modeled in future work. }. 
For comparison, we also plot the current constraints on $A_\mathrm{gauge}$ from galaxy clustering four-point correlation function (4PCF) analysis with the BOSS DR12 sample \citep{Philcox2022:PV4PCF_BOSS}, and from the CMB temperature $T$ and $E$-mode polarization trispectrum analysis with Planck PR4 data \citep{Philcox&Shiraishi2024:Constraints_PV_CMBTE}, shown as horizontal lines. 
Overall, our results indicate that the 3D $EB$ power spectrum can place competitive constraints on $A_\mathrm{gauge}$ with DESI Y5.  
With the fiducial scale cut, the expected constraint is of the same order as the current CMB limit, though slightly weaker.  
In contrast, the optimistic scale cut could yield a constraint that is potentially tighter than the CMB bound and comparable to current limits from the LSS analysis. 
The specific values are summarized in Table~\ref{tab:constraints_params}. 

Note that, naively, imaging systematics are expected to affect angular modes most strongly, i.e., the $\mu = 0$ modes, which are perpendicular to the line of sight \citep{Hand+2017:estimator}. 
On the other hand, as shown in Eq.~\eqref{eq:EB_power_k_mu}, $P_{EB}$ is an odd function of $\mu$, which suggests that it may be less sensitive to contaminants that strongly affect the $\mu = 0$ modes. 
While this needs to be tested with realistic mock data since we need to take into account a nontrivial mode-coupling due to the survey window function, if confirmed, it could indicate that $P_{EB}$ is less affected by imaging systematics than the galaxy clustering power spectrum, potentially allowing access to larger scales (smaller $k_\mathrm{min}$).

The right panel of Fig.~\ref{fig:fisher_1d_DESI} shows the constraint on the amplitude parameter of the $n=1$ component of the $d_n^\mathrm{odd}$ template defined in Eq.~\eqref{eq:def_dn_odd_template}, i.e.,  $d_1^\mathrm{odd}$. 
As discussed in Section~\ref{subsec:case_study}, the constraint on $A_\mathrm{gauge}$ from IA measurements can be directly translated into a constraint on $d_1^\mathrm{odd}/3$ via \eq{eq:dn_odd_parametrization}, since the $n=0$ component ($d_0^\mathrm{odd}$) does not contribute in the collapsed limit. 
A similar situation arises in the CMB analysis of Ref.~\cite{Philcox&Shiraishi2024:Constraints_PV_CMBTE}, where the trispectrum configurations in the collapsed limit (with internal momenta $L \leq 10$) have limited sensitivity to $d_0^\mathrm{odd}$, and the constraint on $A_\mathrm{gauge}$ is therefore dominated by $d_1^\mathrm{odd}$ (see Table~\ref{tab:constraints_params}). 
In contrast, the galaxy 4PCF analysis behaves differently: the constraint on $A_\mathrm{gauge}$ is mainly determined by $d_0^\mathrm{odd}$. This is because the analysis in Ref.~\cite{Philcox2022:PV4PCF_BOSS} uses galaxy pair separations in the range $20 < r < 160~\hiMpc$, corresponding roughly to wavenumbers $0.04 < k < 0.3~\hMpci$. These scales are more sensitive to equilateral trispectrum configurations than to the collapsed limit. 
Since $d_0^\mathrm{odd}$ is suppressed in the collapsed limit but contributes significantly in equilateral configurations, it becomes the dominant contributor to the 4PCF signal on those scales. 
Given the relation among parameters (Eq.~\ref{eq:dn_odd_parametrization}),
\begin{align*}
    d_0^\mathrm{odd} = -d_1^\mathrm{odd}/3 = -A_\mathrm{gauge},
\end{align*}
probes that are sensitive to the $n=0$ component effectively constrain $d_0^\mathrm{odd}$, leading to $\sigma(A_\mathrm{gauge}) \simeq \sigma(d_0^\mathrm{odd})$. 
Conversely, probes that are sensitive to the $n=1$ component constrain $d_1^\mathrm{odd}$, yielding $\sigma(A_\mathrm{gauge}) \simeq \sigma(d_1^\mathrm{odd})/3$.
The former case applies to the galaxy clustering 4PCF analysis, while the latter applies to the CMB and IA measurements.

For $d_1^\mathrm{odd}$, we find that the IA $EB$ power spectrum provides constraints that are comparable to, or potentially tighter than, current limits from galaxy clustering and CMB analyses. 
Note that, for the reasons mentioned above, while both galaxy clustering and IA are observables of the late-time universe using galaxies, they probe different scales and are therefore complementary. 
Moreover, our analysis is based on the power spectrum (i.e., two-point statistics), whose covariance properties are well understood. 
This provides a practical advantage over the 4PCF analysis, where accurate estimation of the covariance matrix remains a challenging and actively studied issue \citep[see e.g.,][]{Hou+2023:PV4PCF_BOSS,Krolewski+2024:NoEvidence_PV_BOSS,Philcox&Ereza2025:Cov_PV_BOSS}.

\subsection{Photometric sample} 
\label{subsec:photoz_sample}

\subsubsection{Angular power spectrum: \texorpdfstring{$C_{EB}(\ell)$}{CEB(ell)}} 
We next consider the case in which only photometric redshifts are available and no spectroscopic redshift information is provided. 
In this case, the most natural observable is the angular power spectrum. 
The angular power spectrum for spin-2 observables on the curved sky is derived based on the formalism developed in Ref.~\cite{Schmidt&Jeong2012:Cosmic_Ruler} (see also Refs.~\cite{Schmidt&Jeong2012:LSSwithGW_2,Schmidt+2014:LSSwithGW_3,Biagetti&Orlando2020:PVGW_IA}). 
Below, we directly present the final expression. 

The full-sky $EB$ angular power spectrum between two shape samples located at redshifts $z$ and $z'$ (for simplicity, we first assume delta-function distributions in redshift ) is given by
\begin{widetext}
\begin{align}
    C_{EB}(\ell;z,z') 
    = 
    \pi 
    \frac{(\ell-2)!}{(\ell+2)!} 
    \sum_{\lambda=1}^2 
    N_C^{(\lambda)} 
    \frac{(\ell+\lambda)!}{(\ell-\lambda)!} 
    \int_k 
    P_-^{(\lambda)} (k;z,z') 
    \left.F_{E,\ell}^{(\lambda)}\right|^\rmI\!\!(k\chi(z)) \;
    \left.F_{B,\ell}^{(\lambda)}\right|^\rmI\!\!(k\chi(z')), 
    \label{eq:angular_ps_II_EB_zfix}
\end{align}
\end{widetext}
where $F_{X,\ell}^{(\lambda)}$ denotes the kernel function for the $X\in\{E,B\}$ mode sourced by a mode with helicity $\lambda$. 
$N_C^{(1)} = 2$ and $N_C^{(2)} = 1/2$ are normalization factors for each helicity. 
The superscript ``I'' indicates the intrinsic contribution to the observed shear (see Appendix~\ref{app:angular_EB_power_spectrum} for further details). 
We confirm that only the parity-odd three-dimensional power spectra, $P_-^{(\lambda)}$, contribute to the parity-odd angular power spectrum. 
The helicity-0 component does not contribute to parity-odd statistics.
Notice that $C_{EB}(\ell;z,z')$ is suppressed at high $\ell$ for $z=z'$, as the 3D power spectrum $P_{EB}$ vanishes for transverse $k$-modes [i.e., $\mu=0$; \eq{eq:EB_power_k_mu}]. 

From \eq{eq:angular_ps_II_EB_zfix}, the tomographic $EB$ spectrum averaged over redshift bins $a$ and $b$ is then computed as 
\begin{align}
    C_{EB}^{ab}(\ell) 
    \equiv 
    \int_0^\infty \rmd \chi\,p_a(\chi)
    \int_0^\infty \rmd \chi'\,p_{b}(\chi') 
    C_{EB}(\ell;z,z'), 
    \label{eq:angular_ps_II_EB_binned} 
\end{align}
where $p_a$ is the redshift (radial) distribution of galaxies in bin $a$, normalized as $\int_0^\infty \rmd \chi\, p_a(\chi) = 1$. 
In this section, we assume the overall redshift distribution of source galaxies to follow
\begin{align}
    p(z) \propto \left(\frac{z}{z_0}\right)^\alpha \exp\left[{- \left(\frac{z}{z_0}\right)^\beta}\right], 
\end{align}
and adopt source galaxy parameters based on the Rubin Observatory Legacy Survey of Space and Time (LSST), with $(z_0, \alpha, \beta) = (0.11, 2, 0.68)$ \citep{Mandelbaum+2018:LSST_DESC_Science_Requirement}. 
We divide this distribution into $N_\mathrm{tomo} = 8$ tomographic bins such that each bin has the same integrated redshift probability for a Fisher analysis. 
The effective redshift of each bin corresponds to 
$z_\mathrm{eff} = [0.23, 0.44, 0.60, 0.78, 0.98, 1.23, 1.60, 2.55]$. 

\begin{figure}
    \centering
    \includegraphics[width=1.0\columnwidth]{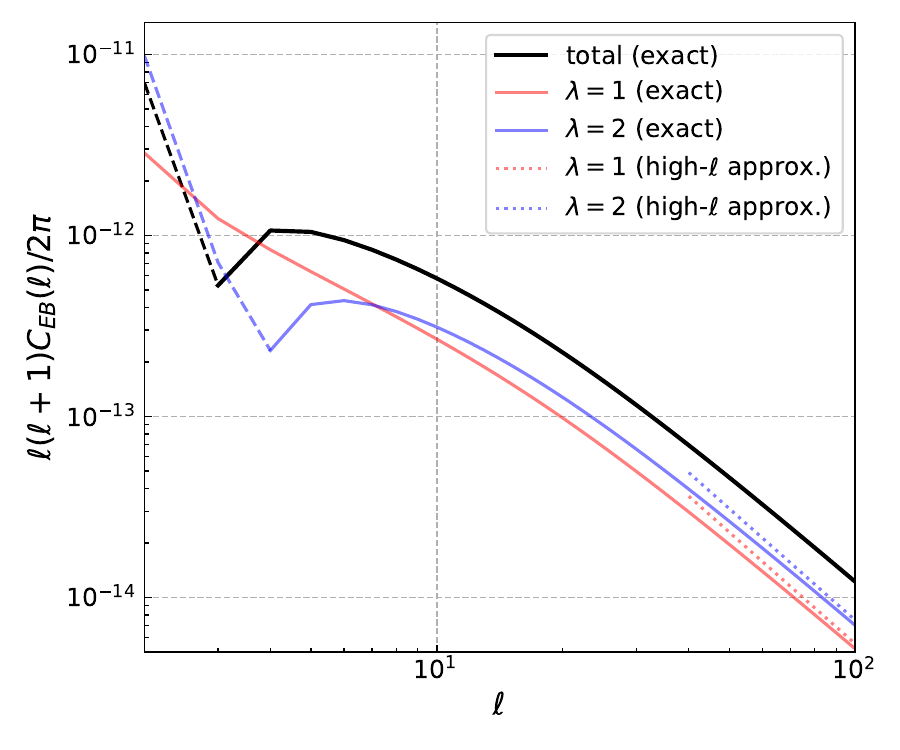}
    \caption{
    Angular $EB$ power spectrum of galaxy IA induced by the $U(1)$-gauge inflationary model, assuming $A_\mathrm{gauge} = 1$.
    The plot shows the auto-spectrum for the tomographic bin with effective redshift $z_\mathrm{eff} = 0.44$. 
    The red and blue lines indicate contributions from helicity $\lambda=1$ and $\lambda=2$, respectively, while the black line shows the total signal.
    Solid and dashed lines represent positive and negative values, respectively. 
    Dotted lines of each color correspond to the high-$\ell$ approximation results computed from \eq{eq:angular_ps_II_EB_binned_highl_lim}.
    }
    \label{fig:CEB}
\end{figure}
Fig.~\ref{fig:CEB} shows the angular $EB$ power spectrum of galaxy IA for the $U(1)$-gauge inflationary model, focusing on the auto-spectrum of the second tomographic bin with effective redshift $z_\mathrm{eff} = 0.44$.  
We set $A_\mathrm{gauge} = 1$ and assume \eq{eq:bK_2D_assumption} for the amplitude of bias parameters. 
The sign flip at low $\ell$ in the helicity-2 contribution is consistent with the results of Ref.~\cite{Biagetti&Orlando2020:PVGW_IA}, which studied parity-violating tensor fossil effects on IA. 
The high-$\ell$ (Limber) approximation of the tomographic $EB$ power spectrum (see Appendix~\ref{subapp:high_ell} for derivation) is given by
\begin{align}
    C_{EB}^{aa}(\ell) 
    \simeq 
    \frac{3}{2\ell} 
    \sum_{\lambda=1}^2 
    \frac{1}{\lambda}
    \int_0^\infty \rmd \chi\,
    \frac{p_a^2(\chi)}{\chi^2} 
    P_-^{(\lambda)} \left(k=\frac{\ell}{\chi};\chi\right), 
    \label{eq:angular_ps_II_EB_binned_highl_lim}
\end{align}
which shows that the spectrum scales as $\ell^{-1} P_-^{(\lambda)}(\ell/\chi) \propto \ell^{-4}$ at high $\ell$, leading to a strong suppression on small angular scales, consistent with the Limber limit ($\mu=0$) of $P_{EB}$. 

\subsubsection{Fisher analysis: LSST Y10} 
We perform a Fisher analysis to estimate the expected constraint on the amplitude parameter $A_\mathrm{gauge}$ using the IA angular power spectra. 
Throughout this analysis, we assume survey specifications consistent with the final year of LSST, hereafter referred to as LSST Y10. 
The Fisher matrix is given by
\begin{widetext}
\begin{align}
    \sigma^{-2}\left(A_\mathrm{gauge}\right) 
    = 
    \sum_{a,b,c,d}
    \sum_\ell 
    \frac{\partial C_{EB}^{ab}(\ell)}{\partial A_\mathrm{gauge}} 
    \left[\mathrm{Cov}\left[ C_{EB}^{ab}, C_{EB}^{cd}\right](\ell) \right]^{-1}
    \frac{\partial C_{EB}^{cd}(\ell)}{\partial A_\mathrm{gauge}}, 
    \label{eq:fisher_mat_Agauge_CEB}
\end{align}
with the Gaussian covariance per multipole $\ell$ given by
\begin{align}
    \mathrm{Cov}\left[ C_{EB}^{ab}, C_{EB}^{cd} \right](\ell) 
    = 
    \frac{1}{(2\ell+1)f_\mathrm{sky}}
    \left[
    \hat{C}_{EE}^{ac}(\ell) \hat{C}_{BB}^{bd}(\ell)
    +
    \hat{C}_{EB}^{ad}(\ell) \hat{C}_{EB}^{cb}(\ell)
    \right], 
    \label{eq:Cov_CEB}
\end{align}
\end{widetext}
where $f_\mathrm{sky}$ denotes the sky coverage fraction. 
We adopt $f_\mathrm{sky}=0.44$ for LSST Y10 dataset. 
The observed (noise-included) angular spectra are given by
\begin{align}
    \hat{C}_{EE}^{ab}(\ell)
    &= 
    \left.C_{EE}^{ab}\right|^{\rmGG+\rmGI+\rmIG+\rmII}(\ell) 
    + \frac{\sigma_{\gamma,a}^2}{\bar{n}_{\rmg,\mathrm{red}}^a} \delK_{ab}, \label{eq:Cl_hat_EE}\\
    \hat{C}_{BB}^{ab}(\ell)
    &= 
    \frac{\sigma_{\gamma,a}^2}{\bar{n}_{\rmg,\mathrm{red}}^a} \delK_{ab}, \\
    \hat{C}_{EB}^{ab}(\ell)
    &= 
    0.
\end{align}
The first term in Eq.~\eqref{eq:Cl_hat_EE}, accounting for both weak lensing (``G'') and IA (``I'') contributions, is explicitly given in \eq{eq:angular_ps_TOT_EE_tomo}.

We estimate the shape noise as follows. 
Assuming the LSST Y10 total galaxy number density to be 
$\bar{n}_{\rmg,\mathrm{tot}} = 27~\mathrm{arcmin}^{-2}$, 
and recalling that the tomographic bins are constructed to have equal galaxy counts, the number density in each bin is 
$\bar{n}_\rmg^a = \bar{n}_{\rmg,\mathrm{tot}} \int_{z\in\mathrm{bin}\,a} p(z) = \bar{n}_{\rmg,\mathrm{tot}}/N_\mathrm{tomo}$. 
Furthermore, since direct measurements of IA are typically restricted to red galaxies that exhibit strong IA signals, we introduce a fraction of red galaxies $f_\mathrm{red}^a$ for each bin. 
The effective number density used in the covariance is then given by
$\bar{n}_{\rmg,\mathrm{red}}^a = f_\mathrm{red}^a \bar{n}_\rmg^a$, 
which corresponds to the number of red galaxies per steradian in tomographic bin $a$. 
Although the precise value of $f_\mathrm{red}^a$ is currently uncertain, we consider $f_\mathrm{red}=0.1$ as a fiducial redshift-independent value in our analysis, which is consistent with Ref.~\cite{Schmidt+2015:IA_PNG}. 
For the variance of intrinsic ellipticities in the numerator, we adopt $\sigma_\epsilon = 0.26$. 
Assuming a shear responsivity factor of ${\cal R} = 1$, this corresponds to a shear variance of $\sigma_\gamma = 0.13$.

\begin{figure*}
    \centering
    \includegraphics[width=2.0\columnwidth]{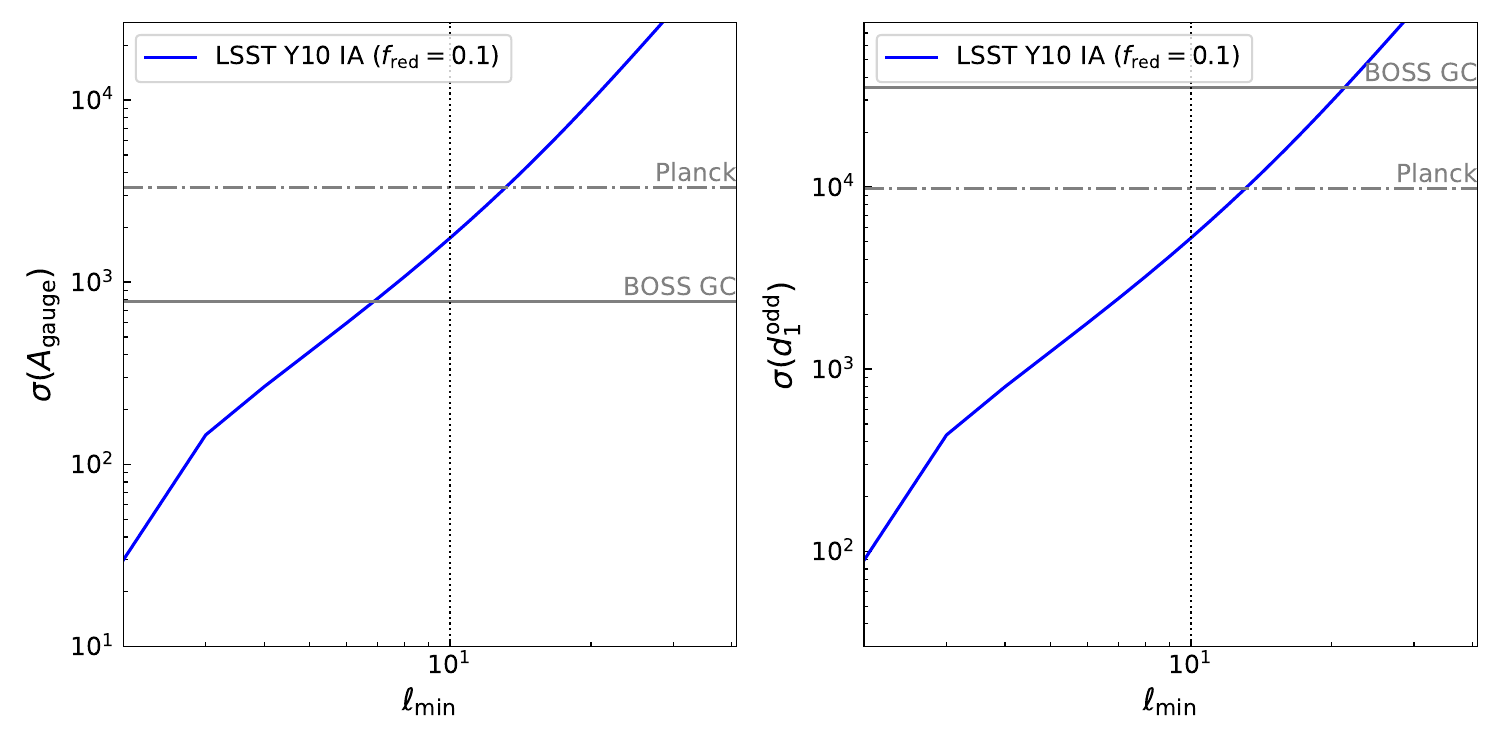}
    \caption{
    Same as Fig.~\ref{fig:fisher_1d_DESI}, but for LSST Y10 using the angular IA power spectrum from red galaxies (blue curves). 
    The forecasted $1\sigma$ constraints on $A_\mathrm{gauge}$ (left) and $d_1^\mathrm{odd}$ (right) are shown as functions of the minimum multipole $\ell_\mathrm{min}$. 
    Horizontal lines indicate current constraints from BOSS 4PCF (solid) and Planck CMB trispectrum (dot-dashed), and the vertical dotted line at $\ell_\mathrm{min} = 10$ corresponds to our fiducial scale cut. 
    The specific values are also summarized in Table~\ref{tab:constraints_params}. 
    }
    \label{fig:fisher_1d_LSST}
\end{figure*}
Fig.~\ref{fig:fisher_1d_LSST} shows the forecasted $1\sigma$ errors from LSST Y10 data: 
the left panel shows the constraint on $A_\mathrm{gauge}$, and the right panel on $d_1^\mathrm{odd}$, as functions of the minimum multipole $\ell_\mathrm{min}$ used in the analysis. 
As in the case of the 3D power spectrum (Fig.~\ref{fig:fisher_1d_DESI}), access to larger scales significantly improves the constraint, although observational systematics at these scales remain a challenge. 
Compared to the DESI case, LSST provides tighter constraints mainly due to its higher galaxy number density, which reduces the shape noise. 
In addition, since the LSST sample covers higher redshifts, it probes a larger volume. 
In this case, these two effects win out over the loss in information due to the projection to angular statistics. 
In both analyses, the covariance is dominated by the shape noise contribution, and thus the expected error bars approximately scale inversely with the shape noise. 
For example, at $\ell_\mathrm{min} = 10$, the improvement factor relative to the constraint obtained with $f_\mathrm{red} = 0.1$ is in the range $[0.53,1.85]$ for $f_\mathrm{red} = [0.05,0.2]$. 
Under this fiducial scale cut, the constraint on $A_\mathrm{gauge}$ is already comparable to current limits from the CMB and galaxy clustering, while the constraint on $d_1^\mathrm{odd}$ is tighter than both.
With the optimistic scale cut (e.g., $\ell_\mathrm{min} \simeq 4$), LSST has the potential to improve upon current constraints by up to an order of magnitude.

\section{Conclusion}
In this work, we investigated, for the first time, how IA of galaxies can probe parity-violating primordial non-Gaussianity (PNG). 
Combining EFT modeling (Section~\ref{sec:theory}), $N$-body simulations (Section~\ref{sec:nbody_sim}), and forecast analyses (Section~\ref{sec:projection_effects}), we established IA as a novel and competitive probe of parity-violating physics in the early universe. 

We derived analytic expressions for the IA parity-odd power spectrum within the EFT framework up to one-loop order, introducing projectors that isolate the parity-odd components and facilitate the calculation. 
The parity-odd contributions only come in through non-zero helicity contributions, which means only the $(22)$-type loop integrals contribute to the parity-odd two-point functions. We also find new structures for the stochastic (shot noise) contributions to the shape statistics, which involve odd powers of wavenumber $k$, the leading contribution scaling as $k$. 

We showed that the IA power spectrum is especially sensitive to collapsed-type parity-odd trispectra, which generate a strong scale-dependent enhancement in large-scale signals. 
This behavior can be physically interpreted as large-scale correlations of helical modulations in the small-scale power spectrum, arising from the collapsed limit of the parity-odd trispectrum. 
These correlations coherently distort galaxy shapes and give rise to large-scale IA correlations. 

To validate these predictions, we developed a flexible and efficient method to implement collapsed-type parity-odd trispectra into $N$-body initial conditions. 
These simulations are useful beyond shape statistics and could be used to investigate the halo four-point function as well, for example. 
The simulations confirmed the expected scale-dependent enhancements and enabled a precise determination of the PNG-induced bias parameters that cannot be predicted from the EFT or first principles.

We further showed that, when the projection of galaxy shapes in observations on the sky is taken into account, the parity-odd signal appears as the cross power spectrum of $E$ and $B$ modes. 
Fisher forecasts for DESI \citep{DESI_Collab2016:DESI_PartI} and LSST \citep{LSST2009:LSST} indicate that IA can provide constraints comparable to, or potentially tighter than, those from existing CMB and galaxy four-point correlation analyses.
Note that the covariance of these large-scale two-point statistics is essentially determined only by the shape noise and is thus very well known. 
This contrasts with the case for the galaxy four-point function on small scales. 
On the other hand, the signal is proportional to a new alignment bias coefficient, which needs to be estimated from simulations. 

Beyond two-point statistics, higher-order IA correlations such as the bispectrum offer a complementary probe of parity violation \citep[e.g.,][]{Pyne+2022:IA_Bistectrum,Linke+2024:IA_3pt_LOWZ,Bakx+2025:IA_Bistectrum_I,Bakx+2026:IA_Bistectrum_II}. 
Unlike galaxy density, tensor quantities like galaxy shapes can still be sensitive to parity-odd signals at the three-point level, and probe configurations beyond the collapsed limit. 
A further open issue is the calculation of wide-angle effects on the three-dimensional shape statistics, which could become relevant on the very large scales where the signal is significant. 
On these very large scales, additive shear systematics such as PSF leakage could generate spurious parity-odd signals. 
Detailed modeling and mitigation of these effects are beyond the scope of this work and are left for future study. 

Ongoing and future galaxy surveys such as DESI, Euclid \citep{Laureijs+2011:Euclid}, Subaru PFS \citep{Takada+2014:PFS}, LSST, the 4-metre Multi-Object Spectroscopic Telescope (4MOST) survey \citep{deJong+2019:4MOST}, and the Roman Space Telescope \citep{Akeson+2019:Roman} will deliver high-quality data that can substantially improve IA measurements. 
They will broaden the range of galaxy samples available and enable more precise and robust tests of parity-violating physics with galaxy shapes.

\acknowledgments 
We thank Jiamin Hou for insightful discussions. 
The $N$-body simulations were carried out on the \texttt{freya} and \texttt{orion} clusters, maintained by the Max Planck Computing and Data Facility. 
We also thank the organizers and participants of \textit{Parity Violation from Home 2024}\footnote{\url{https://parity.cosmodiscussion.com/parity-violation-from-home-2024/}} for useful discussions and feedback. 
This work was supported in part by JSPS KAKENHI Grant Nos.~JP20H05850 and JP20H05859, and the Deutsche Forschungsgemeinschaft (DFG, German Research Foundation) under Germany's Excellence Strategy---EXC-2094---390783311. 
This work has also received funding from the European Union’s Horizon 2020 research and innovation programme under the Marie Skłodowska-Curie Grant Agreement No.~101007633. 
The Kavli IPMU is supported by World Premier International Research Center Initiative (WPI), MEXT, Japan. 

\clearpage
\onecolumngrid
\appendix
\section{Operator degeneracy} 
\label{app:operator_degeneracy} 
We derive the parity-odd power spectra given in \eqs{eq:pop_spectrum_lambda1}{eq:pop_spectrum_lambda2}. 
To simplify the expression for $P^{(22)}$, we utilize the degeneracies among the second-order kernels (Eqs.~\ref{eq:kernel_nqp_211}--\ref{eq:kernel_nqp_223}) derived in Ref.~\citep{Vlah+2020:IA_EFT}. 
Since the argument regarding the longitudinality of \eq{eq:Oij_11_Fourier} holds at all orders, the second-order contribution from $\calO_{ij}^{(1,1)}$ in $P^{(22)}$ is also irrelevant, as seen in \eq{eq:kernel_nqp_211} with $\hbq_{12}=\hbk$. 
Additionally, for $\calO_{ij}^{(2,1)}$, the first term in \eq{eq:kernel_nqp_221} can be ignored for our purposes, as it is purely longitudinal. 
Since the second term in \eq{eq:kernel_nqp_221} is identical to \eq{eq:kernel_nqp_222}, applying a transverse projection to $\calO_{ij}^{(2,1)}$, i.e., the projection onto helicity-1 and helicity-2 components, yields exactly the same result as $\calO_{ij}^{(2,2)}$. 
As a result, the parity-odd power spectra depend only on the sum of the bias coefficients, $c_{2,1} + c_{2,2}$, for both helicities. 
We are thus left with only three relevant operator correlations which are sensitive to the parity-violating signals: 
\begin{align}
    \avrg{\calO_{ij}^{(2,2)} \calO_{kl}^{(2,2)}},~
    \avrg{\calO_{ij}^{(2,2)} \calO_{kl}^{(2,3)}}_\mathrm{sym.},~
    \avrg{\calO_{ij}^{(2,3)} \calO_{kl}^{(2,3)}}, 
    \label{eq:op_corr_relevant}
\end{align}  
where 
$\avrg{\calO_{ij} \calO_{kl}'}_\mathrm{sym.} = \avrg{\calO_{ij} \calO_{kl}' + \calO_{ij}' \calO_{kl}}/2$ 
denotes the symmetrization for the cross term with respect to the operator labels (\textit{not} for the momenta). 
We explicitly show the parity-odd bias kernels for helicity $\lambda=1$ (Eq.~\ref{eq:projected_kernels}):
\begin{align}
    K_{\calO^{(2,2)}\calO^{(2,2)}}^{(1,-)}(\bq_1,\bq_2,\bq_3,\bq_4)
    &= \frac{i}{4} \left[\hbk\cdot(\hbq_1\times\hbq_3) \right] 
    \left[ \mu_{k1} - \frac{q_1}{q_2}\mu_{k2} \right]
    \left[ \mu_{k3} - \frac{q_3}{q_4}\mu_{k4} \right], \label{eq:prj_kernel_lambda1_22_22}\\
    K_{\left(\calO^{(2,2)}\calO^{(2,3)}\right)}^{(1,-)}
    &= K_{\calO^{(2,2)}\calO^{(2,2)}}^{(1,-)}, \label{eq:prj_kernel_lambda1_22_23}\\
    K_{\calO^{(2,3)}\calO^{(2,3)}}^{(1,-)}
    &= K_{\calO^{(2,2)}\calO^{(2,2)}}^{(1,-)}, \label{eq:prj_kernel_lambda1_23_23}
\end{align}
and for $\lambda=2$:
\begin{align}
    K_{\calO^{(2,2)}\calO^{(2,2)}}^{(2,-)}(\bq_1,\bq_2,\bq_3,\bq_4)
    &= \frac{i}{2} \left[\hbk\cdot(\hbq_1\times\hbq_3) \right] 
    (\mu_{13} - \mu_{k1}\mu_{k3})
    \frac{q_1}{q_2}\mu_{12}
    \frac{q_3}{q_4}\mu_{34}, \label{eq:prj_kernel_lambda2_22_22}\\
    K_{\left(\calO^{(2,2)}\calO^{(2,3)}\right)}^{(2,-)}(\bq_1,\bq_2,\bq_3,\bq_4)
    &= -\frac{i}{8} \left[\hbk\cdot(\hbq_1\times\hbq_3) \right] 
    (\mu_{13} - \mu_{k1}\mu_{k3})
    \left[\frac{q_1}{q_2}\mu_{12}\left(1+\frac{q_3^2}{q_4^2}\right) + \left(1+\frac{q_1^2}{q_2^2}\right)\frac{q_3}{q_4}\mu_{34}\right], \label{eq:prj_kernel_lambda2_22_23}\\
    K_{\calO^{(2,3)}\calO^{(2,3)}}^{(2,-)}(\bq_1,\bq_2,\bq_3,\bq_4)
    &= \frac{i}{8} \left[\hbk\cdot(\hbq_1\times\hbq_3) \right] 
    (\mu_{13} - \mu_{k1}\mu_{k3})
    \left(1+\frac{q_1^2}{q_2^2}\right)
    \left(1+\frac{q_3^2}{q_4^2}\right). \label{eq:prj_kernel_lambda2_23_23}
\end{align} 
Note that the expressions for the cross term have already been symmetrized. 
From Eqs.~\eqref{eq:prj_kernel_lambda1_22_22}--\eqref{eq:prj_kernel_lambda1_23_23}, we find that the contributions to the helicity-1 spectrum from $\calO_{ij}^{(2,2)}$ and $\calO_{ij}^{(2,3)}$ are identical, which means it depends only on $c_{2,1} + c_{2,2} + c_{2,3}$. 
Considering the following algebla: 
\begin{align*}
    S_{ij} 
    &\supset c_{2,1}\calO_{ij}^{(2,1)} + c_{2,2}\calO_{ij}^{(2,2)} + c_{2,3}\calO_{ij}^{(2,3)} \\
    &= c_{2,1}\calO_{ij}^{(2,1)} + \left(c_{2,2}+c_{2,3}\right)\calO_{ij}^{(2,2)} + c_{2,3}\left(\calO_{ij}^{(2,3)} - \calO_{ij}^{(2,2)}\right), 
\end{align*}
and introducing the new operator labels: 
$Q_{ij}\equiv \calO_{ij}^{(2,2)}$ and $R_{ij}\equiv \calO_{ij}^{(2,3)} - \calO_{ij}^{(2,2)}$, 
we obtain the expression for the spectra in \eqs{eq:pop_spectrum_lambda1}{eq:pop_spectrum_lambda2} and the corresponding kernels in Eqs.~\eqref{eq:prj_kernel_lambda1_QQ}--\eqref{eq:prj_kernel_lambda2_RR}.

\section{Squeezed-type loop calculations} 
\label{app:calc_detail_squeezed} 

\subsection{Implementation of loop integral} 
\label{subapp:implementation_of_loop_int_squeezed} 
We describe in detail how we numerically implemented the two-loop integral given by \eq{eq:pop_spectrum_XY_explicit}. 
In principle, since the integral involves two momentum variables, it is originally a six-dimensional integral. 
However, due to azimuthal symmetry, it reduces to a four-dimensional integral. Furthermore, because of the separability of the squeezed trispectrum, it factorizes into a product of two two-dimensional integrals. 
That is, the shapes of the two triangles containing the diagonal $\bk$ can be integrated independently. 
As a result, the computational complexity is reduced to $\mathcal{O}(N^2)$.
The choice of how to perform this two-dimensional integration depends on the parameterization of the triangle, leading to different integration variables. 
One approach, which analytically performs the angular integrals using the spherical harmonic expansion of the Dirac delta function \citep[see, e.g., Ref.~][]{Jamieson+2024:POP}, uses the lengths of the two other sides $(q_1, q_2)$ as variables. 
Alternatively, one can instead take the length of one side, $q$, and the cosine of the angle between $\bq$ and $\bk$, $\mu = \hbq \cdot \hbk$, as integration variables, i.e., $(q, \mu)$.
In this study, we adopt the latter parameterization, as it simplifies the angular integration over $\mu$ after taking the UV limit of the integrand.
(We have checked that the integration results shown in Fig.~\ref{fig:pt} of the main text agree between both methods.) 

If the given trispectrum is independent of the diagonal and separable with respect to each momentum, meaning that it can be expressed as a product of individual functions as in Eq.~\eqref{eq:def_f-_squeezed}, the sum over permutations for the antisymmetric part (Eq.~\ref{eq:def_f-}) can be rewritten as follows:
\begin{align}
    T_\delta^{(-)}(\bq_1,\bq_2,\bq_3,\bq_4) 
    &= 
    \left[-i \bQ\cdot\left(\bq_1\times\bq_3\right) \right] 
    \prod_{i=1}^4\left[\calM(q_i)\right]
    \sum_{\sigma \in S_4} \mathrm{sgn}(\sigma) 
    f_-(q_{\sigma(1)},q_{\sigma(2)},q_{\sigma(3)},q_{\sigma(4)}) \\
    &\equiv 
    -g_-\left[-i \bQ\cdot\left(\bq_1\times\bq_3\right) \right] 
    \sum_{\sigma \in S_4} \mathrm{sgn}(\sigma) 
    f_{\sigma(1)}(q_1) 
    f_{\sigma(2)}(q_2)
    f_{\sigma(3)}(q_3)
    f_{\sigma(4)}(q_4) \\
    &= 
    -g_-\left[-i \bQ\cdot\left(\bq_1\times\bq_3\right) \right] 
    \sum_{\sigma \in S_4} \mathrm{sgn}(\sigma) 
    f_{\left[\sigma(1)\right.}(q_1) 
    f_{\left.\sigma(2)\right]}(q_2)
    f_{\left[\sigma(3)\right.}(q_3)
    f_{\left.\sigma(4)\right]}(q_4)
    \label{eq:def_matter_trisp_fsigma}
\end{align}
where we defined the following four functions in the second line: 
\begin{align}
    f_1(q) &\equiv q^\alpha \calM(q) P_\phi(q),~\label{eq:def_f1}\\
    f_2(q) &\equiv q^\beta \calM(q) P_\phi(q),~\label{eq:def_f2}\\
    f_3(q) &\equiv q^\gamma \calM(q) P_\phi(q),~\label{eq:def_f3}\\
    f_4(q) &\equiv \calM(q). \label{eq:def_f4}
\end{align}
Thus, the previous permutations of the momenta have been equivalently replaced by permutations of the functions. 
In the third line, we defined the antisymmetrization with respect to the function for later convenience: 
$f_{\left[a\right.}(q) f_{\left.b\right]}(p) \equiv (f_a(q)f_b(p) - f_b(q)f_a(p))/2$. 
By eliminating $\bq_2$ and $\bq_4$ with the triangle condition: $\bk=\bq_{12}=-\bq_{34}$, the parity-odd kernel function with helicity $\lambda=1$ (Eq.~\ref{eq:prj_kernel_lambda1_QQ}) and the trispectrum (Eq.~\ref{eq:def_matter_trisp_fsigma}) can be expressed as 
\begin{align}
    K_{QQ}^{(1,-)}(\bq,\bq';\bk)
    &= -\frac{i}{4} \left[\hbk\cdot(\hbq\times\hbq') \right] 
    \frac{(k\mu-q)(2q\mu-k)}{|\bk-\bq|^2}
    \frac{(k\mu'+q')(2q'\mu'+k)}{|\bk+\bq'|^2}, 
    \label{eq:prj_kernel_lambda1_QQ_kqp_notation} \\
    T_\delta^{(-)}(\bq,\bq';\bk) 
    &= ig_-k^3 \left[\hbk\cdot(\hbq\times\hbq') \right] \frac{q}{k} \frac{q'}{k}
    \sum_{\sigma \in S_4} \mathrm{sgn}(\sigma) 
    f_{\left[\sigma(1)\right.}(q) 
    f_{\left.\sigma(2)\right]}(|\bk-\bq|)
    f_{\left[\sigma(3)\right.}(q')
    f_{\left.\sigma(4)\right]}(|\bk+\bq'|), 
    \label{eq:matter_trisp_fsigma_kqp_notation}
\end{align}
where we relabel $\bq=\bq_1,\bq'=\bq_3$ and $\mu \equiv \hbk\cdot\hbq, \mu' \equiv \hbk\cdot\hbq'$. 
For later convenience, we note 
\begin{align}
    \left[\hbk\cdot(\hbq\times\hbq') \right]^2 
    = 1+2\mu\mu'\tilde{\mu}-(\mu^2+\mu'^2+\tilde{\mu}^2),
    \label{eq:def_stp2}
\end{align}
with $\tilde{\mu} \equiv \hbq\cdot\hbq'$. 
Consequently, the parity-odd power spectrum (Eq.~\ref{eq:pop_spectrum_XY_explicit}) simplifies to 
\begin{align}
    P_{QQ}^{(1,-)}(k) 
    &= g_-\frac{k^3}{4} 
    \sum_{\sigma \in S_4} \mathrm{sgn}(\sigma) 
    \int_{\bq,\bq'} 
    \left[ 1+2\mu\mu'\tilde{\mu}-(\mu^2+\mu'^2+\tilde{\mu}^2) \right] \nonumber\\
    &\times 
    \left[
    \frac{q}{k} 
    \frac{(k\mu-q)(2q\mu-k)}{|\bk-\bq|^2} 
    f_{\left[\sigma(1)\right.}(q) f_{\left.\sigma(2)\right]}(|\bk-\bq|)
    \right]
    \left[
    \frac{q'}{k} 
    \frac{(k\mu'+q')(2q'\mu'+k)}{|\bk+\bq'|^2}
    f_{\left[\sigma(3)\right.}(q') f_{\left.\sigma(4)\right]}(|\bk+\bq'|)
    \right]. 
    \label{eq:pop_spectrum_h1_QQ_squeezed_with_mu-tilde}
\end{align}
Here, we note that the integration over the two momenta $(\bq, \bq')$ can be expressed in a separable form using some isotropic functions $g$ and $h$, i.e.,  $g(\bk,\bq)=g(k,q,\mu)$, except for the coupling that arises from the scalar triple product through $\tilde{\mu}$:
\begin{align}
    \int_{\bq,\bq'}
    \left[ 1+2\mu\mu'\tilde{\mu}-(\mu^2+\mu'^2+\tilde{\mu}^2) \right]
    g(\bk,\bq)h(\bk,\bq'). 
    \label{eq:def_stp2_gh}
\end{align}
In such cases, the integral, which involves cross-talk due to the dependence on $\tilde{\mu}$, can be rewritten in a separable form using the formula given in Appendix~\ref{app:angular_integral_formula}, as shown below. 
For the second and last term in \eq{eq:def_stp2_gh}, since  
$\mu\mu'\tilde{\mu} = \mu\mu' \hq_i \hq'_i$ 
and 
$\tilde{\mu}^2 = \hq_i \hq_j \hq'_i \hq'_j$, 
we can compute them as follows: 
\begin{align*}
    \left[\int_{\bq}  \hq_i \mu g\right]  
    \left[\int_{\bq'}  \hq'_i \mu' h\right] 
    &= 
    \left[\hk_i \int_{\bq} \mathcal{L}_1(\mu) \mu g\right]  
    \left[\hk_i \int_{\bq} \mathcal{L}_1(\mu') \mu' h\right] \\
    &= 
    \int_{\bq} \mu^2 g \int_{\bq'} \mu'^2 h, \\
    \left[\int_{\bq}  \hq_i \hq_j g\right]  
    \left[\int_{\bq'}  \hq'_i \hq'_j h\right] 
    &=  
    \left[  
    \left( \hk_i \hk_j - \frac{\delK_{ij}}{3} \right) \int_{\bq} \mathcal{L}_2(\mu) g  
    + \frac{\delK_{ij}}{3} \int_{\bq} g  
    \right]  
    \left[  
    \left( \hk_i \hk_j - \frac{\delK_{ij}}{3} \right) \int_{\bq'} \mathcal{L}_2(\mu') h  
    + \frac{\delK_{ij}}{3} \int_{\bq'} h 
    \right] \\
    &=
    \frac{3}{2} \int_{\bq} \mu^2 g \int_{\bq'} \mu'^2 h  
    +\frac{1}{2} 
    \left[ \int_{\bq} \mu^2 g \int_{\bq'} h + \int_{\bq} g \int_{\bq'} \mu'^2 h \right]
    + \frac{1}{2} \int_{\bq} g \int_{\bq'} h. 
\end{align*}
Therefore, \eq{eq:def_stp2_gh} can be reduced to the following form:
\begin{align}
    \int_{\bq,\bq'} 
    \left[ 1+2\mu\mu'\tilde{\mu}-(\mu^2+\mu'^2+\tilde{\mu}^2) \right] 
    g(\bk,\bq)h(\bk,\bq') 
    = 
    \frac{1}{2}
    \left[  
    \int_{\bq} 
    (1-\mu^2) g(\bk,\bq)
    \right]
    \left[  
    \int_{\bq'} 
    (1-\mu'^2) h(\bk,\bq')
    \right]. 
\end{align}
From the above, Eq.~\eqref{eq:pop_spectrum_h1_QQ_squeezed_with_mu-tilde} can be further reduced, yielding the following simplified expression: 
\begin{align} 
    P_{QQ}^{(1,-)}(k;\Lambda,\Lambda') 
    &= 
    g_-\frac{k^3}{8} 
    \sum_{\sigma \in S_4} \mathrm{sgn}(\sigma) I_{\sigma(1)\sigma(2)}(k;\Lambda) I_{\sigma(3)\sigma(4)}(k;\Lambda') 
    \label{eq:pop_spectrum_h1_QQ_squeezed_final_sgnsum} \\
    \Bigl(&= 
    g_-\frac{k^3}{2} 
    \left[ 
    I_{12}(k;\Lambda)I_{34}(k;\Lambda') 
    + I_{13}(k;\Lambda)I_{42}(k;\Lambda') 
    + I_{14}(k;\Lambda)I_{23}(k;\Lambda')\right]
    + (\Lambda \leftrightarrow \Lambda')
    \Bigr), 
    \label{eq:pop_spectrum_h1_QQ_squeezed_final_explicit}
\end{align}
where we explicitly wrote the dependence on the UV cut-off scale, $\Lambda$, on both sides and defined
\begin{align}
    I_{ab}(k;\Lambda)
    &\equiv
    \int_{q_\mathrm{min}}^\Lambda \frac{q^2\mathrm{d}q}{2\pi^2} 
    \int_{-1}^1 \frac{\mathrm{d}\mu}{2}
    (1-\mu^2)\frac{q}{k}\frac{(k\mu-q)(2q\mu-k)}{|\bk-\bq|^2} 
    f_{\left[a\right.}(q) 
    f_{\left.b\right]}(|\bk-\bq|), 
    \label{eq:def_Iab} 
\end{align}
with $a,b\in\{1,2,3,4\}$. 
Note that the $\mu$ integral was implemented in an IR-safe manner by replacing:
\begin{align*}
    \int_{-1}^1 \frac{\mathrm{d}\mu}{2} \to 2\int_{-1}^{\mathrm{min}(k/2q,1)} \frac{\mathrm{d}\mu}{2}, 
\end{align*}
using the symmetry of the integrand to avoid the pole at $\bq = \bk$, which could potentially lead to an IR divergence.

\subsection{A proof of no helicity-2 spectrum} 
\label{subapp:proof_no_helicity-2} 
We prove that the squeezed-type parity-odd trispectrum, as defined in \eq{eq:def_f-_squeezed}, does not contribute to the parity-odd power spectrum for helicity $\lambda=2$.
Following the same argument as above, the parity-odd kernel functions with helicity $\lambda=2$ (Eqs.~\ref{eq:prj_kernel_lambda2_QQ}--\ref{eq:prj_kernel_lambda2_RR}) take the following functional form:
\begin{align}
    K_{XY}^{(2,-)}(\bq,\bq';\bk)
    \propto i\left[\hbk\cdot(\hbq\times\hbq') \right] 
    (\tilde{\mu} - \mu\mu')
    x(\bk,\bq)y(\bk,\bq'), 
    \label{eq:prj_kernel_lambda2_QQ_kqp_notation}
\end{align}
with $X,Y\in\{Q,R\}$, where $x$ and $y$ are some isotropic functions. 
Thus, the structural difference from the helicity $\lambda=1$ case lies in the additional angular dependence introduced by the common factor $\tilde{\mu} - \mu\mu'$ in all components.
Combining this with the trispectrum expression in \eq{eq:matter_trisp_fsigma_kqp_notation}, we find that the parity-odd power spectrum of helicity-2 (Eq.~\ref{eq:pop_spectrum_XY_explicit}) formally takes the following sum-separable form with some isotropic functions, $g$ and $h$:
\begin{align}
    P_{XY}^{(2,-)}(k) 
    \supset
    \int_{\bq,\bq'} 
    \left[ 1+2\mu\mu'\tilde{\mu}-(\mu^2+\mu'^2+\tilde{\mu}^2) \right] 
    (\tilde{\mu} - \mu\mu')
    g(\bk,\bq)h(\bk,\bq'). 
    \label{eq:pop_spectrum_h2_XY_squeezed_with_mu-tilde}
\end{align}
As in the previous case, using the momentum integration formula in Appendix~\ref{app:angular_integral_formula}, we obtain:
\begin{align*} 
    \int_{\bq,\bq'} 
    \left[ 1+2\mu\mu'\tilde{\mu}-(\mu^2+\mu'^2+\tilde{\mu}^2) \right] 
    (\tilde{\mu} - \mu\mu') 
    g(\bk,\bq)h(\bk,\bq')
    &= 
    \int_{\bq,\bq'} \frac{1}{2}(1-\mu^2)(1-\mu'^2) 
    (\mu\mu' - \mu\mu') 
    g(\bk,\bq)h(\bk,\bq') \\
    &= 0.
\end{align*}
Therefore, the helicity-2 power spectrum always vanishes.

\subsection{UV limit} 
\label{subapp:uvlim_squeezed} 
Here, we consider the UV (hard) limit of the loop momenta to investigate the effects of small-scale modes on large-scale correlations. 
For convenience, we define the integrand in \eq{eq:def_Iab} as $\calI_{ab}$: 
\begin{align}
    I_{ab}(k;\Lambda)
    &\equiv
    \int^\Lambda \mathrm{dln\,}q
    ~\calI_{ab}(q;k). 
    \label{eq:def_calIab}
\end{align} 
We omit the dependence on the IR cut-off scale $q_\mathrm{min}$ because it is irrelevant when $q_\mathrm{min} \ll k$. 
Using this, the power spectrum (Eq.~\ref{eq:pop_spectrum_h1_QQ_squeezed_final_sgnsum}) can be rewritten as 
\begin{align}
    P_{QQ}^{(1,-)}(k;\Lambda,\Lambda') 
    = 
    g_-\frac{k^3}{8} \sum_{\sigma \in S_4} \mathrm{sgn}(\sigma) 
    \int^\Lambda \mathrm{dln\,}q~
    \calI_{\sigma(1)\sigma(2)}(q;k)
    \int^{\Lambda'} \mathrm{dln\,}q'~
    \calI_{\sigma(3)\sigma(4)}(q';k). 
    \label{eq:pop_spectrum_h1_QQ_squeezed_final_sgnsum_calI}
\end{align} 
Equivalently, we obtain 
\begin{align} 
    \frac{\partial^2 P_{QQ}^{(1,-)}}{\partial \mathrm{ln\,}q \partial \mathrm{ln\,}q'}(k;q,q')
    &= 
    g_-\frac{k^3}{8} 
    \sum_{\sigma \in S_4} \mathrm{sgn}(\sigma)
    \calI_{\sigma(1)\sigma(2)}(q;k)
    \calI_{\sigma(3)\sigma(4)}(q';k) 
    \label{eq:pop_spectrum_h1_QQ_squeezed_final_shlim_finite_qp} \\
    &\equiv 
    g_- P_\mathrm{shell}(q,q';k). 
    \label{eq:def_Pshell}
\end{align}
Thus, $\calI_{ab}(k,q)$ represents the contribution to the power spectrum at wavenumber $k$ from a $q$-shell of width $\mathrm{dln\,}q$, while $P_\mathrm{shell}(q, q'; k)$ represents the total contribution from both the $q$-shell and $q'$-shell. 
\begin{figure}
    \centering
    \includegraphics[width=1.0\columnwidth]{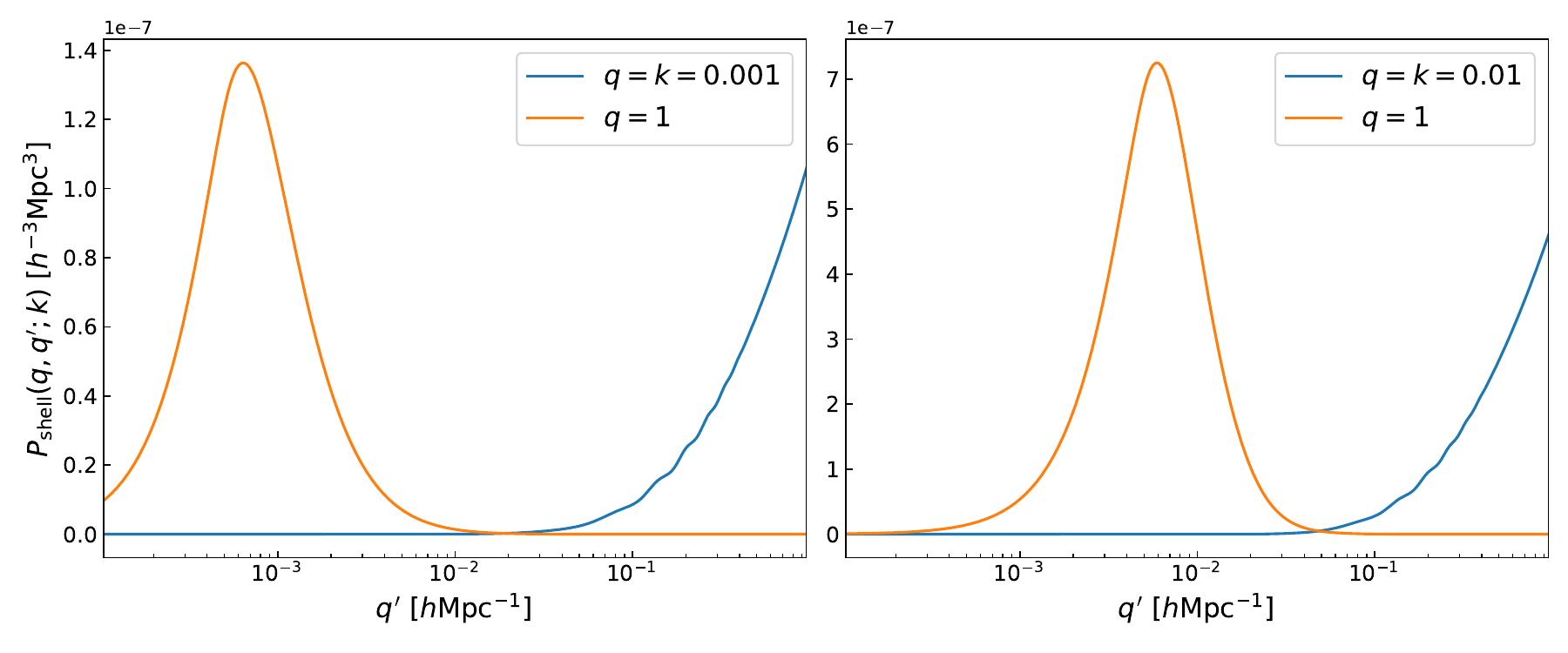}
    \caption{
    Contribution from the $q$-shell and $q'$-shell to the parity-odd power spectrum at wavenumbers $k = 0.001$ (left panel) and $k = 0.01$ (right panel) as a function of $q'$. 
    The blue curve represents the case where $q = k$, meaning that one of the triangles is an \textit{equilateral} triangle. 
    In this case, the contribution is larger for larger values of $q'$. 
    In contrast, the orange curve corresponds to the case where $q = 1$, meaning that one of the triangles is a \textit{squeezed} triangle. 
    Here, the contribution is most significant when $q' \sim k$. 
    Thus, the contribution becomes large when either $q$ or $q'$ equals $k$, while the other is in the UV regime, i.e., in the single-hard limit. 
    }
    \label{fig:Pshell}
\end{figure}

Fig.~\ref{fig:Pshell} shows the contribution of the two shells to the power spectrum.
The dominant contribution arises when one of the wave numbers, $q$ or $q'$, equals $k$ while the other is in the UV regime. This corresponds to the case where the two triangles, sharing the common edge $\bk$, consist of one equilateral triangle and one squeezed triangle.
In other words, this situation represents the UV limit for one of the momenta, i.e., the \textit{single-hard} limit.
To derive an analytical expression for the power spectrum in this regime, we first take the single-hard limit ($k \ll q$), denoted as ``$\textit{sh}$'', of the kernel (Eq.~\ref{eq:prj_kernel_lambda1_QQ_kqp_notation}) and the squeezed trispectrum (Eq.~\ref{eq:matter_trisp_fsigma_kqp_notation}), respectively: 
\begin{align}
    K_{QQ}^{(1,-)}(\bq,\bq';\bk) 
    &\xrightarrow{sh}
    -\frac{i}{4}\left[\hbk\cdot(\hbq\times\hbq') \right]
    \left[ -2\mu + \calO(\epsilon) \right]
    \frac{(k\mu'+q')(2q'\mu'+k)}{|\bk+\bq'|^2}, 
    \label{eq:prj_kernel_lambda1_QQ_kqp_notation_shlim} \\
    T_\delta^{(-)}(\bq,\bq';\bk) 
    &\xrightarrow{sh} 
    ig_-k^3 
    \left[\hbk\cdot(\hbq\times\hbq') \right] 
    \frac{q'}{k}
    \sum_{\sigma \in S_4} \mathrm{sgn}(\sigma)\nonumber\\
    &\times  
    \left[
    -\frac{1}{2} 
    \mu
    f_{\sigma(1)}(q) 
    f_{\sigma(2)}(q)
    \Delta n_{\sigma(2)\sigma(1)}(q)
     + \calO(\epsilon)
    \right]
    f_{\left[\sigma(3)\right.}(q')
    f_{\left.\sigma(4)\right]}(|\bk+\bq'|), 
    \label{eq:matter_trisp_fsigma_kqp_notation_shlim}
\end{align}
where
we defined $\epsilon \equiv k/q$ and $\Delta n_{ba}(q) \equiv n_b(q)-n_a(q)$ with the spectral index of $f_a$: 
$n_a(q) \equiv \rmd \mathrm{ln\,}f_a / \mathrm{ln\,}q$. 
Thus, using \eq{eq:def_calIab}, the contribution from UV modes in a shell with width $\mathrm{dln\,}q$ to the power spectrum can be written as
\begin{align}
    \calI_{ab}(q;k) 
    \xrightarrow{\epsilon\to0} 
    \calI_{ab}^\mathrm{UV}(q) 
    &\equiv 
    \int_{-1}^{1} \frac{\mathrm{d}\mu}{2} (1-\mu^2) \mu^2 
    \frac{q^3}{2\pi^2} 
    f_a(q)f_b(q)\Delta n_{ba}(q) \nonumber\\
    &= 
    \frac{2}{15} 
    \frac{q^3}{2\pi^2} 
    f_a(q)f_b(q)\Delta n_{ba}(q). 
\end{align}
Therefore, the corresponding shell power spectrum at the UV cut-off scale, $\Lambda$, becomes 
\begin{align} 
    \left.
    \frac{\partial^2 P_{QQ}^{(1,-)}}{\partial \mathrm{ln\,}q \partial \mathrm{ln\,}q'}(k;q,q')
    \right|_{q=\Lambda}
    &\xrightarrow{sh} 
    g_-\frac{k^3}{8}
    \sum_{\sigma \in S_4} \mathrm{sgn}(\sigma) 
    \calI_{\sigma(1)\sigma(2)}^\mathrm{UV}(\Lambda)
    \calI_{\sigma(3)\sigma(4)}(q';k). 
    \label{eq:pop_spectrum_h1_QQ_squeezed_shlim_qp_shell}
\end{align}
The low-$k$ behavior of the power spectrum in this limit ($k/q\to0$) is determined by integrating this over the other momentum, $\mathrm{ln\,}q'$. 
From Fig.~\ref{fig:Pshell}, we observe that the integrand $P_\mathrm{shell}(q,q';k)$ has a sharp peak around $q' \sim k$, indicating that the dominant contribution (up to a constant factor) mainly comes from its diagonal: $\calI_{ab}(k,k)$.
Furthermore, among the terms in the summation over permutations of functions in \eq{eq:pop_spectrum_h1_QQ_squeezed_shlim_qp_shell}, the dominant term is given by $\calI_{12}(k,k)$, which corresponds to the combination in Eqs.~\eqref{eq:def_f1}--\eqref{eq:def_f4} that shows the strongest enhancement in the $k\to0$ limit.
Specifically, from \eqs{eq:def_Iab}{eq:def_calIab}, $\calI_{12}(k,k)$ is given by:  
\begin{align}
    \calI_{12}(k,k) 
    &= \frac{2k^6}{\pi^2} \int_0^1 \rmd t~t^3(t^2-1)(4t^2-1) 
    f_{\left[1\right.}(k) 
    f_{\left.2\right]}(2kt) \\
    &=\frac{k^{6+\alpha+\beta}}{\pi^2} \calM(k)P_\phi(k) 
    \int_0^1 \rmd t~t^3(t^2-1)(4t^2-1) 
    \left[
    (2t)^\beta - (2t)^\alpha
    \right]
    \calM(2kt)P_\phi(2kt), 
    \label{eq:pk_shlim}
\end{align}
where we introduced $t\equiv \sqrt{(1-\mu)/2}$. 
Moreover, since $\calM(k) \propto k^2$ where $k\ll k_\mathrm{eq}$, we obtain 
\begin{align}
    \left.\calI_{12}(k,k)\right|_{k\ll k_\mathrm{eq}} = \calI_{12}^{(0)} k^{2n_s-1+\alpha+\beta}, 
\end{align}
where the constant $\calI_{12}^{(0)}$ can be analytically derived from \eq{eq:pk_shlim}. 
Thus, ignoring constant factors and focusing only on the asymptotic scaling, we obtain the following simplified expression for the single-hard limit:  
\begin{align}
    \left[P_{QQ}^{(1,-)}(k)\right]^{sh} 
    &\equiv 
    g_-
    \frac{k^3}{2} 
    \calI_{34}^\mathrm{UV}(\Lambda)
    \left.\calI_{12}(k,k)\right|_{k\ll k_\mathrm{eq}} \\
    &\equiv 
    g_- 
    \mathfrak{S}_{sh}(\Lambda) 
    k^{2n_s+2+\alpha+\beta}, 
    \label{eq:pop_spectrum_h1_QQ_squeezed_final_shlim}
\end{align}
where $\mathfrak{S}_{sh}(\Lambda) \equiv \calI_{12}^{(0)} \calI_{34}^\mathrm{UV}(\Lambda)/2$. 
This function correctly captures the low-$k$ behavior from the two-loop integral (Eq.~\ref{eq:pop_spectrum_h1_QQ_squeezed_final_sgnsum_calI}) as shown in Fig.~\ref{fig:pt}.

Lastly, we consider the double-hard limit ($k\ll q,q'$), denoted as ``$\textit{dh}$''. 
In this regime, the kernel and the trispectrum take the following forms: 
\begin{align}
    K_{QQ}^{(1,-)}(\bq,\bq';\bk) 
    &\xrightarrow{dh}
    -\frac{i}{4}\left[\hbk\cdot(\hbq\times\hbq') \right]
    \left[ -2\mu + \calO(\epsilon) \right]
    \left[ 2\mu' + \calO(\epsilon') \right], 
    \label{eq:prj_kernel_lambda1_QQ_kqp_notation_dhlim} \\
    T_\delta^{(-)}(\bq,\bq';\bk) 
    &\xrightarrow{dh} 
    ig_-k^3 
    \left[\hbk\cdot(\hbq\times\hbq') \right] 
    \sum_{\sigma \in S_4} \mathrm{sgn}(\sigma) \nonumber\\
    &\times 
    \left[
    -\frac{1}{2} 
    \mu
    f_{\sigma(1)}(q) 
    f_{\sigma(2)}(q)
    \Delta n_{\sigma(2)\sigma(1)}(q)
     + \calO(\epsilon)
    \right]
    \left[
    \frac{1}{2} 
    \mu'
    f_{\sigma(3)}(q') 
    f_{\sigma(4)}(q')
    \Delta n_{\sigma(4)\sigma(3)}(q')
     + \calO(\epsilon')
    \right], 
    \label{eq:matter_trisp_fsigma_kqp_notation_dhlim}
\end{align}
where we defined $\epsilon' \equiv k/q'$. 
In the same way, we obtain the shell power spectrum in the double-hard limit: 
\begin{align} 
    \left.
    \frac{\partial^2 P_{QQ}^{(1,-)}}{\partial \mathrm{ln\,}q \partial \mathrm{ln\,}q'}(k;q,q')
    \right|_{q=\Lambda,q'=\Lambda'}
    &\xrightarrow{dh} 
    g_-\frac{k^3}{8} 
    \sum_{\sigma \in S_4} \mathrm{sgn}(\sigma)
    \calI_{\sigma(1)\sigma(2)}^\mathrm{UV}(\Lambda)\calI_{\sigma(3)\sigma(4)}^\mathrm{UV}(\Lambda') \\
    &\equiv 
    g_- \mathfrak{S}_{dh}(\Lambda,\Lambda') k^3, 
    \label{eq:pop_spectrum_h1_QQ_squeezed_final_dhlim}
\end{align}
where 
$\mathfrak{S}_{dh}(\Lambda,\Lambda')$ is the UV cut-off dependent constant. 
This term proportional to $k^3$ is absorbed into the contribution from local stochasticity, which we derived in Section~\ref{subsec:eft_galaxy_shape}. 
However, this does not account for the dominant contribution to the backreaction from UV modes.

\section{Collapsed-type loop calculations} 
\label{app:calc_detail_collapsed} 
\subsection{Implementation of loop integral} 
\label{subapp:implementation_of_loop_int_collapsed} 
In this section, we provide the detailed calculations necessary to derive the results for the collapsed-type trispectrum, as presented in Section~\ref{subsec:case_study}.
Unlike the squeezed trispectrum, the collapsed trispectrum depends on the diagonal of the formed tetrahedron, as shown in \eq{eq:def_f-_collapsed}, which makes the integrals over the two triangles non-separable. (When using the spherical wave expansion of the Dirac delta function, an infinite sum over angular momenta appears.) 
Therefore, for the two-loop integration over $\bq$ and $\bq'$, we directly performed a six-dimensional Monte Carlo integration. 
Given the symmetry of the trispectrum under permutations of momenta, it is useful to decompose it into channels labeled by the following variables: 
\begin{align}
    \bs &\equiv \bq_1 + \bq_2,~\\
    \bt &\equiv \bq_1 + \bq_3,~\\
    \bu &\equiv \bq_1 + \bq_4,
\end{align}
Decomposing into the $s$-, $t$-, and $u$-channels both classifies the structure of the trispectrum and aids the subsequent analysis of its behavior in the collapsed limit. 

Recall the definition of the parity-odd matter trisectrum (Eqs.~\ref{eq:def_trispectrum_delta}, \ref{eq:def_tau-} and \ref{eq:def_f-}):
\begin{align}
    T_\delta^{(-)}(\bq_1,\bq_2,\bq_3,\bq_4) 
    &=
    \left[-i \bq_{12}\cdot\left(\bq_1\times\bq_3\right) \right] 
    \prod_{i=1}^4\left[\calM(q_i)\right]
    \sum_{\sigma \in S_4} \mathrm{sgn}(\sigma) 
    f_-(q_{\sigma(1)},q_{\sigma(2)},q_{\sigma(3)},q_{\sigma(4)},q_{\sigma(1)\sigma(2)},q_{\sigma(1)\sigma(4)}), 
    \label{eq:redef_T_delta}
\end{align}
with the specific collapsed-type trispectrum determined by $f_-$ (Eq.~\ref{eq:def_f-_collapsed}): 
\begin{align}
    f_-(\bq_1,\bq_3,\bq_{12})
    = -\frac{25}{3}A_\mathrm{gauge}
    \calF\left(\hbq_1,\hbq_3,\hbq_{12}\right)
    \frac{P_\phi(q_1)}{q_1}\frac{P_\phi(q_3)}{q_3} \frac{P_\phi(q_{12})}{q_{12}}, 
    \label{eq:redef_f-_collapsed}
\end{align}
and 
$\calF \equiv 1 - \hbq_1\cdot\hbq_3 + \hbq_{12}\cdot\hbq_1 - \hbq_{12}\cdot\hbq_3$. 
We note that $f_-$ is invariant under the simultaneous exchange $\bq_1 \leftrightarrow \bq_3$ and $\bq_2 \leftrightarrow \bq_4$. 
(More precisely, this is a symmetry with respect to the permutation of ``entries'', meaning invariance under the simultaneous exchange of $\bq_{\sigma(1)} \leftrightarrow \bq_{\sigma(3)}$ and $\bq_{\sigma(2)} \leftrightarrow \bq_{\sigma(4)}$ in \eq{eq:redef_T_delta}.) 
Therefore, the sum over the 24 permutations of $\sigma \in S_4$ reduces to 12 terms. 
Furthermore, for these 12 terms, we decompose the dependence on the diagonal into $s$-, $t$-, and $u$-channels. 
That is,
\begin{align}
    T_\delta^{(-)} 
    \equiv
    T_\delta^{(s)} + T_\delta^{(t)} + T_\delta^{(u)},
\end{align}
where $T_\delta^{(s)}$ is defined as the component that depends only on $\bs = \bq_{12}$ as the diagonal variable: 
\begin{align}
    T_\delta^{(s)} 
    = 
    2\left[-i \bq_{12}\cdot\left(\bq_1\times\bq_3\right) \right] 
    \prod_{i=1}^4\left[\calM(q_i)\right]
    \left(
    f_-(\bq_1,\bq_3,\bq_{12}) - (\bq_1 \leftrightarrow \bq_2) - (\bq_3 \leftrightarrow \bq_4) + 
    \begin{pmatrix}
        \bq_1 \leftrightarrow \bq_2\\
        \bq_3 \leftrightarrow \bq_4
    \end{pmatrix}
    \right), \label{eq:def_T-_s-channel}
\end{align}
and $T_\delta^{(t)}$ is obtained by swapping $\bq_2 \leftrightarrow \bq_3$ in each term of $T_\delta^{(s)}$, including the momenta appearing in the scalar triple product, making its diagonal dependence only on $\bt=\bq_{13}$.  
Similarly, $T_\delta^{(u)}$ is defined by swapping $\bq_2 \leftrightarrow \bq_4$ in $T_\delta^{(s)}$, resulting in dependence only on $\bu=\bq_{14}$. 

Recalling that the kernels (Eqs.~\ref{eq:prj_kernel_lambda1_QQ}--\ref{eq:prj_kernel_lambda2_RR}) are always symmetric under the exchanges $\bq_1 \leftrightarrow \bq_2$ and $\bq_3 \leftrightarrow \bq_4$, while being antisymmetric under the simultaneous exchange of $\bq_1 \leftrightarrow \bq_3$ and $\bq_2 \leftrightarrow \bq_4$.  
From these symmetries, in the special case of the (22)-type loop integral (Eq.~\ref{eq:pop_spectrum_XY_explicit}) with $\Lambda=\Lambda'$, all contributions from each term in the $s$-channel become identical.  
Moreover, the contributions from the $t$- and $u$-channels are also identical. 
Additionally, among the four terms in the $t/u$-channel, the first and last terms contribute equally, as do the second and third terms. 
Therefore, due to these symmetries, the number of independent terms to be computed is greatly reduced, and it suffices to evaluate the following three terms (one for the $s$-channel and two for the $t$- and $u$-channels):
\begin{align}
    T_\delta^{(s)} 
    &: 
    8\left[-i \bq_{12}\cdot\left(\bq_1\times\bq_3\right) \right] 
    \prod_{i=1}^4\left[\calM(q_i)\right]
    f_-(\bq_1,\bq_3,\bq_{12}), \label{eq:def_T-_s-channel_reduced} \\
    T_\delta^{(t)} + T_\delta^{(u)}
    &: 
    8\left[-i \bq_{12}\cdot\left(\bq_1\times\bq_3\right) \right] 
    \prod_{i=1}^4\left[\calM(q_i)\right]
    \left(
    f_-(\bq_1,\bq_2,\bq_{13}) + f_-(\bq_3,\bq_2,\bq_{13})
    \right). \label{eq:def_T-_tu-channel_reduced} 
\end{align}

\begin{figure}
    \centering
    \includegraphics[width=1.0\columnwidth]{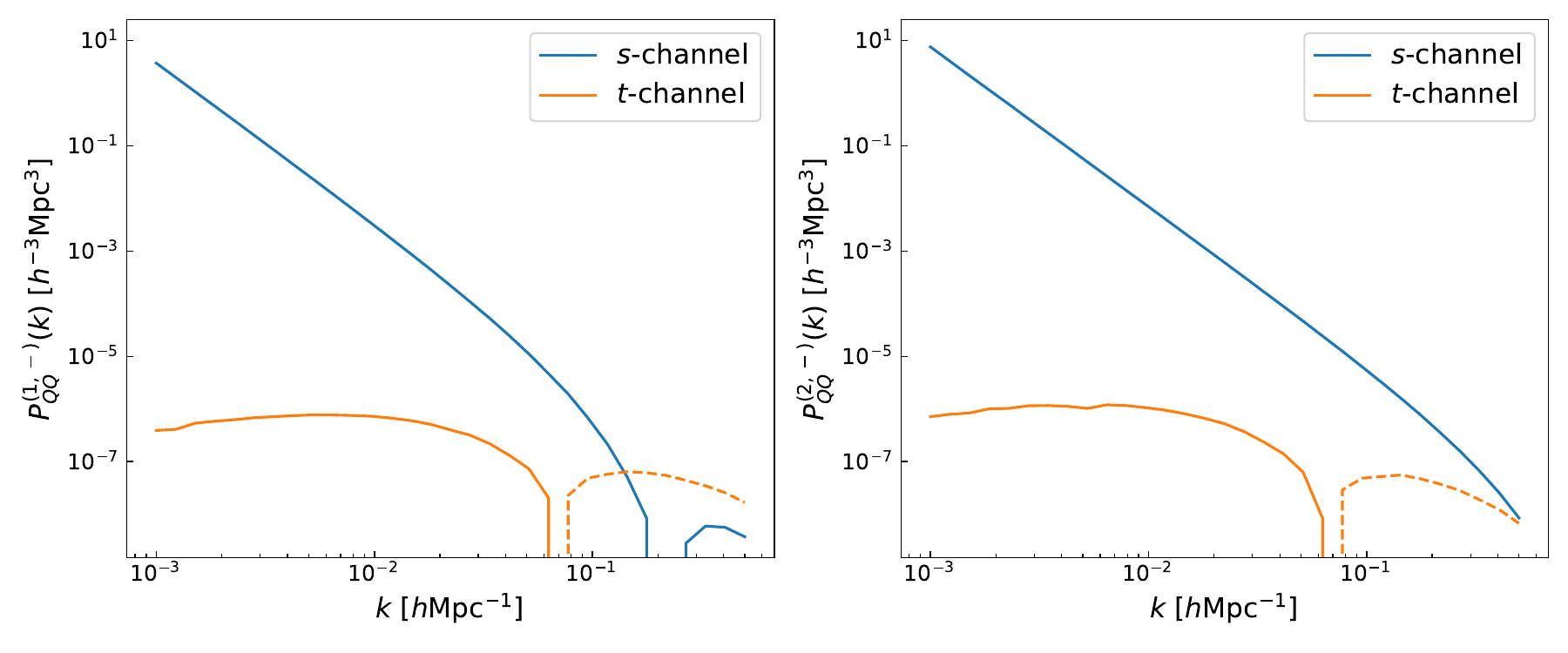}
    \caption{
    Contributions from each channel to the parity-odd power spectrum $P_{QQ}^{(\lambda,-)}$ for helicity $\lambda=1$ (left panel) and $\lambda=2$ (right panel). 
    We omit the contribution from the $u$-channel because it is identical to that from the $t$-channel due to the symmetry of the (22)-type loop integral. 
    In the low-$k$ regime, the dominant contribution comes from the $s$-channel. 
    }
    \label{fig:pt_channel}
\end{figure}
Fig.~\ref{fig:pt_channel} presents the loop integral results for each channel, derived from the expression for the parity-odd power spectrum (Eq.~\ref{eq:pop_spectrum_XY_explicit}).
In the case of the $s$-channel, the direct dependence on the diagonal term $P_\phi(s) = P_\phi(k)$ leads to a significant enhancement in the low-$k$ limit.
In contrast, such behavior is absent in the $t$- and $u$-channels, which makes their contributions subdominant.

\subsection{UV limit} 
\label{subapp:uvlim_collapsed} 
Here, we derive an analytical expression for the UV limit of the parity-odd power spectrum.
As shown in Fig.~\ref{fig:pt_channel}, the dominant contribution in the low-$k$ regime comes from the $s$-channel.
Thus, to capture the leading-order contribution, it is sufficient to focus only on the $s$-channel.
Furthermore, unlike in the squeezed trispectrum case, the form of the collapsed trispectrum matches that of the (22)-type integral, suggesting that the dominant contribution arises from the collapsed (double-hard) limit. 
Thus, we begin by considering the double-hard limit: $k \ll q_1, q_3$. 
Hereafter, as in the previous section, we relabel the variables for convenience. Specifically, we define:
\begin{align*}
    \bk = \bq_{12} = -\bq_{34},~ 
    \bq \equiv \bq_1,~ 
    \bq' \equiv \bq_3,~ 
    \mu \equiv \hbk \cdot \hbq,~ 
    \mu' \equiv \hbk \cdot \hbq',~ 
    \tilde{\mu} \equiv \hbq \cdot \hbq'.
\end{align*}
Additionally, we introduce the small parameters: $\epsilon\equiv k/q,\epsilon'\equiv k/q'$.

The double-hard limit of the kernel with helicity $\lambda=1$ is given in \eq{eq:prj_kernel_lambda1_QQ_kqp_notation_dhlim}. 
In a similar way, we obtain the expressions for the three kernels with $\lambda=2$ (Eqs.~\ref{eq:prj_kernel_lambda2_QQ}--\ref{eq:prj_kernel_lambda2_RR}) in this limit: 
\begin{align}
    K_{QQ}^{(2,-)}(\bq,\bq';\bk) 
    &\xrightarrow{dh}
    \frac{i}{2}\left[\hbk\cdot(\hbq\times\hbq') \right](\tilde{\mu} - \mu\mu') 
    + \calO\left(\{\epsilon,\epsilon'\}\right), \\
    K_{(QR)}^{(2,-)}(\bq,\bq';\bk) 
    &\xrightarrow{dh} 
    \frac{i}{8}\left[\hbk\cdot(\hbq\times\hbq') \right](\tilde{\mu} - \mu\mu') 
    (\epsilon^2+\epsilon'^2) 
    + \calO\left(\{\epsilon,\epsilon'\}^3\right), \\
    K_{RR}^{(2,-)}(\bq,\bq';\bk) 
    &\xrightarrow{dh} 
    \frac{i}{8}\left[\hbk\cdot(\hbq\times\hbq') \right](\tilde{\mu} - \mu\mu')
    \epsilon^2\epsilon'^2 
    + \calO\left(\{\epsilon,\epsilon'\}^5\right). 
\end{align}
From these, $P_{(QR)}$ and $P_{RR}$ have additional suppressions of factor $k^2$ and $k^4$, respectively, compared to $P_{QQ}$ at the kernel level. 
Moreover, since the angular dependence of the leading terms is the same for all three cases, we find that $P_{QQ}$ will dominate in the $k \to 0$ limit, regardless of the assumed form (angular dependence) of the trispectrum (see Fig.~\ref{fig:pt}).

Next, we consider the UV limit of the collapsed trispectrum. 
For convenience, we first rewrite $T_\delta^{(s)}$ from \eq{eq:def_T-_s-channel} in a ``product of differences'' form, expressed as the inner product of unit vectors: 
\begin{align}
    &T_\delta^{(s)} 
    = 
    \frac{50}{3}A_\mathrm{gauge}
    \left[-i \bq_{12}\cdot\left(\bq_1\times\bq_3\right) \right] 
    \prod_{i=1}^4\left[\calM(q_i)\right] 
    \frac{P_\phi(q_{12})}{q_{12}}\nonumber\\
    &\times
    \left[
    \left(\frac{P_\phi(q_1)}{q_1}-\frac{P_\phi(q_2)}{q_2}\right)\hbq_{12}
    + \left(\frac{P_\phi(q_1)}{q_1}\hbq_1-\frac{P_\phi(q_2)}{q_2}\hbq_2\right)
    \right] 
    \cdot
    \left[
    \left(\frac{P_\phi(q_3)}{q_3}-\frac{P_\phi(q_4)}{q_4}\right)\hbq_{34}
    + \left(\frac{P_\phi(q_3)}{q_3}\hbq_3-\frac{P_\phi(q_4)}{q_4}\hbq_4\right)
    \right]. \label{eq:def_T-_s-channel_vectorized}
\end{align}
Taking the double-hard limit, we obtain
\begin{align}
    T_\delta^{(s)} 
    &\xrightarrow{dh} 
    \frac{50}{3}A_\mathrm{gauge}
    \left[-i \bk\cdot\left(\bq\times\bq'\right) \right] 
    \calM^2(q)(1+\calO(\epsilon)) 
    \calM^2(q')(1+\calO(\epsilon'))
    \frac{P_\phi(k)}{k}\nonumber\\
    &\quad\quad\quad\times
    \frac{P_\phi(q)}{q}
    \left[
    \left(\epsilon \mu(n_\phi-1) + \calO(\epsilon^2) \right)\hbk
    + 2\hbq
    \right] 
    \cdot
    \frac{P_\phi(q')}{q'}
    \left[
    \left(\epsilon' \mu' (n_\phi-1) + \calO(\epsilon'^2) \right)\hbk
    + 2\hbq'
    \right] \label{eq:def_T-_s-channel_vectorized_dhlim_mid}\\
    &=
    \frac{200}{3} A_\mathrm{gauge}
    \left[-i \hbk\cdot\left(\hbq\times\hbq'\right) \right] 
    \tilde{\mu}
    P_\phi(k) 
    \calM^2(q) P_\phi(q)
    \calM^2(q') P_\phi(q') 
    + \calO\left(\{\epsilon,\epsilon'\}\right)\nonumber\\
    &=
    \frac{200}{3} A_\mathrm{gauge}
    \left[-i \hbk\cdot\left(\hbq\times\hbq'\right) \right] 
    \tilde{\mu}
    P_\phi(k) 
    P(q)
    P(q') 
    + \calO\left(\{\epsilon,\epsilon'\}\right), \label{eq:def_T-_s-channel_vectorized_dhlim}
\end{align}
where $n_\phi$ is the spectral index of $P_\phi$: 
$n_\phi \equiv \rmd \mathrm{ln\,}P_\phi / \mathrm{ln\,}q$, 
and $P(q)$ is the linear matter power spectrum. 
By performing the following angular integrals with aid of Appendix~\ref{app:angular_integral_formula}, 
\begin{align*}
    \int_{\hbq,\hbq'} 
    \left[ 1+2\mu\mu'\tilde{\mu}-(\mu^2+\mu'^2+\tilde{\mu}^2) \right] 
    \mu\mu'\tilde{\mu}
    &= \frac{2}{225}, \\
    \int_{\hbq,\hbq'} 
    \left[ 1+2\mu\mu'\tilde{\mu}-(\mu^2+\mu'^2+\tilde{\mu}^2) \right] 
    (\tilde{\mu} - \mu\mu') 
    \tilde{\mu}
    &= \frac{8}{225}, 
\end{align*}
we obtain the analytic expressions for the contributions to the parity-odd power spectra at wavenumber $k$ from the $q$-shell and $q'$-shell of the two-loop integral in the double-hard limit: 
\begin{align} 
    \left.
    \frac{\partial^2 P_{QQ}^{(1,-)}}{\partial \mathrm{ln\,}q \partial \mathrm{ln\,}q'}(k;q,q')
    \right|_{q=\Lambda,q'=\Lambda'}
    &\xrightarrow{dh} 
    \frac{16}{27}
    A_\mathrm{gauge} 
    \calP_{(0)}(\Lambda) 
    \calP_{(0)}(\Lambda')
    P_\phi(k)\\
    &\equiv 
    A_\mathrm{gauge} \mathfrak{S}_{QQ}^{(1,-)}(\Lambda,\Lambda') P_\phi(k), 
    \label{eq:pop_spectrum_h1_QQ_collapsed_final_dhlim} \\
    \left.
    \frac{\partial^2 P_{QQ}^{(2,-)}}{\partial \mathrm{ln\,}q \partial \mathrm{ln\,}q'}(k;q,q')
    \right|_{q=\Lambda,q'=\Lambda'}
    &\xrightarrow{dh} 
    \frac{32}{27}
    A_\mathrm{gauge} 
    \calP_{(0)}(\Lambda) 
    \calP_{(0)}(\Lambda')
    P_\phi(k)\\
    &\equiv 
    A_\mathrm{gauge} \mathfrak{S}_{QQ}^{(2,-)}(\Lambda,\Lambda') P_\phi(k), 
    \label{eq:pop_spectrum_h2_QQ_collapsed_final_dhlim} \\
    \left.
    \frac{\partial^2 P_{(QR)}^{(2,-)}}{\partial \mathrm{ln\,}q \partial \mathrm{ln\,}q'}(k;q,q')
    \right|_{q=\Lambda,q'=\Lambda'}
    &\xrightarrow{dh} 
    \frac{8}{27}
    A_\mathrm{gauge} 
    \left(
    \calP_{(-2)}(\Lambda) 
    \calP_{(0)}(\Lambda')
    + (\Lambda \leftrightarrow \Lambda')
    \right)
    k^2P_\phi(k)\\
    &\equiv 
    A_\mathrm{gauge} \mathfrak{S}_{QR}^{(2,-)}(\Lambda,\Lambda') k^2 P_\phi(k), 
    \label{eq:pop_spectrum_h2_QR_collapsed_final_dhlim} \\
    \left.
    \frac{\partial^2 P_{RR}^{(2,-)}}{\partial \mathrm{ln\,}q \partial \mathrm{ln\,}q'}(k;q,q')
    \right|_{q=\Lambda,q'=\Lambda'}
    &\xrightarrow{dh} 
    \frac{8}{27}
    A_\mathrm{gauge} 
    \calP_{(-2)}(\Lambda) 
    \calP_{(-2)}(\Lambda')
    k^4 P_\phi(k)\\
    &\equiv 
    A_\mathrm{gauge} \mathfrak{S}_{RR}^{(2,-)}(\Lambda,\Lambda') k^4P_\phi(k), 
    \label{eq:pop_spectrum_h2_RR_collapsed_final_dhlim} 
\end{align}
where we introduced the following power spectrum: 
\begin{align}
    \calP_{(\alpha)}(q) \equiv \frac{q^{3+\alpha}}{2\pi^2} P(q).
\end{align}

Finally, we parametrize this trispectrum in terms of the $d_n^\mathrm{odd}$-template, as defined in \eq{eq:def_dn_odd_template} recalling \eq{eq:dn_odd_parametrization}: 
\begin{align}
    d_0^\mathrm{odd} = -d_1^\mathrm{odd}/3 = -A_\mathrm{gauge},~
    d_{n\geq2}^\mathrm{odd} = 0, 
\end{align} 
and examine the contributions from $d_0^\mathrm{odd}$ and $d_1^\mathrm{odd}$ separately in the double-hard limit. 
For $n=0$, only the first term exists among the vectorized components in \eqref{eq:def_T-_s-channel_vectorized_dhlim_mid}. Thus, the double-hard limit in this case takes the following form:
\begin{align}
    T_\delta^{(s)} 
    &\xrightarrow{dh} 
    -\frac{50}{3}d_0^\mathrm{odd}
    \left[-i \bk\cdot\left(\bq\times\bq'\right) \right] 
    \calM^2(q)(1+\calO(\epsilon)) 
    \calM^2(q')(1+\calO(\epsilon'))
    \frac{P_\phi(k)}{k}\nonumber\\
    &\quad\quad\quad\times
    \frac{P_\phi(q)}{q}
    \left(\epsilon \mu(n_\phi-1) + \calO(\epsilon^2) \right)\hbk
    \cdot
    \frac{P_\phi(q')}{q'}
    \left(\epsilon' \mu' (n_\phi-1) + \calO(\epsilon'^2) \right)\hbk \nonumber\\
    &= 
     -\frac{50}{3}d_0^\mathrm{odd}
    \left[-i \bk\cdot\left(\bq\times\bq'\right) \right] 
    \mu\mu'
    \epsilon \epsilon'
    (n_\phi-1)^2
    P_\phi(k) P(q) P(q') 
    + \calO\left(\{\epsilon,\epsilon'\}^3\right), 
\end{align}
Due to the lack of angular dependence, a leading-order cancellation occurs, suppressing the collapsed limit by a factor of $k^2$ and thereby reducing the enhancement.
Consequently, the enhancement of the $U(1)$-gauge model in the $k\to0$ limit is driven by the $n=1$ component at leading order.

\section{Initial condition details} 
\label{app:ic_detail} 
In this section, we provide additional details on the initial condition introduced in Section~\ref{subsec:parity-odd_ic}: 
\begin{align}
    \Phi(\bx) 
    = \phi(\bx) + \sum_{\ell m} A_{\ell m} \phi_{\ell m}^{(2)}\left(\bx;\sigma_{\ell m}\right), 
    \label{eq:def_Phix_app}
\end{align}
where the quadratic correction is defined in Fourier space as (Eq.~\ref{eq:phi2_Ylm_FS})
\begin{align}
    \phi_{\ell m}^{(2)}\left(\bk; \sigma_{\ell m}\right)
    \equiv 
    \int_{\bq_1,\bq_2} 
    (2\pi)^3\delD_{\bk-\bq_{12}} 
    i^m
    Y_{\ell m} (\hbq_1;\hbq_2)
    \phi(\bq_1)
    \sigma_{\ell m}(\bq_2). 
    \label{eq:phi2_Ylm_FS_app}
\end{align} 
We define the polar and azimuthal angles for the spherical harmonics,  
$Y_{\ell m} (\hbq_1;\hbq_2) \equiv Y_{\ell m} (\theta_{12},\phi_{12})$,  
as the angles of $\bq_1$ measured in a Cartesian coordinate system,  
where the polar basis associated with $\bq_2$,  
$\{\be_\theta(\hbq_2), \be_\phi(\hbq_2), \hbq_2\}$,  
is treated as the coordinate axes along the $x$-, $y$-, and $z$-directions, respectively (see Fig.~\ref{fig:coordinate_system_for_quad_corr}). 
For each $m$-mode ($m=0,\pm1,\dots,\pm\ell$), we define uncorrelated Gaussian random fields $\sigma_{\ell m}$ such that 
$\avrg{\sigma_{\ell m}\sigma_{\ell' m'}}' = \delK_{mm'} \rho_{\ell\ell'} P_\sigma$.
This condition guarantees statistical isotropy, where $|\rho_{\ell\ell'}|\leq1$ specifies the correlation matrix.

\subsection{Reality condition} 
\label{subapp:reality_condition}
We derive the phase factor $i^m$ in \eq{eq:phi2_Ylm_FS_app} from the reality condition. 
Before that, let us first summarize the properties of the spherical harmonics defined above. 
The standard spherical harmonics are functions of the angles $(\theta, \phi)$ of a single unit vector $\hbn$, measured with respect to the fiducial cartesian coordinate system. 
In contrast, the spherical harmonics defined above take as arguments the angles $(\theta_{12}, \phi_{12})$,  
which are defined in terms of two unit vectors, $\hbq_1$ and $\hbq_2$. 
Specifically, these angles are measured in a local frame associated with $\hbq_2$ in Fourier space. 
As a result, they exhibit nontrivial behavior under inversions such as $\bq_1 \to -\bq_1$ and $\bq_2 \to -\bq_2$. 
When considering the inversion $\bq_1 \to -\bq_1$ with $\bq_2$ fixed,  
the local frame remains unchanged. 
Thus, the angles transform as $(\theta_{12},\phi_{12}) \to (\pi-\theta_{12},\phi_{12}+\pi)$, just as in the case $\hbn\to-\hbn$ for standard spherical harmonics. 
As a result, the transformation of the spherical harmonics follows:  
\begin{align}  
    Y_{\ell m} (-\hbq_1;\hbq_2) = (-1)^\ell Y_\ell^{m} (\hbq_1;\hbq_2).  
    \label{eq:Ylm_12_trs_q1_inv}
\end{align}  
On the other hand, when considering the inversion $\bq_2 \to -\bq_2$ with $\bq_1$ fixed,  
the reference frame for measuring the angles transforms as  
\begin{align*}  
    \{\be_\theta(\hbq_2), \be_\phi(\hbq_2), \hbq_2\}  
    \to \{\be_\theta(-\hbq_2), \be_\phi(-\hbq_2), -\hbq_2\}  
    = \{\be_\theta(\hbq_2), -\be_\phi(\hbq_2), -\hbq_2\}.  
\end{align*}  
In this new frame, the measured angles transform as  
$(\theta_{12},\phi_{12}) \to (\pi-\theta_{12},-\phi_{12})$,  
which leads to the following transformation rule for the spherical harmonics:  
\begin{align}  
    Y_{\ell m} (\hbq_1;-\hbq_2) = (-1)^{\ell+m}Y_\ell^{m*} (\hbq_1;\hbq_2).  
    \label{eq:Ylm_12_trs_q2_inv}
\end{align}  
By combining \eq{eq:Ylm_12_trs_q1_inv} and \eq{eq:Ylm_12_trs_q2_inv},  
we find that the simultaneous inversions $\bq_1 \to -\bq_1$ and $\bq_2 \to -\bq_2$  
result in the angular transformation $(\theta_{12},\phi_{12}) \to (\theta_{12},\pi-\phi_{12})$.  
Accordingly, the spherical harmonics transform as  
\begin{align}  
    Y_{\ell m} (-\hbq_1;-\hbq_2) = (-1)^m Y_\ell^{m*} (\hbq_1;\hbq_2).  
    \label{eq:Ylm_12_trs_q1q2_inv}
\end{align}  

Now, let us determine the condition imposed by the reality condition.  
First, we introduce a phase factor $e^{i\alpha}$ as follows: 
\begin{align}
    \phi_{\ell m}^{(2)}\left(\bk; \sigma_{\ell m}\right)
    \equiv 
    \int_{\bq_1,\bq_2} 
    (2\pi)^3\delD_{\bk-\bq_{12}} 
    e^{i\alpha}
    Y_{\ell m} (\hbq_1;\hbq_2)
    \phi(\bq_1)
    \sigma_{\ell m}(\bq_2). 
    \label{eq:phi2_Ylm_FS_app_phase_alpha}
\end{align} 
Taking the complex conjugate of both sides and using the reality conditions of $\phi$ and $\sigma_{\ell m}$, and \eq{eq:Ylm_12_trs_q1q2_inv}, we obtain 
\begin{align*}
    \left[\phi_{\ell m}^{(2)}\left(\bk; \sigma_{\ell m}\right)\right]^*
    &=
    \int_{\bq_1,\bq_2} 
    (2\pi)^3\delD_{\bk-\bq_{12}} 
    e^{-i\alpha}
    Y_\ell^{m*} (\hbq_1;\hbq_2)
    \phi^*(\bq_1)
    \sigma_{\ell m}^*(\bq_2) \\
    &=
    \int_{\bq_1,\bq_2} 
    (2\pi)^3\delD_{\bk-\bq_{12}} 
    (-1)^me^{-i\alpha}
    Y_\ell^{m} (-\hbq_1;-\hbq_2)
    \phi(-\bq_1)
    \sigma_{\ell m}(-\bq_2) \\
    &=
    (-1)^me^{-2i\alpha}
    \phi_{\ell m}^{(2)}\left(-\bk; \sigma_{\ell m}\right). 
\end{align*} 
Thus, the reality condition for the quadratic correction corresponds to $(-1)^me^{-2i\alpha} = 1$, i.e., $e^{i\alpha}=i^m$.

\subsection{Initial trispectrum} 
\label{subapp:ini_trispectrum}
Here, we provide the details on the trispectrum in the non-Gaussian initial conditions defined in \eqs{eq:def_Phix_app}{eq:phi2_Ylm_FS_app}. 
The leading order trispectrum is given by
\begin{align}
    T_\Phi(\bk_1,\bk_2,\bk_3,\bk_4) 
    &= 
    \sum_{\ell m} \sum_{\ell' m'} \tilde{A}_{\ell m} \tilde{A}_{\ell' m'}
    \avrg{\phi(\bk_1) \phi_{\ell m}^{(2)}(\bk_2;\sigma_{\ell m}) \phi(\bk_3) \phi_{\ell' m'}^{(2)}(\bk_4;\sigma_{\ell' m'})}' 
    + 5~\mathrm{perms} \nonumber\\
    &= 
    \sum_{\ell m} \sum_{\ell' m'} \tilde{A}_{\ell m} \tilde{A}_{\ell' m'} 
    \int_{\bq_1,\bq_2,\bq_3,\bq_4} 
    (2\pi)^3\delD_{\bk_2-\bq_{12}} (2\pi)^3\delD_{\bk_4-\bq_{34}} 
    i^m Y_{\ell m} (\hbq_1;\hbq_2) i^{m'} Y_{\ell' m'} (\hbq_3;\hbq_4)  \nonumber\\
    &\quad\quad\quad\quad\quad\quad\quad\quad\quad\quad\quad\quad\quad\quad\quad\quad\times
    \avrg{\phi(\bk_1) \phi(\bq_1) \sigma_{\ell m}(\bq_2) \phi(\bk_3) \phi(\bq_3) \sigma_{\ell' m'}(\bq_4)}' 
    + 5~\mathrm{perms} \nonumber\\
    &= 
    \sum_{\ell m} \sum_{\ell' m'} 
    \rho_{\ell\ell'} \tilde{A}_{\ell m} \tilde{A}_{\ell' m'} 
    \delK_{mm'} 
    i^{m+m'} Y_{\ell m} (-\hbk_1;\hbk_{12}) Y_{\ell' m'} (-\hbk_3;-\hbk_{12}) 
    P_\phi(k_1) P_\phi(k_3) P_\sigma(k_{12})
    + 11~\mathrm{perms} \nonumber\\
    &= 
    \sum_{\ell\ell'm} 
    \rho_{\ell\ell'} \tilde{A}_{\ell m} \tilde{A}_{\ell' m}
    (-1)^\ell 
    Y_{\ell m} (\hbk_1;\hbk_{12}) 
    Y_{\ell' m}^* (\hbk_3;\hbk_{12})
    P_\phi(k_1) P_\phi(k_3) P_\sigma(k_{12}) + 11~\mathrm{perms}, 
    \label{eq:T_Phi_two_Ylm_app}
\end{align} 
where we used the Gaussianity of $\phi$ and $\sigma$ and the isotropy condition: $\avrg{\sigma_{\ell m}\sigma_{\ell' m'}}' = \delK_{mm'} \rho_{\ell\ell'} P_\sigma$ in the third line, and the properties of the spherical harmonics (Eqs.~\ref{eq:Ylm_12_trs_q1_inv} and \ref{eq:Ylm_12_trs_q1q2_inv}) in the last line. 

In the following, we derive an explicit expression for the product of two spherical harmonics in Eq.~\eqref{eq:T_Phi_two_Ylm_app} as a scalar function composed of four scalar quantities constructed from the three vectors $\hbk_1$, $\hbk_3$, and $\hbk_{12}$, i.e., 
\begin{align}
    \mu_{13} \equiv \hbk_1\cdot\hbk_3,~
    \mu_1 \equiv \hbk_1\cdot\hbk_{12},~
    \mu_3 \equiv \hbk_3\cdot\hbk_{12},~
    \hbk_{12} \cdot \left(\hbk_1 \times \hbk_3\right). 
\end{align}
The resulting expression will be given in Eq.~\eqref{eq:two_Ylm_in_polys}.
We begin by recalling the explicit form of the spherical harmonics: 
\begin{align}
    Y_{\ell m}(\theta,\phi) 
    = \calN_{\ell m} 
    \calL_\ell^m(\mu) e^{im\phi}
    = (-1)^{(m-|m|)/2} 
    \calN_{\ell |m|}
    \calL_\ell^{|m|}(\mu) e^{im\phi}, 
\end{align}
where $\mu\equiv\cos\theta$, $\calN_{\ell m}$ is the normalization factor, 
\begin{align}
    \calN_{\ell m} = (-1)^m \sqrt{\frac{2\ell+1}{4\pi} \frac{(\ell-m)!}{(\ell+m)!}},
\end{align}
and $\calL_\ell^m$ denotes the associated Legendre polynomials, defined for $0 \leq m \leq \ell$ as 
\begin{align}
    \calL_\ell^{m}(\mu) \equiv (-1)^m (1-\mu^2)^{m/2} \frac{\rmd^m}{\rmd \mu^m} \calL_\ell(\mu), 
    \label{eq:def_assoc_Legendre}
\end{align}
with $\calL_\ell$ being the ordinary Legendre polynomials. 
Since both spherical harmonics are defined with respect to the coordinate frame associated with $\hbk_{12}$, i.e., $\{\be_\theta(\hbk_{12}), \be_\phi(\hbk_{12}), \hbk_{12}\}$, the product of the two spherical harmonics takes the form 
\begin{align}
    Y_{\ell m} (\hbk_1;\hbk_{12}) Y_{\ell' m}^* (\hbk_3;\hbk_{12})
    =
    \calN_{\ell |m|} \calN_{\ell' |m|} 
    \calL_\ell^{|m|}(\mu_1) \calL_{\ell'}^{|m|}(\mu_3) 
    e^{im(\phi_1-\phi_3)}, 
    \label{eq:two_Ylm_in_polys_tmp}
\end{align}
where $\phi_1$ and $\phi_3$ are the azimuthal angles of $\hbk_1$ and $\hbk_3$, respectively, measured counterclockwise from the $\be_\theta(\hbk_{12})$-axis (see Fig.~\ref{fig:coordinate_system_for_quad_corr}). 
In other words, we parametrize the three vectors as
\begin{align}
    \hbk_{12} = (0,0,1),~
    \hbk_1 = \left(\sqrt{1-\mu_1^2} \cos\phi_1, \sqrt{1-\mu_1^2} \sin\phi_1, \mu_1\right),~
    \hbk_3 = \left(\sqrt{1-\mu_3^2} \cos\phi_3, \sqrt{1-\mu_3^2} \sin\phi_3, \mu_3\right).
\end{align}
From this parametrization, we obtain expressions for $\cos(\phi_1 - \phi_3)$ and $\sin(\phi_1 - \phi_3)$ using the inner product and scalar triple product: 
\begin{align}
    \sqrt{1-\mu_1^2} \sqrt{1-\mu_3^2} \cos(\phi_1-\phi_3)
    &= \mu_{13} - \mu_1 \mu_3, \label{eq:triangles_cos_phi}\\
    \sqrt{1-\mu_1^2} \sqrt{1-\mu_3^2} \sin(\phi_1-\phi_3)
    &= - \hbk_{12} \cdot \left(\hbk_1 \times \hbk_3\right). \label{eq:triangles_sin_phi}
\end{align} 
Note that the difference $\phi = \phi_1 - \phi_3$ corresponds to the angle between the planes of the two triangles sharing the diagonal $\bk_{12}$ (see Fig.~\ref{fig:tetrahedron_triangles_phi}). 
By substituting these expressions and Eq.~\eqref{eq:def_assoc_Legendre} into Eq.~\eqref{eq:two_Ylm_in_polys_tmp}, we arrive at the final result: 
\begin{align}
    Y_{\ell m} (\hbk_1;\hbk_{12}) Y_{\ell' m}^* (\hbk_3;\hbk_{12})
    =
    \calN_{\ell |m|} \calN_{\ell' |m|}
    \frac{\rmd^{|m|} \calL_\ell(\mu_1)}{\rmd \mu_1^{|m|}} 
    \frac{\rmd^{|m|} \calL_{\ell'}(\mu_3)}{\rmd \mu_3^{|m|}} 
    \left(\mu_{13} - \mu_1\mu_3 
    - i \left[\hbk_{12} \cdot \left(\hbk_1 \times \hbk_3\right)\right] \right)^m. 
    \label{eq:two_Ylm_in_polys}
\end{align}

Let us examine a few of the simplest cases.  
When only $\tilde{A}_{00}$ is nonzero, we recover the standard $\tau_\mathrm{NL}$-type trispectrum with $\tau_\mathrm{NL} = 9\tilde{A}_{00}^2 / 100\pi$, and there is no parity-odd contribution in this case.
When only $\tilde{A}_{1+1}$ is nonzero, the resulting trispectrum contains both parity-even and parity-odd components:
\begin{align}
    \mathrm{Re}\left[T_\Phi\right] &= -\frac{3}{8\pi} \tilde{A}_{1+1}^2 (\mu_{13} - \mu_1\mu_3) P_\phi(k_1) P_\phi(k_3) P_\sigma(k_{12}) + 11~\mathrm{perms}, \\
    \mathrm{Im}\left[T_\Phi\right] &= +\frac{3}{8\pi} \tilde{A}_{1+1}^2 \left[\hbk_{12} \cdot \left(\hbk_1 \times \hbk_3\right)\right] P_\phi(k_1) P_\phi(k_3) P_\sigma(k_{12}) + 11~\mathrm{perms},
\end{align}
where the imaginary part corresponds to the $d_n^\mathrm{odd}$ template defined in \eq{eq:def_dn_odd_template}, with 
$d_0^\mathrm{odd} = -9\tilde{A}_{1+1}^2 / 400\pi$ (which is negative).  
A parity-odd trispectrum with $d_0^\mathrm{odd} > 0$ can be obtained by flipping the sign of $m$.
Next, consider the case where the three coefficients $\tilde{A}_{1+1}$, $\tilde{A}_{2+1}$, and $\tilde{A}_{2+2}$ are nonzero. 
Here we also assume that the correlation matrix satisfies $\rho_{\ell\ell'}=1$ for any $\ell,\ell'=1,2$. 
In this case, the real and imaginary parts of the trispectrum are given by
\begin{align}
    \mathrm{Re}\left[T_\Phi\right] &= 
    \frac{3}{8\pi}
    \Biggl\{ \left(- \tilde{A}_{1+1}^2 
    + \sqrt{5} \tilde{A}_{1+1} \tilde{A}_{2+1} (\mu_1 - \mu_3) 
    + 5 \tilde{A}_{2+1}^2  \mu_1\mu_3 \right) (\mu_{13} - \mu_1\mu_3) \nonumber\\
    &\quad\quad\quad
    + \frac{5}{4} \tilde{A}_{2+2}^2 (-1 + \mu_1^2 + \mu_3^2 + 2\mu_{13}^2 - 4\mu_1\mu_3\mu_{13} + \mu_1^2\mu_3^2)
    \Biggr\} 
    P_\phi(k_1) P_\phi(k_3) P_\sigma(k_{12}) + 11~\mathrm{perms}, \\
    \mathrm{Im}\left[T_\Phi\right] 
    &= 
    \frac{3}{8\pi} 
    \left\{\tilde{A}_{1+1}^2 
    - \sqrt{5} \tilde{A}_{1+1} \tilde{A}_{2+1} (\mu_1 - \mu_3) 
    - 5 \tilde{A}_{2+1}^2  \mu_1\mu_3 
    - \frac{5}{2} \tilde{A}_{2+2}^2 (\mu_{13} - \mu_1\mu_3)
    \right\} \nonumber\\
    &\quad\quad\quad\times
    \left[\hbk_{12} \cdot \left(\hbk_1 \times \hbk_3\right)\right] P_\phi(k_1) P_\phi(k_3) P_\sigma(k_{12}) + 11~\mathrm{perms}. 
\end{align} 
By choosing the parameters such that
\begin{align}
    \tilde{A}^2 \equiv \tilde{A}_{1+1}^2 = 5\tilde{A}_{2+1}^2 = \frac{5}{2}\tilde{A}_{2+2}^2, 
    \quad \text{with} \quad
    \tilde{A}_{1+1} \tilde{A}_{2+1} < 0,
    \label{eq:Alm_tilde_U1gauge_app}
\end{align}
we reproduce the $U(1)$-gauge model template given in \eq{eq:dn_odd_parametrization}, with 
$A_\mathrm{gauge} = 9\tilde{A}^2 / 400\pi > 0$. 
Finally, using the redefinition of the amplitude parameters from \eq{eq:redef_Alm}, we find that the condition in \eq{eq:Alm_tilde_U1gauge_app} corresponds to
\begin{align}
    A^2 \equiv A_{1\pm1}^2 = A_{2\pm1}^2 = A_{2\pm2}^2, 
    \quad \text{with} \quad
    A_{1+1} A_{2+1} > 0,
    \label{eq:Alm_U1gauge_app}
\end{align}
and in this case, $A_\mathrm{gauge} = 3A^2 / 100 > 0$, where we have used $c_{11} = \sqrt{3 / 4\pi}$ in \eq{eq:def_clm}.  
The trispectrum with $A_\mathrm{gauge} < 0$ can again be obtained by flipping the sign of $m$.

\subsection{Correction to initial power spectrum} 
\label{subapp:correction_initial_power} 
The quadratic correction in \eq{eq:def_Phix_app} modifies the initial power spectrum. 
The correction is given by the auto power spectrum of the quadratic term. 
Following the same procedure as in the previous subsection, we obtain the general expression: 
\begin{align}
    \Delta P_\Phi(k) 
    &= 
    \sum_{\ell m} \sum_{\ell' m'} \rho_{\ell\ell'}\tilde{A}_{\ell m} \tilde{A}_{\ell' m'}
    \avrg{\phi_{\ell m}^{(2)}(\bk;\sigma_{\ell m}) \phi_{\ell' m'}^{(2)}(\bk';\sigma_{\ell' m'})}' \nonumber\\
    &= 
    \sum_{\ell\ell'm} \rho_{\ell\ell'} \tilde{A}_{\ell m} \tilde{A}_{\ell' m} 
    \int_{\bq_1,\bq_2} 
    (2\pi)^3\delD_{\bk-\bq_{12}} 
    Y_{\ell m} (\hbq_1;\hbq_2) 
    Y_{\ell' m}^* (\hbq_1;\hbq_2)
    P_\phi(q_1) P_\sigma(q_2). 
    \label{eq:correction_ini_power} 
\end{align}
For the $U(1)$-gauge model, the correction takes the specific form: 
\begin{align}
    \Delta P_\Phi(k) 
    = 
    \frac{A^2}{4} 
    \int_{\bq_1,\bq_2} 
    (2\pi)^3\delD_{\bk-\bq_{12}} 
    (3-4\mu_{12}-2\mu_{12}^2+4\mu_{12}^3-\mu_{12}^4) 
    P_\phi(q_1) P_\sigma(q_2). 
    \label{eq:correction_ini_power_U1gauge} 
\end{align}

\section{Validation tests for initial conditions} 
\label{app:validation_tests_ic} 
To validate our initial conditions, we compare the trispectrum measured from the simulations with its analytical prediction. 
We first review an estimator for the parity-odd trispectrum developed in Ref.~\cite{Coulton+2024:Quijote-Odd}. 
Unlike the squeezed-type trispectrum considered in Ref.~\cite{Coulton+2024:Quijote-Odd}, the collapsed-type trispectrum depends explicitly on the diagonal momenta. 
As a result, when computing the corresponding theoretical prediction for the measurements, the spherical wave expansion of the Dirac delta function used in Ref.~\cite{Coulton+2024:Quijote-Odd}, while still applicable, leads to an expression involving an infinite sum over angular momentum, making it less tractable. 
Therefore, in order to derive a simpler expression in the thin-shell limit that is applicable to more general trispectra, including the collapsed type, we present a more careful review and derivation in the following.

\subsection{Parity-odd trispectrum estimator} 
\label{subapp:parity-odd_trisp_estimator}
We adopt the parity-odd binned trispectrum estimator originally proposed in Ref.~\cite{Coulton+2024:Quijote-Odd}, which is defined as
\begin{align}
    \hat{T}_{\delta,12345}^{(-)}
    &\equiv 
    \hat{T}_\delta^{(-)}(b_1,b_2,b_3,b_4,b_5) \nonumber\\
    &\equiv 
    \frac{V^3}{N^T_{12345}} 
    \sum_{\mybm_1 \in b_1} \sum_{\mybm_2 \in b_2} \sum_{\mybm_3 \in b_3} \sum_{\mybm_4 \in b_4} \sum_{\mybm_5 \in b_5} 
    \left[i\bk_1\cdot(i\bk_2\times i\bk_3) \right]
    \delta(\bk_1)\delta(\bk_2)\delta(\bk_3)\delta(\bk_4) 
    \delK_{\mybm_{12},\mybm_5} \, \delK_{\mybm_{34},-\mybm_5}, 
    \label{eq:def_trispectrum_estimator}
\end{align}
where 
$\mybm_i = (m_{i,x}, m_{i,y}, m_{i,z})$ is a tuple of three integers specifying the $i$-th wavevector as 
$\bk_i = k_\rmF \mybm_i$, with $k_\rmF \equiv 2\pi/L$ denoting the fundamental frequency of the periodic simulation box with volume $V = L^3$. 
The index $b_i$ indicates the bin label of the $i$-th wavevector $\bk_i$ in a spherical shell. 
Here, $b_1,\dots,b_4$ correspond to the bin labels for the four edge momenta, while $b_5$ corresponds to that for the diagonal momentum $\bk_5 = \bk_{12} = -\bk_{34}$. 
The factor $V^3$ on the right-hand side arises from our forward-normalized discrete Fourier transform (DFT), where we define $\delta(\bk) \equiv \delta^\mathrm{DFT}(\bk) / N_\mathrm{grid}^3$, with $\delta^\mathrm{DFT}(\bk)$ obtained from the unnormalized DFT and $N_\mathrm{grid}$ being the number of grid points per side used in the FFT algorithm. 
For normalization, we define the number of tetrahedra formed within a bin as
\begin{align}
    N_{12345}^T
    \equiv 
    N^T (b_1,b_2,b_3,b_4,b_5) 
    \equiv \sum_{\mybm_1 \in b_1} \sum_{\mybm_2 \in b_2} \sum_{\mybm_3 \in b_3} \sum_{\mybm_4 \in b_4} \sum_{\mybm_5 \in b_5} 
    \delK_{\mybm_{12},\mybm_5} \delK_{\mybm_{34},-\mybm_5}. 
    \label{eq:def_trispectrum_normalization}
\end{align}
Note that the scalar triple product in Eq.~\eqref{eq:def_trispectrum_estimator} extracts the parity-odd component (i.e., the imaginary part) of the matter trispectrum. 
Since the normalization defined in Eq.~\eqref{eq:def_trispectrum_normalization} does not include the scalar triple product, the ensemble average of the quantity measured by Eq.~\eqref{eq:def_trispectrum_estimator} is not exactly the parity-odd component of the matter trispectrum itself (see the next subsection for further details).  
We compute $\hat{T}_\delta^{(-)}$ and $N_{12345}^T$ efficiently using FFTs, based on the (discrete) plane wave expansion of the Kronecker delta \citep[see, e.g.,][]{Steele&Baldauf2021:Trisp_Estimator,Coulton+2024:Quijote-Odd}.
\begin{align}
    \hat{T}_{\delta,12345}^{(-)}
    = 
    \frac{V^3}{N^T_{12345}} 
    \sum_{\mybm_5 \in b_5} 
    \varepsilon_{ijk}
    \calF_{b_1b_2}^{ij}(\bk_5) 
    \calG_{b_3b_4}^k(-\bk_5), 
    \quad
    N_{12345}^T
    = 
    \sum_{\mybm_5 \in b_5} \calU_{b_1b_2}(\bk_5) \calU_{b_3b_4}(-\bk_5), 
\end{align}
with 
\begin{align}
    \calF_{bb'}^{ij}(\bk) 
    \equiv 
    \int_{\bk}
    \left[ \partial_i\delta_{b}(\bx) \right]
    \left[ \partial_j\delta_{b'}(\bx) \right]
    e^{i\bk\cdot\bx}, 
    \quad
    \calG_{bb'}^{k}(\bk) 
    \equiv 
    \int_{\bk}
    \left[ \partial_k\delta_{b}(\bx) \right]
    \delta_{b'}(\bx)
    e^{i\bk\cdot\bx}, 
    \quad
    \calU_{bb'}(\bk) 
    \equiv 
    \int_{\bk}
    u_{b}(\bx)
    u_{b'}(\bx)
    e^{i\bk\cdot\bx}, 
    \label{eq:calF_calG_calU}
\end{align} 
where $\delta_b$ is the masked density field which, in Fourier space, takes $\delta_b(\bk)=\delta(\bk)$ when the mode is in the $b$-th spherical-shell bin (otherwize zero), and $u_b$ is the field similarly defined but it takes unity, i.e., $u_b(\bk)=1$, instead. 
Note that the presence of the imaginary unit in front of each $\bk_i$ in \eq{eq:def_trispectrum_estimator} ensures that all calculations can be carried out using real-to-complex Fourier transforms because its real-space counterpart corresponds to the derivative operator, $\partial_i$, as shown in \eq{eq:calF_calG_calU}. 
Similarly, the bin-averaged wavenumber $k_i$ ($i = 1,\dots,5$) for each bin can be computed as
\begin{align}
    k_i (b_1,b_2,b_3,b_4,b_5) 
    \equiv 
    \frac{1}{N^T_{12345}} 
    \sum_{\mybm_1 \in b_1} \sum_{\mybm_2 \in b_2} \sum_{\mybm_3 \in b_3} \sum_{\mybm_4 \in b_4} \sum_{\mybm_5 \in b_5} 
    |\bk_i|
    \delK_{\mybm_{12},\mybm_5} \delK_{\mybm_{34},-\mybm_5}. 
    \label{eq:def_effective_k}
\end{align}

\subsection{Thin-shell limit} 
\label{subapp:thin-shell_limit}
Here we derive the analytical expression for the expectation value of Eq.~\eqref{eq:def_trispectrum_estimator}. 
We consider the continuous limit of the discrete sum and the Kronecker delta as follows: 
\begin{align}
    \sum_{\mybm_i \in b_i} 
    \rightarrow 
    V \int_{\bq_i} u_{b_i}(\bq_i), 
    \quad 
    \delK_{\mybm_{12},\mybm_5} 
    \rightarrow 
    \frac{(2\pi)^3}{V} \delD_{\bq_{12}-\bq_5}, 
    \label{eq:cont_lim}
\end{align}
where $u_b$ is the selection function for the $b$-th shell, defined in Eq.~\eqref{eq:calF_calG_calU}.
We begin by taking the continuous limit of Eq.~\eqref{eq:def_trispectrum_normalization}, we obtain: 
\begin{align}
    N_{12345}^T
    &\rightarrow 
    V^3 
    \prod_{i=1}^5 
    \left[
    \int_{\bq_i} u_{b_i}(\bq_i) 
    \right]
    (2\pi)^3 \delD_{\bq_{12}-\bq_5}
    (2\pi)^3 \delD_{\bq_{34}+\bq_5} \nonumber\\
    &= 
    V^3 
    \int_{\bq_5} u_{b_5}(\bq_5) 
    \left[
    \int_{\bq_1} u_{b_1}(\bq_1) u_{b_2}(\bq_5-\bq_1) 
    \right]
    \left[
    \int_{\bq_3} u_{b_3}(\bq_3) u_{b_4}(-\bq_5-\bq_3) 
    \right]. 
    \label{eq:def_trispectrum_normalization_cont_lim}
\end{align}
Next, adopting the thin-shell limit for the selection function:
\begin{align}
    u_{b_i}(\bq_i) 
    \rightarrow 
    \Delta k \delD(q_i - k_i),
\end{align} 
where $k_i$ is the effective wave number defined in Eq.~\eqref{eq:def_effective_k}, Eq.~\eqref{eq:def_trispectrum_normalization_cont_lim} simplifies to:
\begin{align}
    N_{12345}^T 
    &\rightarrow 
    V^3 (\Delta k)^5 
    \int_{q_5} \delD(q_5 - k_5)
    \int_{\hbq_5} 
    \left[ 
    \int_{\bq_1} 
    \delD(q_1 - k_1)
    \delD(|\bq_5-\bq_1| - k_2)
    \right]
    \left[ 
    \int_{\bq_3} 
    \delD(q_3 - k_3)
    \delD(|\bq_5+\bq_3| - k_4)
    \right] \nonumber\\
    &= 
    V^3 (\Delta k)^5 
    \frac{k_1^2}{2\pi^2}\frac{k_3^2}{2\pi^2}\frac{k_5^2}{2\pi^2} 
    \int_{\hbq_1,\hbq_3,\hbq_5} 
    \delD(|k_5\hbq_5-k_1\hbq_1| - k_2) 
    \delD(|k_5\hbq_5+k_3\hbq_3| - k_4) \nonumber\\
    &= 
    V^3 (\Delta k)^5 
    \frac{k_1^2}{2\pi^2}\frac{k_3^2}{2\pi^2}\frac{k_5^2}{2\pi^2} 
    \left[
    \int_{-1}^1 \frac{\rmd\mu_{1}}{2} 
    \delD(|k_5\hbq_5-k_1\hbq_1| - k_2) 
    \right]
    \left[
    \int_{-1}^1 \frac{\rmd\mu_{3}}{2} 
    \delD(|k_5\hbq_5+k_3\hbq_3| - k_4) 
    \right] 
    \int_{0}^{2\pi} \frac{\rmd\phi}{2\pi}\,1\nonumber\\
    &= 
    V^3 (\Delta k)^5 
    \frac{k_1k_2k_3k_4}{32\pi^6} \Delta(k_1,k_2,k_5) \Delta(k_3,k_4,k_5), 
    \label{eq:def_trispectrum_normalization_thin_lim} 
\end{align}
where $\Delta(a,b,c)$ is a selection function that returns 1 if the arguments satisfy the triangle condition, and 0 otherwise. 
Note that, in the third line, $\mu_i \equiv \hbq_5 \cdot \hbq_i$ denotes the cosine of the angle between the diagonal momentum $\bq_5$ and edge momenta, and $\phi$ is the angle between the two triangles sharing $\bq_5$. 

Similarly, evaluating the expectation value of Eq.~\eqref{eq:def_trispectrum_estimator} under the continuous and thin-shell limits gives:
\begin{align}
    \avrg{\hat{T}_{\delta,12345}^{(-)}} 
    \rightarrow 
    k_5^2 \prod_{i=1}^4 
    \left[
    k_i\sqrt{1-\mu_i^2} 
    \right]
    \int_{0}^{2\pi} \frac{\rmd\phi}{2\pi} 
    \sin^2\phi 
    ~\tau_-^\delta(k_1,k_2,k_3,k_4,k_5,\phi), 
    \label{eq:trispectrum_thin_lim} 
\end{align}
where $\tau_-^\delta$ is the parity-odd component of the matter trispectrum (excluding the scalar triple product as defined in Eq.~\ref{eq:def_tau-}). 
If the trispectrum has no dependence on the diagonal configuration, i.e., no $\phi$ dependence, the $\phi$ integral reduces to a constant factor of $1/2$. 
However, if the trispectrum depends on $\phi$, the full one-dimensional integral over $\phi$ must be evaluated explicitly. 

\begin{figure}
    \centering
    \includegraphics[width=1.0\columnwidth]{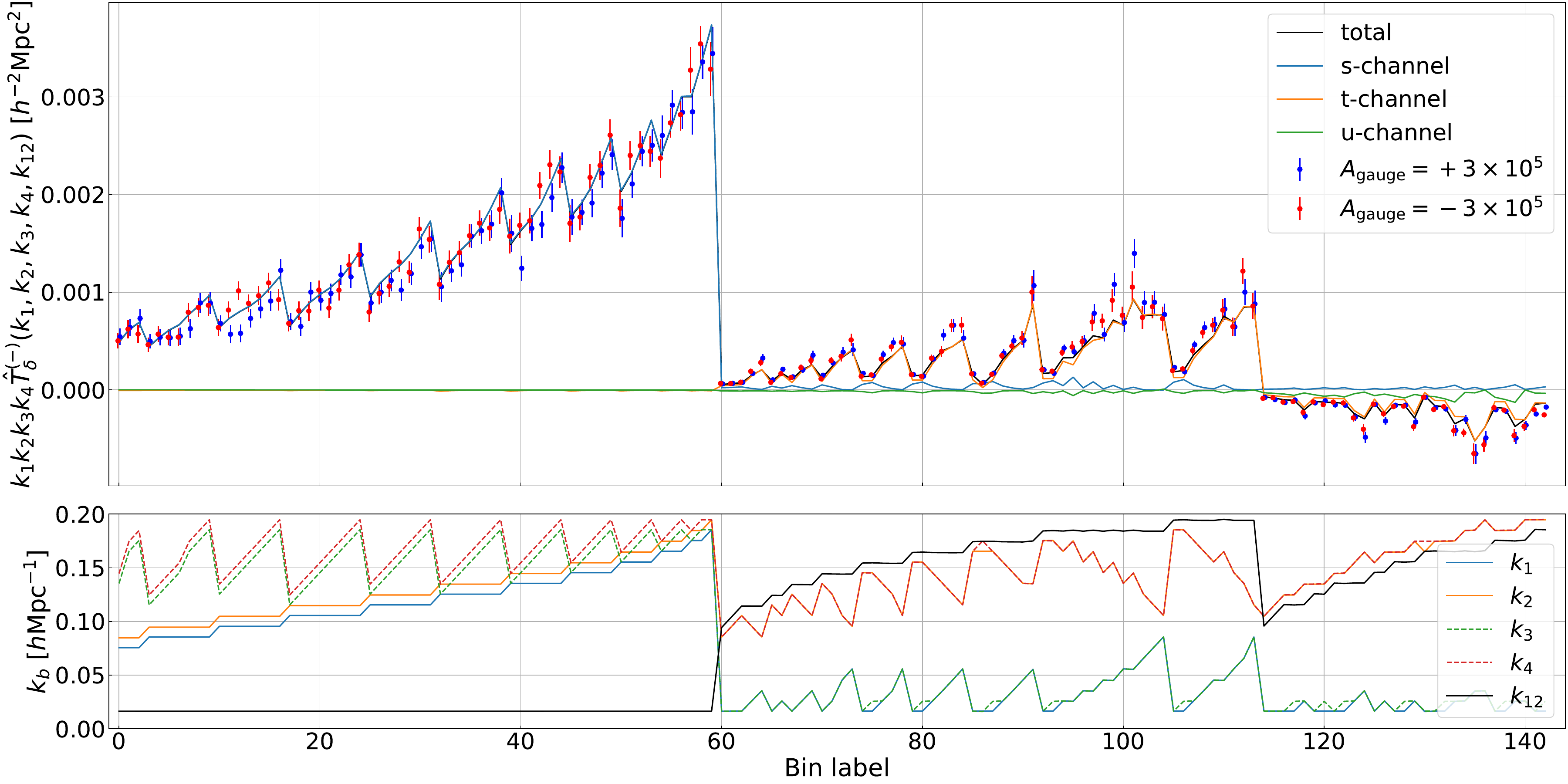}
    \caption{Same as Fig.~\ref{fig:trisp_u1}, but the theoretical prediction is split into individual channels for clarity. 
    The theory curves are computed using a sub-gridding procedure to improve the thin-shell approximation (see text for details). 
    For visual clarity, the sign of the signals with opposite amplitude $A_\mathrm{gauge}$ is flipped. 
    Note that $k_{12} \equiv k_5$ is the diagonal momentum.} 
    \label{fig:trisp_u1_app}
\end{figure}
Fig.~\ref{fig:trisp_u1_app} shows a comparison between the parity-odd trispectrum measured from the initial conditions and the theoretical prediction computed from Eq.~\eqref{eq:trispectrum_thin_lim}. 
To highlight the structure of the theoretical prediction, we decompose the total signal into contributions from individual channels, shown in different colors. 
For the decomposition into the three channels, see Section~\ref{subapp:implementation_of_loop_int_collapsed}. 
The measurement is performed using $k_\mathrm{min} = 0.0$, $k_\mathrm{max} = 0.2$, and $N_\mathrm{bin} = 20$, i.e., with a bin width $\Delta k = 0.01$. 
However, using the effective wavenumbers $k_i$ derived directly from this binning setup leads to a poor approximation in the thin-shell limit. 
To improve accuracy, we employ a sub-gridding method: we increase the number of bins to $N_\mathrm{bin}^\mathrm{HR} = 100$ (i.e., $\Delta k^\mathrm{HR} = 0.002$), compute Eq.~\eqref{eq:trispectrum_thin_lim} for each subgrid configuration with $k_i^\mathrm{HR}$, and then take the weighted average of the resulting values using the corresponding $N_{12345}^{T,\mathrm{HR}}$. 
Since we refine each of the five $k_i$ directions by a factor of 5, this results in $5^5 = 3125$ subgrid points for each original configuration. 
We have checked that this result is well converged for both $N_\mathrm{bin}^\mathrm{HR} = 80$ and $N_\mathrm{bin}^\mathrm{HR} = 100$. 
As shown in Fig.~\ref{fig:trisp_u1_app}, the configurations listed at the beginning correspond to cases where the diagonal momentum is much smaller than the edge momenta, i.e., $k_{12} \ll k_i$. 
In this regime, since $\bs = \bk_1 + \bk_2$ is small, the $s$-channel contribution (blue) dominates. 
On the other hand, for configurations where $k_1 = k_3$ and $k_2 = k_4$, the diagonal momentum $\bt = \bk_1 + \bk_3$ becomes small, and the $t$-channel (orange) provides the dominant contribution. 
Finally, the last group of configurations illustrates how the overall sign of the trispectrum can flip depending on the relative size of $k_{12}$ and $k_2$. 

\section{Dependence on shape measurements} 
\label{app:shape_diff} 
In this section, we further examine how the definitions of halo shape measurements affect the bias parameters. 
Our main analysis in the paper uses the reduced inertia tensor, which employs a radial weighting $w(r) = 1/r^2$ in \eq{eq:def_Iij}:
\begin{align}
    I_{ij,\rmg}^\mathrm{red} = \frac{1}{N_w}\int \rmd\br\, \rho_\rmg(\br) \frac{r_i r_j}{r^2}, 
    \label{eq:def_Iij_reduced}
\end{align}
to focus on the inner regions of halos. 
Here, we additionally present results obtained using the original (unweighted) inertia tensor with $w(r) = 1$:
\begin{align}
    I_{ij,\rmg}^\mathrm{orig} = \frac{1}{N_w}\int \rmd\br\, \rho_\rmg(\br) r_i r_j. 
    \label{eq:def_Iij_original}
\end{align}

\begin{figure}
    \centering
    \includegraphics[width=1.0\columnwidth]{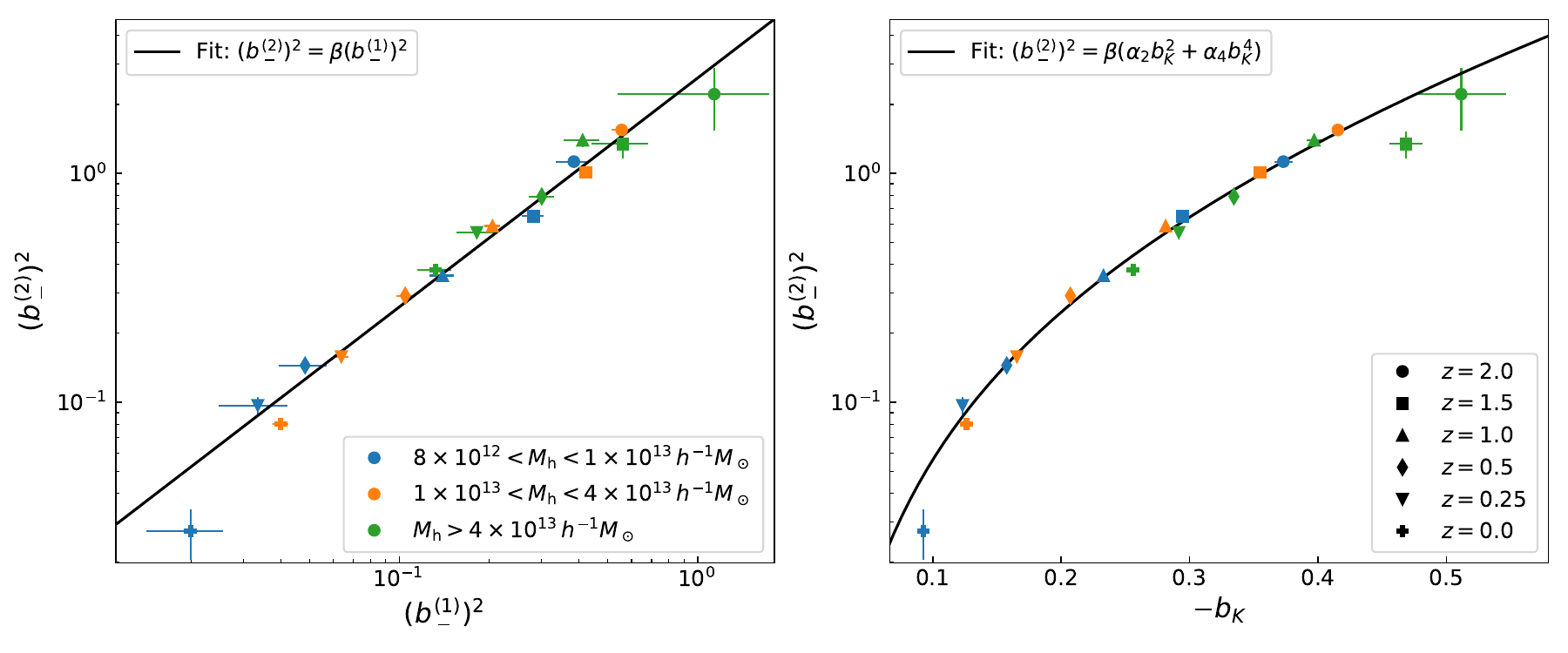}
    \caption{Same as Fig.~\ref{fig:bias_fit}, but based on halo shapes defined using the original (unweighted) inertia tensor instead of the reduced one.}
    \label{fig:bias_fit_app}
\end{figure}
Fig.~\ref{fig:bias_fit_app} shows the bias parameters measured with this alternative shape definition, along with the corresponding best-fit curves. 
The best-fit parameters, assuming the same functional forms as \eqs{eq:bias_relations:bm1_bm2}{eq:bias_relations:bm1_bK}, are given by 
\begin{align}
    \beta = 2.61 \pm 0.11,\quad
    \alpha_2 = 2.08 \pm 0.16,\quad
    \alpha_4 = 7.34 \pm 1.78. 
    \label{eq:bias_best-fit_beta_alpha_app}
\end{align}
While $\beta$ and $\alpha_2$ remain consistent within $1\sigma$, the best-fit values of $\alpha_4$ differ from those obtained using the reduced inertia tensor (see Eqs.~\ref{eq:bias_best-fit_beta} and \ref{eq:bias_best-fit_alpha}). 
This difference likely reflects the fact that the amplitude of the Gaussian linear bias $b_K$ depends on the definition of the inertia tensor used in shape measurements. 
Despite the shift in parameter values, the same polynomial model provides a good fit to the data in both cases.

\section{Three-dimensional \texorpdfstring{$EB$}{EB} power spectrum} 
\label{app:3d_EB_power_spectrum}
We derive \eq{eq:EB_power_k_mu} starting from \eq{eq:projection_gamma_x}, following the method in Ref.~\cite{Kurita&Takada2022:AnalysisIAPS}, but with slightly modified notation for improved clarity. 
The traceless part of the projected tensor has two degrees of freedom, corresponding to spin-2 quantities on the sky, which can be isolated as
\begin{align}
    {}_{\pm2}\gamma(\bx;\hbn) 
    &\equiv 
    \left[m_i^\pm(\hbn) m_j^\pm(\hbn)\right]^\dagger \gamma_{ij}(\bx;\hbn) \\
    &= 
    \left[m_i^\pm(\hbn) m_j^\pm(\hbn)\right]^\dagger S_{ij}(\bx), 
    \label{eq:gamma_x_spin2} 
\end{align} 
where $m_i^\pm$ is a complex unit vector associated with the line-of-sight direction $\hbn$, defined as
\begin{align}
    m_i^\pm(\hbn) 
    \equiv 
    \mp\frac{1}{\sqrt{2}} 
    \left( e_{\theta,i}(\hbn) \mp j \, e_{\phi,i}(\hbn) \right), 
    \label{eq:def_mpm}
\end{align} 
and we have used the identity 
$m_i^\pm m_j^\pm \calP_{ijkl} = m_k^\pm m_l^\pm$ 
in the second equality.
This vector is the complex conjugate of the complex basis vector defined in \eq{eq:def_epm}, i.e. $e_i^\pm$, but with the argument replaced from the Fourier direction $\hbk$ to the line-of-sight direction $\hbn$. 
To distinguish the two, we denote it by a different symbol $m_i^\pm$, although its mathematical properties are identical to $e_i^\pm$. 
Note that the complex notation associated with $m_i^\pm$, i.e., the spin-2 decomposition, is introduced purely for notational convenience. 
To avoid confusion with complex conjugation of the physical field $S_{ij}$ in Fourier space in the following, we use ``$j$'' for the imaginary unit and ``$\dagger$'' for complex conjugation for $m_i^\pm$. 
For example, in our notation, $[m_i^\pm]^\dagger = -m_i^\mp$ and thus $[{}_{\pm2}\gamma]^\dagger = {}_{\mp2}\gamma$. 
The overall sign in \eq{eq:def_mpm} is irrelevant, as the projection is always quadratic in $m_i^\pm$. 

Under the global plane-parallel approximation, the spin-2 decomposition in Fourier space takes an analogous form:
\begin{align}
    {}_{\pm2}\gamma(\bk;\hbn) 
    = 
    \left[m_i^\pm(\hbn) m_j^\pm(\hbn)\right]^\dagger S_{ij}(\bk). 
    \label{eq:gamma_k_spin2} 
\end{align} 
We define coordinate-independent quantities, the so-called $E$-mode and $B$-mode fields, as 
\begin{align}
    E(\bk;\hbn) \pm j B(\bk;\hbn) 
    \equiv 
    {}_{\pm2}\gamma(\bk;\hbn) e^{\mp 2j\phi_{\hbk,\hbn}}
    \quad
    \mathrm{with}
    \quad
    e^{\pm 2j\phi_{\hbk,\hbn}} 
    \equiv 
    \frac{2m_i^\pm m_j^\pm \hk_i\hk_j}{\calP_{ij}(\hbn)\hk_i\hk_j}. 
    \label{eq:def_EB_modes} 
\end{align}
The phase factor cancels the dependence on the coordinate system in the plane perpendicular to $\hbn$. 
Note that these two fields satisfy the reality condition, e.g. $E(-\bk;\hbn) = E^*(\bk;\hbn)$. 
Following the same calculation as in Ref.~\cite{Kurita&Takada2022:AnalysisIAPS}, the coordinate-independent power spectra can be summarized as: 
\begin{align}
    \avrg{\left[E(\bk;\hbn) + j B(\bk;\hbn)\right] \left[E(\bk';\hbn) - j B(\bk';\hbn)\right]}' 
    &= 
    P_{EE}(k,\mu) + P_{BB}(k,\mu) + 2jP_{EB}(k,\mu), \label{eq:P_plus}\\
    \avrg{\left[E(\bk;\hbn) + j B(\bk;\hbn)\right] \left[E(\bk';\hbn) + j B(\bk';\hbn)\right]}' 
    &= 
    P_{EE}(k,\mu) - P_{BB}(k,\mu). \label{eq:P_minus} 
\end{align}
The parity-even part, $P_{EE}$ and $P_{BB}$, has already been derived in Ref.~\cite{Kurita&Takada2022:AnalysisIAPS}, so we omit the details here.  
The parity-odd part in \eq{eq:P_plus}, $P_{EB}$, arises due to parity violation and was absent in Ref.~\cite{Kurita&Takada2022:AnalysisIAPS}. 
Below, we derive its explicit form in detail. 

\begin{table}
    \centering
    \renewcommand{\arraystretch}{2.25}
    \begin{tabular}{cccccc}
    \toprule\midrule
    & $\Lambda_{ij,kl}^{(0,+)}(\hbk)$ & $\Lambda_{ij,kl}^{(1,+)}(\hbk)$ & $\Lambda_{ij,kl}^{(2,+)}(\hbk)$ & $\Lambda_{ij,kl}^{(1,-)}(\hbk)$ & $\Lambda_{ij,kl}^{(2,-)}(\hbk)$ \\
    \midrule
    $\left[m_i^+ m_j^+ m_k^- m_l^-\right]^\dagger(\hbn)$ & $\dfrac{3}{8}(1-\mu^2)^2$ & $\dfrac{1}{8}(1-\mu^2)\left\{(1-\mu)^2 + (1+\mu)^2\right\} $ & $\dfrac{1}{32}\left\{(1-\mu)^4 + (1+\mu)^4\right\}$ & $\dfrac{ij}{2}\mu(1-\mu^2)$ & $\dfrac{ij}{4}\mu(1+\mu^2)$\\
    $\left[m_i^+ m_j^+ m_k^+ m_l^+\right]^\dagger(\hbn)$ & $\dfrac{3}{8}(1-\mu^2)^2 e^{+4j\phi_{\hbk,\hbn}}$ & $-\dfrac{1}{4}(1-\mu^2)^2 e^{+4j\phi_{\hbk,\hbn}}$ & $\dfrac{1}{16}(1-\mu^2)^2 e^{+4j\phi_{\hbk,\hbn}}$ & $0$ & $0$\\
    \midrule\bottomrule
    \end{tabular}
    \caption{Summary of the tensor contractions between the projection vectors, $m_i^\pm$, defined with respect to the line-of-sight direction $\hbn$, and the Fourier-space projection tensor $\Lambda_{ij,kl}^{(\lambda,s)}$ corresponding to the tensor field mode $\hbk$. 
    The parity-even components are reproduced from Ref.~\cite{Kurita&Takada2022:AnalysisIAPS}. 
    }
    \label{tab:mmmm_x_Lambda}
\end{table}
From \eq{eq:def_EB_modes}, the imaginary part of the left hand side of \eq{eq:P_plus} in $j$ becomes 
\begin{align}
    -\avrg{E(\bk;\hbn) B(\bk';\hbn)} + \avrg{B(\bk;\hbn) E(\bk';\hbn)}. 
    \label{eq:Im_P_plus}
\end{align}
The first term in \eq{eq:Im_P_plus} reduces to 
\begin{align}
    \avrg{E(\bk;\hbn) B(\bk';\hbn)} 
    = 
    -\frac{j}{4} 
    \Bigl\{
    &+\avrg{{}_{+2}\gamma(\bk;\hbn) {}_{+2}\gamma(\bk';\hbn)} e^{- 2j\phi_{\hbk,\hbn}} e^{- 2j\phi_{\hbk',\hbn}}  
    -\avrg{{}_{+2}\gamma(\bk;\hbn) {}_{-2}\gamma(\bk';\hbn)} e^{- 2j\phi_{\hbk,\hbn}} e^{+ 2j\phi_{\hbk',\hbn}} \nonumber\\ 
    &+\avrg{{}_{-2}\gamma(\bk;\hbn) {}_{+2}\gamma(\bk';\hbn)} e^{+ 2j\phi_{\hbk,\hbn}} e^{- 2j\phi_{\hbk',\hbn}} 
    -\avrg{{}_{-2}\gamma(\bk;\hbn) {}_{-2}\gamma(\bk';\hbn)} e^{+ 2j\phi_{\hbk,\hbn}} e^{+ 2j\phi_{\hbk',\hbn}} 
    \Bigr\}. 
    \label{eq:EB_4terms} 
\end{align}
Using Eqs.~\eqref{eq:gamma_k_spin2}, \eqref{eq:def_Pijkl}, and \eqref{eq:def_projection}, the first term in \eq{eq:EB_4terms} reduces to 
\begin{align}
    \avrg{{}_{+2}\gamma(\bk;\hbn) {}_{+2}\gamma(\bk';\hbn)} 
    &= 
    \left[m_i^+ m_j^+ m_k^+ m_l^+\right]^\dagger \avrg{S_{ij}(\bk) S_{kl}(\bk')} \\
    &= 
    (2\pi)^3 \delD_{\bk+\bk'} 
    \sum_{\lambda,s} 
    \calN_{\lambda}^{-1} 
    \left[m_i^+ m_j^+ m_k^+ m_l^+\right]^\dagger \Lambda^{(\lambda,s)}_{ij,kl}(\hbk) P_s^{(\lambda)}(k). 
\end{align} 
Here we encounter the tensor contraction between the projection vectors with respect to the line-of-sight, $[m_i^+ m_j^+ m_k^+ m_l^+]^\dagger(\hbn)$, and the tensor projectors for each Fourier mode $\Lambda^{(\lambda,s)}_{ij,kl}(\hbk)$. 
These contractions are summarized in Table~\ref{tab:mmmm_x_Lambda}.
We observe that, since the line-of-sight projector $m_i^+ m_j^+ m_k^+ m_l^+$ is totally symmetric in this case, only the parity-even components ($s=+$) contribute. 
However, these contributions cancel exactly with the fourth term in Eq.~\eqref{eq:EB_4terms}. 
In a similar manner, the second term in \eq{eq:EB_4terms} reduces to  
\begin{align}
    \avrg{{}_{+2}\gamma(\bk;\hbn) {}_{-2}\gamma(\bk';\hbn)} 
    &= 
    \left[m_i^+ m_j^+ m_k^- m_l^-\right]^\dagger \avrg{S_{ij}(\bk) S_{kl}(\bk')} \\
    &= 
    (2\pi)^3 \delD_{\bk+\bk'} 
    \sum_{\lambda,s} 
    \calN_{\lambda}^{-1} 
    \left[m_i^+ m_j^+ m_k^- m_l^-\right]^\dagger \Lambda^{(\lambda,s)}_{ij,kl}(\hbk) P_s^{(\lambda)}(k). 
\end{align} 
According to Table~\ref{tab:mmmm_x_Lambda}, both parity-even and parity-odd components become nonzero here. 
The parity-even contributions cancel with those from the third term, while the parity-odd contributions from the second and third terms are identical. 
Therefore, we can simply compute one of them and double the result. 
We thus obtain 
\begin{align} 
    P_{EB}(k,\mu) 
    \equiv
    \avrg{E(\bk;\hbn) B(\bk';\hbn)}' 
    &= 
    \frac{j}{2} \avrg{{}_{+2}\gamma(\bk;\hbn) {}_{-2}\gamma(\bk';\hbn)}' e^{- 2j\phi_{\hbk,\hbn}} e^{+ 2j\phi_{\hbk',\hbn}} \nonumber\\
    &= 
    \frac{j}{2} 
    \sum_{\lambda=1,2} 
    \calN_{\lambda}^{-1} 
    \left[m_i^+ m_j^+ m_k^- m_l^-\right]^\dagger \Lambda^{(\lambda,-)}_{ij,kl}(\hbk) P_-^{(\lambda)}(k) 
    e^{- 2j\phi_{\hbk,\hbn}} e^{+ 2j\phi_{-\hbk,\hbn}} \nonumber\\
    &= 
    \frac{j}{2}
    \left\{
    ij\mu(1-\mu^2)P_-^{(1)}(k)  + \frac{ij}{2}\mu(1+\mu^2)P_-^{(2)}(k) 
    \right\} \nonumber\\
    &= 
    -\frac{i}{2}\mu(1-\mu^2)P_-^{(1)}(k) - \frac{i}{4}\mu(1+\mu^2)P_-^{(2)}(k), 
    \label{eq:P_EB_app} 
\end{align}
where we used $e^{2j\phi_{-\hbk,\hbn}} = e^{2j\left(\phi_{\hbk,\hbn}+\pi\right)} = e^{2j\phi_{\hbk,\hbn}}$. 
This corresponds to \eq{eq:EB_power_k_mu} in the main text.

\section{Angular \texorpdfstring{$EB$}{EB} power spectrum} 
\label{app:angular_EB_power_spectrum} 
\subsection{Full-sky formulae} 
\label{subapp:full-sky} 
In this appendix, we summarize the expressions for the angular power spectra required in the Fisher analysis presented in Section~\ref{subsec:photoz_sample}. 
All formulae are derived based on the formalism developed in Ref.~\cite{Schmidt&Jeong2012:Cosmic_Ruler} (see also Refs.~\cite{Schmidt&Jeong2012:LSSwithGW_2,Schmidt+2014:LSSwithGW_3,Biagetti&Orlando2020:PVGW_IA,Philcox+2024:NewPhys_GalaxyShape}), and we only present the final results here.
The primary signal we consider is the IA $EB$ angular power spectrum. 
Since the analysis is performed on linear scales, the sample variance in the covariance includes contributions from the $EE$ auto power spectra of IA (``I'' component) and weak lensing (``G'' component), and their cross-spectrum. 

The auto angular power spectrum for IA can be expressed as
\begin{align}
    \left.C_{XY}\right|^\rmII (\ell; z,z')
    = 
    \pi
    \frac{(\ell-2)!}{(\ell+2)!} 
    \sum_\lambda 
    N_C^{(\lambda)} 
    \frac{(\ell+\lambda)!}{(\ell-\lambda)!} 
    \int_k 
    P_{s(XY)}^{(\lambda)} (k;z,z') 
    \left.F_{X,\ell}^{(\lambda)}\right|^\rmI(x)
    \left.F_{Y,\ell}^{(\lambda)}\right|^\rmI(x'), 
    \label{eq:angular_ps_II_zfix}
\end{align}
where $x \equiv k\chi(z)$ and $x' \equiv k\chi(z')$, and $X,Y \in \{E,B\}$. 
Here, $s(XY)=+$ for parity-even spectra ($XY=EE$ or $BB$), and $s(XY)=-$ for the parity-odd spectrum ($XY=EB$). 
The helicity index $\lambda$ runs over $\{0,1,2\}$ for parity-even spectra and $\{1,2\}$ for parity-odd spectra. 
The normalization factor is given by 
$N_C^{(0)} = 3/2$, $N_C^{(1)} = 2$ and $N_C^{(2)} = 1/2$. 
The (real) kernel function for the intrinsic contribution 
$\left.F_{X,\ell}^{(\lambda)}\right|^\rmI$ 
is defined via 
\begin{align}
    \left.F_{E,\ell}^{(\lambda)}\right|^\rmI(x) + i\left.F_{B,\ell}^{(\lambda)}\right|^\rmI(x) 
    \equiv 
    \hat{Q}^{(\lambda)}(x) \left[ \frac{j_\ell(x)}{x^\lambda}\right], 
    \label{eq:def_kernels_FE_EB}
\end{align}
where $\hat{Q}^{(\lambda)}$ are (complex) differential operators acting on the spherical Bessel function $j_\ell$. 
The explicit forms of these operators are given by
\begin{align}
    \hat{Q}^{(0)}(x) &= 
    4 + 8x\partial_x + x^2 + 12\partial_x^2 + 8x\partial_x^3 + 2x^2\partial_x^2 + x^2\partial_x^4, \\
    \hat{Q}^{(1)}(x) &= 
    \left(
    x^2 + 4x\partial_x + x^2\partial_x^2
    \right)
    + 
    i\left(
    4x + 12\partial_x + x^2\partial_x + 8x\partial_x^2 + x^2\partial_x^3
    \right), \\
    \hat{Q}^{(2)}(x) &= 
    \left(
    12 - x^2 + 8x\partial_x + x^2\partial_x^2
    \right)
    - 
    i\left(
    8x + 2x^2\partial_x
    \right). 
\end{align} 
Applying these operators explicitly, we obtain the right-hand side of \eq{eq:def_kernels_FE_EB} in closed form: 
\begin{align}
    \hat{Q}^{(0)}(x) \left[j_\ell(x)\right]
    &= 
    (\ell-1)\ell(\ell+1)(\ell+2) \frac{j_\ell(x)}{x^2}, \label{eq:act_Q0}\\
    \hat{Q}^{(1)}(x) \left[ \frac{j_\ell(x)}{x}\right]
    &= 
    (\ell^2+\ell-2) \frac{j_\ell(x)}{x}
    +i (\ell^2+\ell-2) 
    \left[
    \frac{j_\ell'(x)}{x} + \frac{j_\ell(x)}{x^2}
    \right], \label{eq:act_Q1}\\ 
    \hat{Q}^{(2)}(x) \left[ \frac{j_\ell(x)}{x^2}\right]
    &= 
    (\ell^2+\ell+2-2x^2)\frac{j_\ell(x)}{x^2} 
    + 2\frac{j_\ell'(x)}{x} 
    -2i
    \left[
    j_\ell'(x) + 2\frac{j_\ell(x)}{x}
    \right]. \label{eq:act_Q2}
\end{align}

For photometric redshift samples, the angular power spectrum between tomographic bins with redshift (radial) distributions $p_a(\chi)$ and $p_{b}(\chi')$ is obtained by replacing the integrand in \eq{eq:angular_ps_II_zfix} as:
\begin{align} 
    P_{s(XY)}^{(\lambda)} (k;z,z') 
    \left.F_{X,\ell}^{(\lambda)}\right|^\rmI(x)
    \left.F_{Y,\ell}^{(\lambda)}\right|^\rmI(x')
    \to 
    \int_0^\infty \rmd \chi\,p_a(\chi)
    \int_0^\infty \rmd \chi'\,p_{b}(\chi')
    P_{s(XY)}^{(\lambda)} (k;z,z') 
    \left.F_{X,\ell}^{(\lambda)}\right|^\rmI(x)
    \left.F_{Y,\ell}^{(\lambda)}\right|^\rmI(x'). 
\end{align} 
For the parity-odd power spectrum based on Eq.~\eqref{eq:eft_model_Agauge}, assuming the geometric approximation $P(k;z,z') \simeq \sqrt{P(k;z)P(k;z')}$ and slowly varying bias parameters $b_-^{(\lambda)}(z)$ within each bin, the tomographic $EB$ power spectrum becomes:
\begin{align}
    \left.C_{EB}^{ab}\right|^\rmII (\ell)
    = 
    A_\mathrm{gauge} 
    \pi
    \frac{(\ell-2)!}{(\ell+2)!} 
    \sum_{\lambda=1}^2 
    N_C^{(\lambda)}
    \frac{(\ell+\lambda)!}{(\ell-\lambda)!} 
    \int_k 
    P_\phi(k) 
    \left[
    \left|
    \bar{b}_-^{(\lambda)}(a) 
    \right|
    \left.\bar{F}_{E,\ell}^{(\lambda)}\right|^\rmI(k;a)
    \right]
    \left[
    \left|
    \bar{b}_-^{(\lambda)}(b)
    \right|
    \left.\bar{F}_{B,\ell}^{(\lambda)}\right|^\rmI(k;b)
    \right], 
    \label{eq:angular_ps_II_EB_tomo}
\end{align}
with the averaged quantities over redshift:
\begin{align}
    \left|
    \bar{b}_-^{(\lambda)}(a) 
    \right|
    \equiv 
    \int_0^\infty \rmd \chi\,p_a(\chi) 
    \sqrt{
    \left(b_-^{(\lambda)}(\chi)\right)^2
    }, 
    \quad
    \bar{F}_{X,\ell}^{(\lambda)}(k;a) 
    \equiv 
    \int_0^\infty \rmd \chi\,p_a(\chi) F_{X,\ell}^{(\lambda)}\left(k\chi\right). 
\end{align}

For the sample variance in the covariance matrix, we consider the linear alignment contribution for IA, which suffices for large-scale signals beyond the flat-sky regime. 
The tomographic $EE$ power spectrum from IA is then:
\begin{align}
    \left.C_{EE}^{ab}\right|^\rmII (\ell)
    = 
    \pi
    \frac{(\ell-2)!}{(\ell+2)!} 
    \int_k 
    P(k;0) 
    \left[\bar{\tilde{b}}_K(a) \left.\bar{F}_{E,\ell}^{(0)}\right|^\rmI(k;a)\right]
    \left[\bar{\tilde{b}}_K(b) \left.\bar{F}_{E,\ell}^{(0)}\right|^\rmI(k;b)\right], 
    \label{eq:angular_ps_II_EE_tomo}
\end{align}
where $\tilde{b}_K(z) \equiv b_K(z)\tilde{D}(z)$ with $\tilde{D}(0)=1$, and
\begin{align}
    \bar{\tilde{b}}_K(a) 
    \equiv 
    \int_0^\infty \rmd \chi\,p_a(\chi) 
    \tilde{b}_K(\chi). 
\end{align}
Note that we assumed that $\tilde{b}_K(z) \propto A_\mathrm{IA}(z)$ varies slowly within each bin.

In addition, we consider the weak lensing contribution to the covariance, which is a scalar (helicity-0) component. 
Under the geometric approximation:
\begin{align}
    \left.C_{EE}\right|^\rmGG (\ell; z,z')
    = 
    \pi
    \frac{(\ell-2)!}{(\ell+2)!} 
    \int_k 
    P(k;0) 
    \left.F_{E,\ell}^{(0)}\right|^\rmG(x)
    \left.F_{E,\ell}^{(0)}\right|^\rmG(x'), 
    \label{eq:angular_ps_GG_zfix}
\end{align}
where
\begin{align}
    \left.F_{E,\ell}^{(0)}\right|^\rmG(x) 
    \equiv 
    3\Omega_\rmm H_0^2 \int_0^\chi \rmd \chi' (\chi-\chi')\frac{\chi'}{\chi} \frac{\tilde{D}(\chi')}{a(\chi')} 
    \hat{Q}^{(0)}(x') \left[ j_\ell(x')\right], 
\end{align}
with $a(\chi)$ being the scale factor. 
The tomographic weak lensing power spectrum becomes:
\begin{align}
    \left.C_{EE}^{ab}\right|^\rmGG (\ell) 
    = 
    \pi
    \frac{(\ell-2)!}{(\ell+2)!} 
    \int_k 
    P(k;0) 
    \left.\bar{F}_{E,\ell}^{(0)}\right|^\rmG(k;a)
    \left.\bar{F}_{E,\ell}^{(0)}\right|^\rmG(k;b).  
    \label{eq:angular_ps_GG_EE_tomo}
\end{align}

In summary, we obtain the weak lensing and IA contributions to the covariance at linear scales: 
\begin{align}
    \left.C_{EE}^{ab}\right|^{\rmGG+\rmGI+\rmIG+\rmII} (\ell)
    = 
    \pi
    \frac{(\ell-2)!}{(\ell+2)!} 
    \int_k 
    P(k;0)  
    \left[
    \left.\bar{F}_{E,\ell}^{(0)}\right|^\rmG(k;a)
    +
    \bar{\tilde{b}}_K(a) \left.\bar{F}_{E,\ell}^{(0)}\right|^\rmI(k;a)
    \right]
    \left[
    \left.\bar{F}_{E,\ell}^{(0)}\right|^\rmG(k;b)
    +
    \bar{\tilde{b}}_K(b) \left.\bar{F}_{E,\ell}^{(0)}\right|^\rmI(k;b)
    \right].  
    \label{eq:angular_ps_TOT_EE_tomo}
\end{align}

\subsection{High-\texorpdfstring{$\ell$}{ell} limit} 
\label{subapp:high_ell} 
The angular power spectrum presented in Eq.~\eqref{eq:angular_ps_II_EB_tomo} is exact for all $\ell$ and thus suitable for a full-sky analysis including low-$\ell$ modes. 
However, its numerical evaluation is computationally expensive. 
In cases where a flat-sky approximation is valid, it is useful to derive the high-$\ell$ limit expressions. 
The leading terms of the kernels in Eqs.~\eqref{eq:act_Q0}–\eqref{eq:act_Q2} in the high-$\ell$ limit are given by: 
\begin{align}
    \hat{Q}^{(0)}(x) \left[j_\ell(x)\right]
    &\simeq
    \frac{\ell^4}{x^2} j_\ell(x), \label{eq:act_Q0_high_ell}\\
    \hat{Q}^{(1)}(x) \left[ \frac{j_\ell(x)}{x}\right]
    &\simeq
    \frac{\ell^2}{x}j_\ell(x)
    +i 
    \left[
    \frac{\ell^2}{x}j_\ell'(x) + \frac{\ell^2}{x^2}j_\ell(x)
    \right], \label{eq:act_Q1_high_ell}\\ 
    \hat{Q}^{(2)}(x) \left[ \frac{j_\ell(x)}{x^2}\right]
    &\simeq 
    \left(\frac{\ell^2}{x^2}-2\right) j_\ell(x)
    -2i 
    \left[
    j_\ell'(x) + \frac{2}{x}j_\ell(x)
    \right]. \label{eq:act_Q2_high_ell}
\end{align}
Note that in this limit, $j_\ell'(x)$ is of the same order as $j_\ell(x)/x$, so we keep both.

The $EB$ cross-power spectrum involves integrals of the real and imaginary parts of these kernels, leading to terms of the form $j_\ell(x)j_\ell(x')$ and $j_\ell'(x)j_\ell(x')$.
The first type can be evaluated using the well-known Limber approximation,
\begin{align}
    j_\ell(x) \simeq \sqrt{\frac{\pi}{2\nu}}\,\delD(x - \nu),
    \quad \text{where} \quad \nu = \ell + 1/2.
\end{align}
Since we are interested only in the leading-order behavior, we will set $\nu \simeq \ell$ in the following. 
Using the Limber approximation, we find 
\begin{align}
    \calI_-^{(\lambda)}[f,g] (\ell)
    &\equiv 
    \int_0^\infty \frac{k^2\rmd k}{2\pi^2} 
    P_\phi(k)
    \int_0^\infty \rmd \chi\,p_a(\chi) 
    b_-^{(\lambda)}(\chi) f(x) j_\ell(x) 
    \int_0^\infty \rmd \chi'\,p_a(\chi')
    b_-^{(\lambda)}(\chi') g(x') j_\ell(x') \nonumber\\
    &\simeq 
    \frac{1}{4\pi} 
    \int_0^\infty \rmd \chi\, 
    \frac{p_a^2(\chi)}{\chi^2} 
    \left(b_-^{(\lambda)}(\chi)\right)^2 P_\phi\left(k=\frac{\ell}{\chi}\right) 
    f(\ell) g(\ell). 
\end{align}
For the second type involving derivatives, we perform integration by parts and define 
\begin{align}
    \calJ_-^{(\lambda)}[f,g] (\ell)
    &\equiv 
    \int_0^\infty \frac{k^2\rmd k}{2\pi^2} 
    P_\phi(k)
    \int_0^\infty \rmd \chi\,p_a(\chi)
    b_-^{(\lambda)}(\chi) f(x) j_\ell'(x) 
    \int_0^\infty \rmd \chi'\,p_a(\chi')
    b_-^{(\lambda)}(\chi') g(x') j_\ell(x') \nonumber\\
    &\simeq 
    \int_0^\infty \frac{k^2\rmd k}{2\pi^2} 
    P_\phi(k)
    \left(-\frac{1}{k}\right)
    \int_0^\infty \rmd \chi\,j_\ell(x) 
    \frac{\rmd }{\rmd \chi}\left[p_a(\chi) b_-^{(\lambda)}(\chi) f(x) \right]
    \int_0^\infty \rmd \chi'\,p_a(\chi') 
    b_-^{(\lambda)}(\chi') g(x') j_\ell(x') \nonumber\\
    &\simeq 
    \frac{3 - n_s}{2\ell} \calI_-^{(\lambda)}[f,g](\ell) - \calI_-^{(\lambda)}[f',g](\ell),
\end{align}
where in the first approximation we dropped the boundary term from integration by parts, and in the second we applied the Limber approximation, assumed weak redshift dependence of the bias, and used $P_\phi(k) \propto k^{n_s-4}$.

Using the above results, the contributions from each helicity mode to the high-$\ell$ limit of Eq.~\eqref{eq:angular_ps_II_EB_tomo} are given by: 
\begin{align}
    \left.C_{EB}^{aa}\right|^{(\lambda=1)} (\ell)
    &\simeq 
    A_\mathrm{gauge} 
    \pi
    \cdot 
    \frac{1}{\ell^4}
    \cdot 
    2 
    \cdot 
    \ell^2 
    \cdot 
    \left(
    \calJ_-^{(1)}\left[\frac{\ell^2}{x},\frac{\ell^2}{x}\right] (\ell) 
    + 
    \calI_-^{(1)}\left[\frac{\ell^2}{x},\frac{\ell^2}{x^2}\right] (\ell) 
    \right)
    = 
    A_\mathrm{gauge} \frac{7-n_s}{\ell} \calI_-^{(1)}[1,1], \\
    \left.C_{EB}^{aa}\right|^{(\lambda=2)} (\ell)
    &\simeq
    A_\mathrm{gauge} 
    \pi
    \cdot 
    \frac{1}{\ell^4}
    \cdot 
    \frac{1}{2} 
    \cdot 
    \ell^4 
    \cdot 
    \left(
    2\calJ_-^{(2)}\left[1,1\right] (\ell) 
    + 
    4\calI_-^{(2)}\left[1,\frac{1}{x}\right] (\ell) 
    \right)
    = 
    A_\mathrm{gauge} \frac{7-n_s}{2\ell} \calI_-^{(2)}[1,1], 
\end{align}
which yields the expression in \eq{eq:angular_ps_II_EB_binned_highl_lim} of the main text by substituting $n_s \simeq 1$ for simplicity.

\section{Angular integral of product of momenta}
\label{app:angular_integral_formula} 
We derive a useful formula of the angular integral of the product of momenta: 
\begin{align}
    \int_{\bq} \hq_{i_1} \cdots \hq_{i_\ell} f(\bk,\bq)
    = \sum_{k=0}^{[\ell/2]} a_{\ell,k} \delta_{(i_1 i_2} \dots \delta_{i_{2k-1} i_{2k}} \left\{ \hk_{i_{2k+1}} \cdots \hk_{i_{\ell})} \right\}^\mathrm{TF} \int_{\bq} \mathcal{L}_{\ell-2k}(\mu) f(\bk,\bq), 
    \label{eq:int_of_product}
\end{align}
for any isotropic function $f(\bk,\bq)=f(k,q,\mu)$ with $\mu\equiv\hbk\cdot\hbq$, where $a_{\ell,k}$ will be defined in Eq.~\eqref{eq:TF_decomp}. 
We denote the symmetric part of a tensor using parentheses around its indices, e.g. $x_{(i}y_{j)} = (x_iy_j+x_jy_i)/2$, and the trace-free part using curly brackets and/or a superscript ``$\mathrm{TF}$'', e.g., $\left\{ x_{(i}y_{j)}\right\}^\mathrm{TF} = (x_iy_j+x_jy_i)/2 - x_ky_k\delta_{ij}/3$. 

We begin with the formula for constructing a symmetric trace-free rank-$\ell$ tensor from a symmetric rank-$\ell$ tensor $T_{i_1 i_2 \cdots i_\ell}$ \citep[e.g.,][]{Thorne1980:Multipole_GW,Spencer1970:STF_decomp,Matsubara2024:IA_iPT_I}:
\begin{align}
    T_{i_1 i_2 \cdots i_\ell}^\mathrm{TF} = \sum_{k=0}^{[\ell/2]} \tilde{a}_{\ell,k} \delta_{(i_1 i_2} \dots \delta_{i_{2k-1} i_{2k}} T_{i_{2k+1} \cdots i_{\ell})j_1j_1 \cdots j_kj_k}, 
    \quad
    \mathrm{with} 
    \quad
    \tilde{a}_{\ell,k} = \frac{\ell!}{(2\ell-1)!!} \frac{(-1)^k (2\ell-2k-1)!!}{2^k k! (\ell-2k)!}.
    \label{eq:TF_construct}
\end{align}
Applying Eq.~\eqref{eq:TF_construct} recursively, we derive its inverse formula, i.e., the decomposition formula of $T_{i_1 i_2 \cdots i_\ell}$ into the sum of its trace-free parts: 
\begin{align}
    T_{i_1 i_2 \cdots i_\ell} = \sum_{k=0}^{[\ell/2]} a_{\ell,k} \delta_{(i_1 i_2} \dots \delta_{i_{2k-1} i_{2k}} T_{i_{2k+1} \cdots i_{\ell})j_1j_1 \cdots j_kj_k}^\mathrm{TF}, 
    \quad
    \mathrm{with} 
    \quad
    a_{\ell,k} = \frac{\ell!}{(2\ell-2k+1)!!} \frac{(2\ell-4k+1)!!}{2^kk!(\ell-2k)!}. 
    \label{eq:TF_decomp}
\end{align}
For $T_{i_1 i_2 \cdots i_\ell} \equiv \hq_{i_1} \cdots \hq_{i_\ell}$, Eq.~\eqref{eq:TF_decomp} reduces to 
\begin{align}
    \hq_{i_1} \cdots \hq_{i_\ell} = \sum_{k=0}^{[\ell/2]} a_{\ell,k} \delta_{(i_1 i_2} \dots \delta_{i_{2k-1} i_{2k}} \left\{\hq_{i_{2k+1}} \cdots \hq_{i_{\ell})}\right\}^\mathrm{TF}. 
    \label{eq:TF_decomp_ps}
\end{align}
Next, we recall the expansion of the Legendre polynomials in terms of the symmetric trace-free tensors \citep{Thorne1980:Multipole_GW}: 
\begin{align}
    \mathcal{L}_\ell(\hbk \cdot \hbq) = \frac{(2\ell-1)!!}{\ell!} 
    \hk_{i_1} \cdots \hk_{i_\ell} 
    \left\{ \hq_{i_1} \cdots \hq_{i_\ell} \right\}^\mathrm{TF}. 
    \label{eq:legendre_TF}
\end{align}
Using Eq.~\eqref{eq:legendre_TF}, we derive the formula for the angular integral of \textit{trace-free part} of the product of momenta:
\begin{align}
    \int_{\bq} \left\{ \hq_{i_1} \cdots \hq_{i_\ell} \right\}^\mathrm{TF} f(\bk,\bq)
    = \left\{ \hk_{i_1} \cdots \hk_{i_\ell} \right\}^\mathrm{TF} 
    \int_{\bq} \mathcal{L}_{\ell}(\mu) f(\bk,\bq), 
    \label{eq:int_of_TF_product}
\end{align}
where we have used the symmetric trace-free condition on the left-hand side to restrict possible tensors on the right-hand side to $\left\{ \hk_{i_1} \cdots \hk_{i_\ell} \right\}^\mathrm{TF}$ alone, and $\mathcal{L}_\ell (1)=1$ for any $\ell$ to match the normalization.
Substituting \eq{eq:int_of_TF_product} into \eq{eq:TF_decomp_ps}, we obtain \eq{eq:int_of_product}.

In the following, we present some examples of angular integrals. 
We introduce a shorthand notation for the multipole moment of the isotropic function $f(\bk,\bq)$: 
\begin{align}
    F_\ell(k) \equiv \int_{\bq} \mathcal{L}_\ell(\mu) f(\bk,\bq). 
\end{align}

$\ell=0$:
\begin{align}
    \int_{\bq} f = F_0. 
\end{align}

$\ell=1$:
\begin{align}
    \int_{\bq} \hq_{i} f = \hk_{i} F_1. 
\end{align}

$\ell=2$:
\begin{align}
    \hq_{i}\hq_{j} 
    &= \left\{\hq_{i}\hq_{j}\right\}^\mathrm{TF} + \frac{1}{3}\delta_{i j},\\
    \int_{\bq} \hq_{i}\hq_{j} f 
    &= \left\{\hk_{i}\hk_{j}\right\}^\mathrm{TF} F_2 
    + \frac{1}{3}\delta_{i j} F_0. 
\end{align}

$\ell=3$:
\begin{align}
    \hq_{i}\hq_{j}\hq_{k} 
    &= \left\{\hq_{i}\hq_{j}\hq_{k}\right\}^\mathrm{TF} + \frac{1}{5} \left(\delta_{i j}\hq_{k} + 2\mathrm{~perms}\right), \\
    \int_{\bq} \hq_{i}\hq_{j}\hq_{k} f 
    &= \left\{\hk_{i}\hk_{j}\hk_{k}\right\}^\mathrm{TF} F_3
    + \frac{1}{5} \left(\delta_{i j}\hk_{k} + 2\mathrm{~perms}\right) F_1. 
\end{align}

$\ell=4$:
\begin{align}
    \hq_{i}\hq_{j}\hq_{k}\hq_{l} 
    &= \left\{\hq_{i}\hq_{j}\hq_{k}\hq_{l}\right\}^\mathrm{TF} 
    + \frac{1}{7} \left(\delta_{i j}\left\{\hq_{k}\hq_{l}\right\}^\mathrm{TF}  + 5\mathrm{~perms}\right) 
    + \frac{1}{15} \left(\delta_{i j}\delta_{k l} + 2\mathrm{~perms}\right), \\
    \int_{\bq} \hq_{i}\hq_{j}\hq_{k}\hq_{l} f
    &= \left\{\hk_{i}\hk_{j}\hk_{k}\hk_{l}\right\}^\mathrm{TF} F_4 
    + \frac{1}{7} \left(\delta_{i j}\left\{\hk_{k}\hk_{l}\right\}^\mathrm{TF} + 5\mathrm{~perms}\right) F_2 
    + \frac{1}{15} \left(\delta_{i j}\delta_{k l} + 2\mathrm{~perms}\right) F_0. 
\end{align}

\bibliography{main} 
\end{document}